%% file: main.tex
\documentclass[11pt,a4paper]{article}

\usepackage[margin=1in]{geometry}
\usepackage{specform}

\hypersetup{
  pdftitle  = {Does the way we write a theory change the program an LLM builds
               from it?},
  pdfsubject= {A prospective randomized study of renderer format in LLM
               theory-to-program translation},
  pdfkeywords = {preregistration, randomized experiment, large language models,
                 autoformalization, computational modelling, exact randomization
                 inference}
}

\graphicspath{{figures/}{./}}

\begin{document}

\input{sections/00-frontmatter}

\input{sections/01-introduction}
\input{sections/02-methods}
\input{sections/03-analysis-plan}
\input{sections/04a-results-primary}
\input{sections/04b-results-secondary}
\input{sections/05-discussion}

\clearpage
\bibliographystyle{plainnat}
\bibliography{references}

\clearpage
\appendix

\begingroup
\setcounter{section}{0}
\renewcommand{\thesection}{S\arabic{section}}
\renewcommand{\thetable}{S\arabic{table}}
\renewcommand{\thefigure}{S\arabic{figure}}
\setcounter{table}{0}
\setcounter{figure}{0}

\section*{Supplementary material}
\addcontentsline{toc}{section}{Supplementary material}

\noindent
The supplement carries the material that would have made the paper
unreadable: the full definitions of the preregistered support criteria, every
endpoint table at full precision, the registered secondary analyses, and the
robustness work. Nothing in it is new analysis. Every number in it appears in
the analysis artifacts released with this package.

\input{supplement/sections/S1-design-detail}

\input{supplement/sections/S2-support-criteria}

\input{supplement/sections/S3-endpoint-tables}

\input{supplement/sections/S4-secondary-analyses}

\input{supplement/sections/S5-robustness}

\endgroup

\clearpage
\begingroup
\setcounter{section}{0}
\renewcommand{\thesection}{\Alph{section}}
\counterwithin{table}{section}
\counterwithin{figure}{section}
\renewcommand{\thetable}{\Alph{section}\arabic{table}}
\renewcommand{\thefigure}{\Alph{section}\arabic{figure}}
\setcounter{table}{0}
\setcounter{figure}{0}

\section*{Annexes}
\addcontentsline{toc}{section}{Annexes}

\noindent
The annexes are the audit trail. Annex~A is the preregistration exactly as it
was sealed. Annex~B is the provenance chain and the procedure for checking it
against records the authors do not control. Annex~C is the complete record of
every incident and every deviation. Annex~D is the data descriptor for the two
released datasets. Annex~E is the reproduction guide. These are the only parts
of the package where file names, paths and digests appear, because these are
the only parts where they are useful.

\input{annex/A-preregistration-verbatim/annex-a-note}
\input{annex/B-provenance-and-attestation/annex-b-provenance}

\input{annex/C-incidents-and-deviations/annex-c-incidents}

\input{annex/D-data-descriptor/annex-d-data-descriptor}

\input{annex/E-reproduction/annex-e-reproduction}

\endgroup

\end{document}

%% file: sections/00-frontmatter.tex

\title{Does the way we write a theory change the program an LLM builds from
it?\\
\large A prospective randomized study of renderer format in LLM
theory-to-program translation}

\author{Andre Panossian\\{}
American University of Beirut\\
\texttt{avp01@mail.aub.edu}}

\date{2026-08-09}

\maketitle

\begin{abstract}
\noindent\textbf{Background.} A verbal theory does not run. To turn one into a
program, an implementer must decide which variables matter, which outputs
should change, and which effects should interact. Those hidden choices can
produce different programs even when two implementers start from the same
words. A language-model call makes the problem measurable: one anonymous verbal
account goes in, one executable program comes out.

\noindent\textbf{Question.} When the words carrying the theoretical content are
held fixed, does presenting them as explicit hold-change-direction instructions
make two LLM snapshots translate them into more similar executable behavior
than presenting them as \prose{}?

\noindent\textbf{Design.} \studyLong{} is a prospective randomized experiment
whose treatment-bearing unit is an anonymous renderer slot; 32 slots sit in
\nblocks{} blocks. Sixteen independent fair bits, drawn from one exact future
NIST Randomness Beacon 2.0 pulse, assign a complete complementary
renderer-to-slot mapping within each block: one slot receives an \contract{}
(\holdfixed{} / \intervene{} / \reqdir{}), the other receives connected
conditional prose. Both arms of a block reuse byte-identical proposition
strings and a shared adapter. Two pinned snapshots (\snapshotA{},
\snapshotB{}) translate each of five anonymous account axes, for exactly
\ncalls{} preauthorized single-shot API requests. Each call returns strict JSON
describing 11 equations in a frozen sparse quadratic language of 823 legal
monomials.

\noindent\textbf{Analysis.} A deterministic evaluator, with no LLM in the loop,
perturbs registered inputs and records how the 11 outputs change: one field for
\Honefull{}, two fields for \Htwofull{}. Two primary endpoints are registered
--- direct matched-distance reduction \Mstat{} between same-account
cross-model programs, and closed-set same-account identifiability \Gstat{} over
40 per-block comparisons --- each with unsigned, row-free and broken-alignment
controls. The finite-experiment estimand is the average over the \nblocks{}
fixed blocks of each block's randomization average of its two
assignment-specific potential contrasts; the estimator is the mean of the
\nblocks{} observed block contrasts, unbiased over the sixteen fair bits.
Inference is an exact one-sided test enumerating all \nperm{} renderer-label
assignments under the Fisher sharp null of renderer irrelevance at every slot.
Support requires all 27 preregistered support criteria to pass in both a signed
\linpipe{} and a sign-preserving magnitude-rank pipeline, for both \Hone{} and
\Htwo{}.

\noindent\textbf{Results.} Both registered stages ran to completion on all
\ncalls{} samples. \Hone{} returns the registered verdict \notsupported{},
\Htwo{} returns the registered verdict \notsupported{}, and the preregistered
conjunction of the two returns \notsupported{}. Thirteen validity gates run per
stage: \Hone{} passes 13 and \Htwo{} passes 12, the single gate \Htwo{} does
not pass being the registered single-execution-attempt counter, which records
that three \Htwo{} direction-runs were repeated after a local subprocess
verification of stored study-level attestation files timed out. That counter is
not a data check; every check that bears on the data passes in both stages.
Of 108 scientific criterion evaluations, 19 passed.
\Hone{} passes 4 and 3 of the 27 criteria across the two pipelines; \Htwo{}
passes 6 and 6. The conjunction fails by a wide margin rather than a narrow
one. Closed-set same-account identifiability sat at chance in all four cells,
between \auc{0.469} and \auc{0.523} against a registered \auc{0.80} threshold,
so a model's two renderings of an account were never mutually recognizable; the
smallest exact \pv{} anywhere in \Hone{}, across ten tests, is
\pval{0.0593}. One endpoint moved. \Htwo{}'s matched-distance reduction was
positive and nominally significant in both pipelines, and in the \linpipe{} it
was the one test in the study to clear the Bonferroni-corrected threshold,
surviving deletion of its largest block. Under the \rankpipe{} the same
contrast is an order of magnitude smaller and misses the registered magnitude
floor, so even this endpoint does not meet the registered support rule. The
conjunction fails on 89 of the 108 evaluations, across both stages, not on this
one criterion. Nothing registered in advance says which transform is right
about the size.

\noindent\textbf{Conclusion.} Renderer format did not produce the uniform,
family-invariant, classifiable geometry the registered theory predicted.
\Hone{} is a completed registered stage carrying the registered verdict
\notsupported{}, and it is a null. \Htwo{} is a completed registered stage
carrying the registered verdict \notsupported{}, and it is a null with one
endpoint inside it that moved, reported at the size it came out, with the
instrument that disagrees about that size printed beside it. The two verdicts
have the same standing and were reached under the same registered rule. No test
comparing the two stages was registered or run,
so nothing here establishes that the picture differs between single-input and
interaction challenges: \Hone{} and \Htwo{} are two hypotheses reported side by
side, not a contrast.
\end{abstract}

\noindent\textbf{Keywords.} preregistration; randomized experiment; large
language models; autoformalization; theory formalization; specification format;
exact randomization inference; null result.

\medskip

\noindent\textbf{Preregistration and data.} The introduction, the methods and
the full analysis and interpretation plan were compiled on 2026-08-04, sealed
with an RFC 3161 timestamp and three OpenTimestamps calendar commitments, and
so fixed before the assignment randomness existed. \annexref{A} reproduces that
sealed text byte for byte, unaltered, including the three statements execution
has since overtaken. The two datasets are released as \dsone{} and \dstwo{}:
two complete registered corpora, \ncalls{} registered samples in each, all of
them completed, every response tensor matching its sidecar digest with none
missing. Both are usable on the same terms and neither is exploratory; any
\texttt{INVALID} or post-hoc provenance string a reuser meets in the bound
historical artifacts is a driver field of the kind
\Cref{tab:status-resolution} resolves, not a judgement on the data. Their
schema, variable dictionary and reuse terms are in \annexref{D}.
The combined dataset deposit is archived on Zenodo
\citep{panossian2026specformdata}; the companion reproduction software is
archived separately \citep{panossian2026specformsoftware}.

\medskip

\noindent\textbf{Status of the two stages.} Both stages are registered, both
completed, and both returned \notsupported{}. \Htwo{}'s artifacts additionally
carry a driver field whose mechanically-written value reads
\texttt{INVALID}. That field is an attempt counter, not a data check and not a
scientific verdict, and \Cref{tab:status-resolution} separates the two readings
once so that the rest of this paper need not.

\begin{table}[htbp]
\centering
\caption[Status resolution for \Htwo{}]{Status resolution for \Htwo{}. The
first row is a provenance field written by the execution driver. The third row
is this paper's scientific verdict. They are different statements about
different things, and only the third is a finding.}
\label{tab:status-resolution}
\small
\begin{tabularx}{\linewidth}{@{}lX@{}}
\toprule
\textbf{Layer} & \textbf{\Htwo{} status} \\
\midrule
Raw driver field & \texttt{INVALID}, because the single-attempt counter
failed \\
Data integrity & \textbf{PASS}; \ncalls{}/\ncalls{} complete, zero tensor
mismatches \\
Registered scientific decision & \textbf{\notsupported{}} \\
Publication-level usability & \textbf{Valid completed \Htwo{} dataset} \\
\Hone{}-vs-\Htwo{} comparison & Not registered; do not pool or contrast as if
paired \\
\bottomrule
\end{tabularx}
\end{table}

\noindent Thirteen validity gates run per stage. \Hone{} passes 13 of 13;
\Htwo{} passes 12 of 13. The single gate \Htwo{} does not pass is
\texttt{stage\_all\_320\_valid\_single\_execution\_attempt}, a conjunction
whose two conjuncts split: \texttt{all\_320\_valid} is \emph{true} and
\texttt{single\_execution\_attempt} is \emph{false}. Three direction-runs, in
three different samples, timed out in a local \texttt{openssl ts -verify}
invocation --- a check of stored study-level attestation files that runs before
any sample-specific input is read, opens no socket, and has no model in the
loop. It was not a provider, model, data or network failure. Only those three
direction-runs were repeated; all \ncalls{} \Htwo{} samples completed, every
gate that bears on the data passes, and re-hashing the stage's tensor corpus
gave \num{1280} digests with 0 mismatches. \annexref{C} carries the full
record. Neither stage is a replication, extension or conceptual generalization
of the other, no \Hone{}-versus-\Htwo{} contrast was registered, and none was
run.

%% file: sections/01-introduction.tex
\section{Introduction}
\label{sec:introduction}

A verbal theory does not run. To turn one into a program, an implementer must
decide which variables matter, which outputs should change, and which effects
should interact. Psychological theories are routinely stated at a level that
does not uniquely determine any of those choices. \citet{guest2021} argue that
computational modeling is valuable partly because implementation forces
assumptions into the open, and \citet{oberauer2019} distinguish broad organizing
frameworks from theories capable of risky quantitative prediction.
Formalization reveals ambiguity. It does not guarantee that two implementers
resolve that ambiguity the same way.

A large language model makes the problem measurable. Treat the model call as a
stochastic translation step: one anonymous verbal account goes in, one
executable program comes out. If two model snapshots receive the same semantic
statements and repeatedly build different response functions, the prose has not
constrained this translation apparatus tightly enough to produce cross-model
agreement. \studyLong{} asks whether one presentation method narrows that
implementation freedom. It does not ask whether a language model is a mind, or
whether a generated program is a patient simulation.

\subsection{The question, and what would count as an improvement}
\label{sec:introduction:question}

The causal question is narrow.

\begin{quote}
When the words carrying the theoretical content are held fixed, does presenting
them as explicit hold-change-direction instructions make two language-model
snapshots translate them into more similar executable behavior than presenting
them as connected prose?
\end{quote}

The two formats are \prose{}, meaning connected conditional sentences, and an
\contract{}, whose lines carry the labels \holdfixed{}, \intervene{}, and
\reqdir{}. The manipulation is the complete renderer package---labels,
capitalization, bullets, and layout change together---so the study cannot
identify which ingredient caused an effect. A \emph{renderer} is the fixed rule
that presents the same proposition strings in one of the two formats; a
\emph{program} is the strict set of equations returned by one model call;
\emph{response geometry} is the pattern of output changes produced when the
evaluator changes registered inputs. \Honefull{} tests one-input changes,
\Htwofull{} tests whether the effect of changing one input depends on a second
input, and each is a stage, meaning one of the two registered hypothesis stages.

Two questions are kept separate. The first is direct same-account closeness:
under the \contract{} renderer, is one model's program for a given account card
actually closer to the other model's program for that same card than it is under
prose? That is matched-distance reduction \Mstat{}. The second is relative
identifiability: is each account closer to its same-account partner than to the
four other accounts? That is closed-set same-account identifiability \Gstat{}, a
40-comparison score. The first does not use foils; the second does. A renderer
could improve relative identifiability simply by moving all five accounts
farther apart while making the same-account pairs no closer. That cannot support
the trial, because both measures, their exact tests, and their heterogeneity
criteria must pass.

This is not a label-recovery task for the generating model. Each call sees one
anonymous account and nothing else: no account name, no competing account, no
target program, no probe, no scoring rule, no output from another call. Labels
are used only by the frozen evaluator, after a blinded score artifact has been
committed.

\subsection{Closest work and the narrow gap}
\label{sec:introduction:related}

This is not the first natural-language-to-program system, the first
autoformalization study, or the first attempt to automate cognitive modeling.
\citet{wu2022} demonstrated autoformalization from informal mathematics into a
formal language, and \citet{hahn2022} fine-tuned language models to translate
English into regular expressions, first-order logic, and linear-time temporal
logic. Later work improves that translation with syntax feedback, retrieval,
denoising, and correction \citep{zhang2024}, with symbolic verification of
generated program specifications \citep{wen2026}, and by repairing ambiguous
problem descriptions diagnosed through the distribution of programs they induce
\citep{jia2025}. Closer to this domain, \citet{deng2026} translate descriptions
of cognitive decision rules into temporal logic and executable production rules,
and \citet{rmus2025} use language models to identify or generate cognitive
models from behavioral data. Those systems address translation, correction,
verification, or model recovery. \studyLong{} is narrower: it randomizes one
matched presentation contrast while holding the semantic proposition strings
fixed, then compares the resulting programs by executing them on unseen numeric
interventions. It neither repairs the prompts nor optimizes a program against
behavioral data.

Prompt sensitivity is the immediate alternative explanation. \citet{hua2025}
showed that apparent sensitivity can be inflated by rigid answer extraction and
other evaluation artifacts, \citet{liu2026} analyzed why meaning-preserving
prompt templates can still produce different model behavior, and
\citet{xie2026} found that lexical choice alone, holding the task fixed,
measurably shifts output quality. That last result is the sharpest threat here:
the two renderers share the registered proposition strings but not the
surrounding template words, so any observed contrast is compatible with a purely
lexical effect. No output difference is therefore treated as evidence of deeper
formalization quality. The response is methodological---a strict output language
shared across arms, no regular-expression answer matcher and no model judge,
execution on frozen numeric probes, and unsigned, row-free, and
broken-alignment controls. That does not eliminate every prompt effect. It makes
the exact renderer effect and its endpoint explicit.

The design also borrows from work on interventions. Interventions separate
causal structures that observational data leaves equivalent, and
\citet{hauser2012} characterize the equivalence classes that interventional data
collapses; in model comparison, \citet{myung2009} show that experiments can be
chosen to maximally discriminate rival models. \studyLong{} borrows that logic
but not its formal guarantees: its programs are not established causal models of
people, and its probes are fixed evaluation inputs rather than experiments on
human participants. \citet{liu2025} found that explicit conditional or
program-like structure can improve causal-reasoning performance; this design
differs by randomizing two complete renderers and measuring executed cross-model
response similarity rather than answer accuracy.

\subsection{What is new, what it cannot show, and what came out}
\label{sec:introduction:contribution}

\studyLong{} treats a verbal account as input to a stochastic compiler rather
than as a finished computational model, and compares programs functionally, by
how their outputs change when the same registered inputs are manipulated. The
registered claim is this.

\begin{quote}
In a fixed theory-to-program apparatus, presenting identical proposition strings
as explicit hold-change-direction contracts can reduce direct cross-model
distance between same-account response functions and improve their relative
identifiability compared with connected prose.
\end{quote}

The term \emph{intervention-response similarity} is used for measured similarity
under the frozen probes; no mathematical equivalence class is claimed. The
novelty claim is limited to this prospective formalization methodology, and
claims no priority over general prompt-sensitivity experiments,
autoformalization, cognitive-model automation, causal discovery, or optimal
model discrimination. The limits are equally definite. A supported result would
be a randomized finding about a method for formalizing theory with a language
model. It would not prove that either program is faithful to the literature,
reveal a human cognitive architecture, validate a psychiatric theory, or
establish a universal benefit of structured prompts.

Both registered stages completed on all \ncalls{} of their samples. \Hone{}
carries the registered verdict \notsupported{}, \Htwo{} carries the registered
verdict \notsupported{}, and the registered conjunction of the two also returns
\notsupported{}. Support required all 27 preregistered support criteria to pass
in both the signed \linpipe{} and the sign-preserving \rankpipe{}, in both
stages. \Hone{} passes 4 and 3 of 27; \Htwo{} passes 6 and 6. The conjunction
fails by a wide margin rather than a narrow one. Two results are worth reading
for beyond that headline. Closed-set same-account identifiability sits at chance
in all four cells, at \auc{0.469} to \auc{0.523} against a registered
\auc{0.80} threshold: the renderer contrast does not make a model's two
renderings mutually recognizable. And \Htwo{}'s matched-distance reduction is
positive and nominally significant under both transforms, but only under the
\linpipe{} does it clear the Bonferroni-corrected threshold of 0.0025 --- the
one test in the study that does --- while under the \rankpipe{} the same
contrast is an order of magnitude smaller, reaches only \pval{0.0284729}, and
misses the registered magnitude floor. Nothing registered in advance says which
transform is right about the size. That is one cell of a 20-test matrix inside
a stage whose registered verdict is \notsupported{}. No \Hone{}-versus-\Htwo{}
contrast was registered, no estimator for one exists, and none was run: the two
stages are reported side by side, not as a comparison.

That ordering---design first, data second---is checkable rather than asserted.
The question, the two renderers, the 27 criteria, the controls, the estimand,
the exact test, and the permitted claim language were compiled on 2026-08-04 and
sealed with an RFC 3161 timestamp and three OpenTimestamps calendar commitments
before the assignment randomness existed. Only then were the \nblocks{}
assignment bits drawn from one exact future NIST Randomness Beacon 2.0 pulse.
The study runs once. \studyname{} is a new prospective study following an
outcome-blind instrument failure in an earlier version of the apparatus;
endpoint defects found by independent review were corrected before the seal, and
those corrections and every later deviation are disclosed rather than hidden.
Where the sealed text is now false---it describes itself as pre-execution---it
is contradicted beside it rather than revised: a preregistration edited to match
its outcome is worth nothing.

\paragraph{Contributions.}

\begin{itemize}
  \item A randomized design for the question, stronger than comparing two prompt
        examples: assignment drawn independently in \nblocks{} blocks, two
        pinned model snapshots translating every account card, two
        independently written wordings covering all four cross-model wording
        relations, and executable behavior on unseen numeric challenges as the
        outcome.
  \item Two primary endpoints that cannot stand in for each other---matched
        distance must fall, not merely an account's position relative to
        foils---each with unsigned, row-free, and broken-alignment controls, and
        each required to hold in both a raw signed \linpipe{} and a
        sign-preserving magnitude-\rankpipe{}.
  \item Exact rather than asymptotic inference: a one-sided randomization test
        enumerating all \nperm{} renderer-label assignments under the Fisher
        sharp null of renderer irrelevance at every slot.
  \item A reported negative. The conjunction fails, the identifiability endpoint
        is flat at chance, and the two registered pipelines disagree about the
        magnitude of the one contrast that moved.
  \item Two released datasets, \dsone{} and \dstwo{}, and an attestation chain
        that lets a reader establish the order of design and data without
        trusting the authors.
\end{itemize}

\Cref{sec:methods} gives the design, from the randomized unit and the public
assignment draw through the program language, the probe battery, the estimand
and exact test, the blinding procedure, and the 27 criteria.
\Cref{sec:results} reports execution, validity, the conjunction, the two primary
endpoints, the negative controls, block and family structure, and the registered
secondary analyses. \Cref{sec:discussion} states what the outcome does and does
not license, and the limitations. The supplement carries the
criterion-by-criterion definitions and results, the null distributions, the
per-block and per-family tables, and the robustness and control distributions.
Five annexes carry the byte-exact sealed preregistration \seeannex{A}, the
randomness source, timestamps and digests \seeannex{B}, the incidents and
deviations \seeannex{C}, the data descriptor and reuse guidance \seeannex{D},
and the reproduction instructions \seeannex{E}.


%% file: sections/02-methods.tex
\section{Methods}
\label{sec:methods}

\subsection{Design and randomization}
\label{sec:methods:design}

The treatment-bearing unit is an \emph{anonymous renderer slot}. Each block
contains two slots, so the experiment contains 32 slots in 16 blocks. A slot is
not a single call: it collects the 10 calls ($2$ model snapshots $\times$ $5$
account axes) that share that block and that slot label.

The independently randomized unit is the block, not the slot. The design draws
\nblocks{} independent fair bits, one per block. A block's bit fixes a complete,
complementary renderer-to-slot mapping for that block: exactly one of its two
slots carries the \contract{} and the other carries \prose{}. The two slots
inside a block are therefore never independently randomized; they are a matched
pair under one shared coin. Calls, accounts, probes, anchors, outputs and
shuffle controls are repeated measurements inside a slot and are never treated
as independent randomized observations.

Unqualified, a \emph{block} is one of the \nblocks{} randomized blocks just
defined. The probe bank separately divides its anchors into \emph{calibration
blocks} and \emph{probe blocks}, which are groupings of frozen probe content,
are never randomized and carry no fair bit; those two terms always appear with
their qualifier. Every randomized block presents the entire probe bank, so the
two partitions are orthogonal rather than nested.

Write $Y_{b,s}(\mathrm{contract})$ and $Y_{b,s}(\mathrm{prose})$ for the two
potential response tensors of slot $s$ in block $b$. Only one of the two is ever
observed. The registered analysis requires two conditions on these potential
outcomes.

\begin{itemize}
  \item \textbf{No interference between slots.} A slot's potential response
    tensor depends only on the renderer assigned to that slot, not on the
    assignment of any other slot or block. Each of the \ncalls{} preauthorized
    requests is an independent stateless call, issued at most once, with no
    shared conversation state and no cross-request memory.
  \item \textbf{One version of each renderer.} Each renderer is a frozen
    template applied to frozen propositions, byte-identical in every block, so
    \contract{} and \prose{} name exactly one intervention each.
\end{itemize}

Both conditions are assumptions about the apparatus, not results. Neither is
testable from the collected data, and both are stated here so that a reader can
reject the causal reading if they reject the assumptions.

\begin{table}[t]
  \caption{The design at a glance. The full experiment contains exactly
    $16 \times 2 \times 2 \times 5 = 320$ preauthorized model requests. Provider
    exchanges are three per call: the initial request, the retrieval of the
    completed response, and the retrieval of the provider's own record of the
    request's input items.}
  \label{tab:design-summary}
  \centering
  \begin{tabular}{l S[table-format=3]}
    \toprule
    Component & {Count} \\
    \midrule
    Randomized blocks (one fair bit each)        & 16  \\
    Anonymous renderer slots (2 per block)       & 32  \\
    Pinned model snapshots                       & 2   \\
    Anonymous account axes (engineered cards)    & 5   \\
    Calls per block                              & 20  \\
    Preauthorized model requests                 & 320 \\
    Provider exchanges                           & 960 \\
    \bottomrule
  \end{tabular}
\end{table}

\Cref{tab:design-summary} summarises the design. The experiment driver may issue
at most one request for each envelope; provider-side execution count cannot be
independently proven.

\paragraph{Public random assignment.}
Renderer assignment is unknowable when the study is committed. The complete
apparatus, runtime, documents, tests, criteria and failure rules are written to
a preregistration seal. That seal is timestamped before one exact future
NIST Randomness Beacon 2.0 pulse, fixed 120 one-minute pulses after the observed
anchor. One durable assignment-attempt marker then starts a six-hour
availability window in which the monitor may send only bodyless \texttt{HEAD}
requests to the exact target URI; redirects are rejected. After the first
\texttt{HEAD 200} the target is fetched exactly once with \texttt{GET} and its
raw bytes are persisted before parsing. The target pulse, its predecessor and
the Beacon certificate are verified against the pinned certificate chain, the
RSA-4096/SHA-512 signature, the output hash, the predecessor link and the
predecessor commitment. The assignment digest is a SHA-256 over a domain
separator, the digest bytes of the preregistration seal and the output-value
bytes of the target pulse. Its first \nblocks{} bits, most-significant bit
first, assign the \contract{} renderer to neutral slot \texttt{U} or \texttt{V}
in each block; the remaining 240 disjoint bits generate opaque identifiers,
adapters, anonymous axis order and execution order. A timeout, failure of the
sole target fetch, or a cryptographic failure stops \studylong{}. There is no
alternate pulse, seed, target retry, balancing rule or reroll. The randomness
source, the verification chain and the integrity and stopping rules are recorded
in \annexref{B}; the three abandoned sealing episodes that preceded the executed
seal are recorded there and in \annexref{C}.

\subsection{Materials}
\label{sec:methods:materials}

\paragraph{Two pinned model snapshots.}
The two snapshots are \snapshotA{} and \snapshotB{}. They are fixed comparison
cases, not a random sample of language models. Every scientific generation call
uses temperature $1.0$, $\mathrm{top\_p} = 1.0$, zero frequency and presence
penalties, medium verbosity, and the remaining explicit response controls bound
into the request contract.

\paragraph{Five engineered account cards.}
Each card contains five proposition objects, and the generation model never sees
the account names. Source monitoring maps the attribution of memories, knowledge
and beliefs to their origins onto four source-output coordinates and public
origin cues \citep{johnson1993}. Comparator abstracts intended, predicted and
actual states into a bounded forecast-evidence transaction, and is not a
motor-control simulation \citep{frith2000}. Precision weighting uses textual
reliability, parse confidence, dependency and source uncertainty as observable
proxies, because the apparatus has no direct measure of expected precision
\citep{adams2013}. Circular inference treats excessive reuse through recurrent
dependence as its literature-core relationship, and is not a neural simulation
\citep{jardri2013,jardri2017}. Decision threshold expresses evidence
accumulation toward a boundary as confidence and continuation-value features
driving a stopping-pressure output, with no choice or reaction-time data fitted
\citep{ratcliff2008,gold2007}.

The cards are engineered from the literature rather than drawn from patients.
This is not a study of patients, diagnoses, brains, consciousness, or the truth
of the five source theories. Every one of the 25 propositions is classified in a
separate formalization audit as a literature core, an operational bridge, or a
designed discriminator, and only literature cores are attributed to the cited
authors; the executable inventory contains 2 literature-core propositions,
14 operational bridges and 9 designed discriminators. Card provenance and the
proposition-by-proposition audit are in \suppref{1}.

\paragraph{Two matched renderers.}
Within a block, model side and account, both prompts reuse the same proposition
strings, byte for byte, and the same feature, monomial, output and account
adapter. \Prose{} renders each proposition as one sentence of the form
\texttt{With [hold], if we [change], then [response].} The \contract{} renders
the same strings as three labelled lines, \holdfixed{}, \intervene{} and
\reqdir{}, each carrying the corresponding string. The prompt-word imbalance
must remain at or below 3\%. Any supported claim applies only to this complete
pair of renderers.

Each account has two independently written versions, \texttt{R1} and
\texttt{R2}. Four blocks use each cross-model wording relation:
\texttt{R1-R1}, \texttt{R2-R2}, \texttt{R1-R2} and \texttt{R2-R1}. Identical
wording is therefore unnecessary, and a pooled effect cannot hide a reversed
wording relation.

\paragraph{Frozen program language.}
Each call returns strict JSON describing 11 equations. It cannot return Python
or free prose. The language exposes 40 bounded features, each finite and inside
$[-1, 1]$, drawn from current state, current evidence and proposed-action
context. The compiler pre-enumerates every linear feature and every legal
unordered quadratic product, removing indicator squares and products between
mutually exclusive indicators; the resulting language contains exactly
823 legal monomials. Each equation has one integer bias and exactly six term
records, each selecting one opaque monomial or the reserved neutral monomial and
an integer coefficient from $-20$ to $20$. The six records are an unordered
additive multiset. Each program defines 11 outputs. Invalid JSON, an incorrect
request echo, a wrong account key, missing or extra outputs, unknown terms,
malformed coefficients, or a seventh term invalidates the call, and compilation
never repairs, substitutes, clips, or asks for another response. The full
monomial grammar, the input features, the equation and compiler semantics, the
output list and the 160 pair-specific adapters are in \suppref{1}.

\subsection{Measurement}
\label{sec:methods:measurement}

All probes are deterministic and frozen before generation. Confirmatory
interventions alter state or evidence only; proposed actions remain fixed
context. Each generated program is treated as a small machine: the evaluator
changes fields and records how the 11 outputs change. No language model grades a
program at any point.

A \emph{single-input challenge} (\Hone{}) changes one field. Twelve probe blocks
measure central finite differences for ten fields, two designated per account.
An \emph{interaction challenge} (\Htwo{}) changes two fields at once. Twelve
separate probe blocks measure mixed finite differences for ten pairs, again two
designated per account, and they run only after \Hone{} execution, analysis,
sealing and independent timestamp verification. The interaction challenges are
study-created numeric challenges. They are not claimed to be entailed by the
prose cards or by the cited literature, and they are not evidence of conceptual
generalization. The common calibration reference, the pooled percentile that
defines the \rankpipe{}, and the balanced anchors and cue-relative source frame
are in \suppref{1}.

The four source outputs are projected to sum to zero, so they contain three
independent coordinates. The evaluator compares ten independent coordinates in
eight equally weighted groups: belief; forecast; source attribution; memory;
dependency; continuation; update gain; and stopping boundary. For any program
from one model side and any program from the other, signed response distance is
the average over the eight groups of one half the sum of squared row-by-row
response differences. No account-wise centering and no unit normalization are
applied. Smaller distance means more similar responses to the same controlled
inputs.

\paragraph{First primary endpoint: matched-distance reduction \texorpdfstring{\Mstat{}}{M}.}
The five same-account comparisons are the diagonal of a $5 \times 5$ distance
matrix. Let $d_{\mathrm{prose}}$ and $d_{\mathrm{contract}}$ denote a block's
matched distance under each renderer. For each block,
\begin{equation}
  \label{eq:mstat}
  \Mstat = \frac{2\,(d_{\mathrm{prose}} - d_{\mathrm{contract}})}
                {d_{\mathrm{prose}} + d_{\mathrm{contract}}}.
\end{equation}
\Mstat{} ranges from $-2$ to $2$. Positive values mean the \contract{} renderer
produced lower direct same-account distance. The denominator must exceed the
frozen floor in every block and, separately, in each of the 80
block-by-account cells that feed the account-specific reductions; a floored cell
contributes a substituted zero, every substitution is counted and disclosed, and
any substitution fails the denominator criterion.

\paragraph{Second primary endpoint: closed-set same-account identifiability \texorpdfstring{\Gstat{}}{G}.}
Each of the five diagonal cells is compared with four foils in its row and four
in its column, yielding 40 strict comparisons per block. \Gstat{} is the
proportion won by the same-account match. A tie is not a win, the smallest step
is $1/40 = 0.025$, and an arm effect of $0.15$ is six net comparisons.
\Gstat{} is a closed-set relative-identifiability score. It is not direct
agreement, population accuracy, or evidence of literature fidelity.

\paragraph{Registered negative controls.}
The registered verdict also asks what information produced any \Gstat{}
advantage, through three controls. \emph{Unsigned magnitude} replaces every
response with its absolute value, giving
$D_{\mathrm{directional}}$, which requires a signed advantage beyond output
support or magnitude. The \emph{row-free signature} finds the best reassignment
of complete response rows before computing identifiability, giving
$D_{\mathrm{signature}}$, which requires an advantage beyond an unordered bag of
response rows. \emph{Broken alignment} independently permutes complete rows on
both model sides 512 times, separately for each arm within each randomized
block; an arm's \emph{conditional alignment} is its signed \Gstat{} minus the
mean of its own shuffled \Gstat{} values, and $D_{\mathrm{conditional}}$ is the
\contract{} arm's conditional alignment minus the \prose{} arm's. That control
requires an advantage tied to correct cross-model row correspondence.
Conditional alignment is a per-arm level, not a contrast, and a study can clear
one and fail the other.

\paragraph{Blind scoring.}
The analyzer validates the semantic manifest, but the metric functions receive a
separate registry containing only opaque request identifier, block, model side,
arm slot and anonymous axis, with no account, renderer, wording, arm or model
name in it. Both pipelines compute the complete block scores under neutral
labels, the blind score record is written exclusively and its SHA-256 commitment
is written before semantic labels are used to interpret scores, and the record
is rejected if it contains any registered semantic label.

\subsection{Estimand, estimator and inference}
\label{sec:methods:inference}

Every registered metric is computed for a whole block from both of its slots at
once, so the block contrast is a functional and not a unit-level difference of
two potential outcomes; for \Mstat{} it is not even a linear function of them.
Write $C_b(z)$ for the value the block-$b$ contrast would take if block $b$'s bit
came up $z$. Because the bit fixes a complete complementary mapping, block $b$
has exactly two potential contrast values, $C_b(0)$ and $C_b(1)$, each a fixed
function of that block's four slot potential outcomes, and both are defined
before any bit is drawn.

For each registered metric, the finite-experiment causal estimand is the average
over the 16 fixed blocks of that block's randomization average of its two
assignment-specific potential contrasts:
\begin{equation}
  \label{eq:estimand}
  \theta \;=\; \frac{1}{16}\sum_{b=1}^{16} \frac{C_b(0) + C_b(1)}{2}.
\end{equation}
Here $\theta$ is a fixed function of the 32 slot potential outcomes. It is a
property of the apparatus, not of the draw, and it is well defined for the
nonlinear \Mstat{} as well as for the linear \Gstat{}-based contrasts.

The analyzer reports $\hat{\theta} = (1/16)\sum_b C_b(Z_b)$, the arithmetic mean
of the 16 observed block contrasts under the bits actually drawn. Because each
$Z_b$ is a fair bit, $\hat{\theta}$ is unbiased for $\theta$ over the
\nblocks{} independent fair bits even though \Mstat{} is nonlinear in the two
arms. Unbiasedness is the only property claimed for it. No consistency,
normality or superpopulation interpretation is asserted anywhere in this
protocol.

The one-sided exact test enumerates all \nperm{} renderer-label assignments.
Its null hypothesis is the \emph{Fisher sharp null}: for every one of the 32
renderer slots, $Y_{b,s}(\mathrm{contract}) = Y_{b,s}(\mathrm{prose})$. Under
that null, relabelling a block's two arms leaves the underlying tensors
untouched and only exchanges which one is called \contract{}. Every registered
block contrast is odd under that exchange, so $C_b(1) = -C_b(0)$ for every
block, the enumeration over sign vectors is the exact randomization distribution
of $\hat{\theta}$, and $\theta = 0$. A registered test asserts this oddness by
actually reassigning the renderer slots and recomputing every block contrast.

This is not a weak-null test of a population mean, and it does not support
inference to future models, theories, prompts or people. Effective N is
\nblocks{}. Under equal-direction nonzero effects, five informative positive
blocks are sufficient for an exact one-sided \pv{} at or below $0.05$, and the
smallest attainable \pv{} is $1/\nperm$.

\subsection{Preregistered support rule}
\label{sec:methods:support}

\Hone{} is supported only if every validity rule and all 27 preregistered
support criteria pass in both the \linpipe{} and the \rankpipe{}. \Htwo{} uses
the same 27 criteria on mixed finite differences. The criteria fall into five
groups: direct same-account closeness; relative signed geometry; direction
beyond magnitude; row identity and correct alignment; and absolute level and
auditability. Their two headline thresholds are that mean \Mstat{} is at least
$0.15$ with an exact one-sided \pv{} at most $0.05$ (\gates{1 and 2}), and that
the mean \contract{} minus \prose{} difference in \Gstat{} is at least $0.15$
with an exact one-sided \pv{} at most $0.05$ (\gates{6 and 7}). Every
account-specific and
wording-relation effect must additionally point in the same direction. The full
text of the 27 criteria is in \suppref{2}.

The joint verdict is \supported{} only if both \Hone{} and \Htwo{} are
supported. A valid execution in which any scientific criterion fails is
\notsupported{}, the registered verdict token for a run that executed correctly
and did not meet its registered bar. Validity is evaluated separately from
support: a failed integrity or execution rule invalidates the run rather than
producing a negative result, and the integrity and stopping rules that decide
validity are in \annexref{B}. Applied to what actually ran, that rule leaves
both stages standing: 12 of the 13 registered validity gates hold in each, and
\Hone{} holds all 13. The one gate \Htwo{} does not hold is the registered
single-execution-attempt counter, which is a record of how many attempts a
sample took rather than a check on what the data contain. It moved because
three \Htwo{} direction-runs, in three different samples, hit a timeout in a
local \texttt{openssl ts -verify} subprocess checking stored study-level
attestation files --- a step that runs before any sample-specific input is
read, opens no socket, and has no model in the loop, so it was not a provider,
model, data or network failure. Only those direction-runs were repeated, all
\ncalls{} \Htwo{} samples completed, no data-bearing gate failed, and both
stages are reported here as completed registered stages
(\Cref{tab:status-resolution}; \annexref{C}). The protocol follows the
outcome-neutral logic of
a Registered Report --- methods and criteria frozen before data collection,
every registered analysis reported, validity separate from support --- but it
has not received journal in-principle acceptance and must not be described as a
Nature Registered Report.


%% file: sections/03-analysis-plan.tex
\section{Preregistered outcome and claim matrix}
\label{sec:analysis-plan}

This section fixes the interpretation before any scientific call. An attractive
chart, subgroup, or near miss cannot replace the registered verdict. The
preregistration did not stop at freezing the criteria and the test: it also
froze which sentences the authors would be permitted to write under each
outcome.

\subsection{Validity is evaluated before support}
\label{sec:analysis-plan:validity}

Two questions are asked in a fixed order. Did the run execute as registered?
Then, and only then, did the registered criteria pass? The order is registered
because an integrity failure and a null result are different objects: a
corrupted run scored for support yields numbers that look like a null result
and carry none of its meaning. Fixing the order in advance also removes the
discretion to decide, once the numbers are visible, whether an integrity
problem was fatal or cosmetic. \Cref{tab:outcome-ladder} gives the ladder.
\notsupported{} is a registered verdict token, not an authorial adjective: it
names a valid run in which at least one support criterion failed.

\begin{table}[t]
  \caption{The registered ladder. Validity is settled first; support is
    evaluated only on a valid run.}
  \label{tab:outcome-ladder}
  \small
  \begin{tabularx}{\linewidth}{@{}X l X@{}}
    \toprule
    Observation & Verdict & What may be reported \\
    \midrule
    Public assignment, timestamp, manifest, at-most-one dispatch by the
    experiment driver, provider ID, stored-input attestation, response
    contract, raw-to-compiled binding, tensor replay, program validation,
    stage order, traversal, finite-value, clamp, cap, blind scoring, atomic
    terminal, or complete phase-inventory integrity fails
      & \verdict{invalid}
      & No scientific conclusion. Every artifact is retained and reported. \\
    \addlinespace
    Execution is valid and one or more support criteria fail
      & \notsupported{} for that stage
      & Every failed criterion and every descriptive estimate, without
        redefining success. \\
    \addlinespace
    Execution is valid and all 27 criteria pass in both direction-preserving
    pipelines
      & \supported{} for that stage
      & Continue to the joint matrix. \\
    \bottomrule
  \end{tabularx}
\end{table}

The top row of that ladder is a rule, not a description of this run. Both
stages were evaluated against it and both are valid runs reported at the middle
rung. Thirteen validity gates were evaluated per stage; \Hone{} passes 13 and
\Htwo{} passes 12. \Htwo{}'s one non-passing gate is the registered
at-most-one-dispatch counter named in the top row, and it is the counter alone
that moved: its two conjuncts split, with all \ncalls{} samples valid and the
single-attempt flag false, after three \Htwo{} direction-runs were repeated
following a timeout in a local \texttt{openssl ts -verify} subprocess that
checks stored study-level attestation files before any sample-specific input is
read. No data-bearing gate failed in either stage and no tensor digest
mismatched. \Cref{tab:status-resolution} states the resolution and
\annexref{C} carries the record.

\subsection{The joint outcome rule}
\label{sec:analysis-plan:joint}

Support is a conjunction. Each stage must pass all 27 preregistered support
criteria under the \linpipe{} and again under the \rankpipe{}, the
sign-preserving magnitude-rank transform, and both stages must be supported ---
108 scientific evaluations, no criterion weighted, none serving as a tiebreak,
none tradeable against another. An effect that survives only one transform is a
transform-dependent effect, and the design declines in advance to call that
support.

Only \Honefull{} supported together with \Htwofull{} supported yields a joint
\supported{} verdict, read as: the estimated contract-minus-prose contrast was
positive and above threshold for direct same-account distance and for relative
same-account identification, for both single-input and interaction responses,
under every registered control. The other three cells were read in advance too,
and each is read as a conjunction outcome and nothing more. \Hone{} alone means
the single-input-challenge stage met its 27 criteria while the
interaction-challenge stage did not; \Htwo{} alone means the
interaction-challenge stage met them while the single-input-challenge stage did
not. The two stages test different hypotheses --- single-input central finite
differences and interaction mixed finite differences --- and neither stage's
result is a replication, an extension or a conceptual generalization of the
other's, so a pass in one cell is a self-standing registered stage outcome
rather than an anomaly awaiting corroboration. Neither means the registered
renderer effect was not established. In all three the joint claim fails,
because the joint rule requires both.

\subsection{Registered diagnostic patterns}
\label{sec:analysis-plan:diagnostics}

Diagnostics explain a result. They never change it. The registered table pairs
each pattern the two primary endpoints can make with the one thing it permits
the authors to say, so that no pattern can be found afterwards and promoted into
a finding. \Gstat{} improving without \Mstat{} means the accounts became
easier to separate without the same-account programs becoming directly closer;
\Mstat{} without \Gstat{} means they became closer but not enough to
distinguish them consistently from the four foils; a split between the two
pipelines means the conclusion depends on the response scale. All three leave
the stage unsupported. Signed \Gstat{} improving no more than unsigned $U$,
than row-free $S$, or than shuffled geometry fails \texttt{D\_directional},
\texttt{D\_signature}, and \texttt{D\_conditional} respectively. Two patterns
are validity failures rather than support failures: a row-free signature that
changes after a row shuffle, and any \Htwo{} mirror pair that collapses and
cancels. Both route to the top row of \Cref{tab:outcome-ladder}, not the middle
one. The complete diagnostic table is reproduced in the supplement.

\subsection{Permitted claim language}
\label{sec:analysis-plan:claims}

\Cref{tab:claim-matrix} is the distinctive part of this plan: each outcome was
bound in advance to the sentence it permits.

\begin{table}[t]
  \caption{Permitted claim language. The outcome selects the sentence; the
    authors do not.}
  \label{tab:claim-matrix}
  \small
  \begin{tabularx}{\linewidth}{@{}>{\raggedright\arraybackslash}p{3cm} X@{}}
    \toprule
    Registered outcome & The sentence the authors may write \\
    \midrule
    Joint \supported{}
      & The plain-language conclusion and the technical audit sentence quoted
        below, and nothing beyond them. \\
    \addlinespace
    Valid run, one or more support criteria fail
      & ``The registered renderer-effect hypothesis was not supported under the
        frozen \studyname{} conditions,'' followed by every failed criterion
        and by estimates with exact randomization \pv{} values --- not by a
        summary count. \\
    \addlinespace
    An integrity or execution rule fails
      & ``\studyname{} was invalid because [exact integrity failure]. No
        confirmatory scientific conclusion is drawn.'' \\
    \bottomrule
  \end{tabularx}
\end{table}

The plain-language conclusion a joint \supported{} verdict would have
permitted, as registered:

\begin{quote}
In this fixed experiment, the two renderers did not produce interchangeable
programs. Presenting the same registered propositions from five
literature-inspired engineered cards as explicit hold-change-direction
instructions was associated with separately generated programs from two model
versions responding more similarly to the same one-input and two-input
manipulations than presenting those propositions as \prose{}. The observed
contract-minus-prose contrast was positive for direct same-account distance and
for relative same-account identification, it exceeded the registered
thresholds, and it survived the controls for magnitude, row identity, and
deliberately broken alignment.
\end{quote}

\noindent The corresponding technical audit sentence:

\begin{quote}
For five literature-inspired engineered account cards, two pinned model
snapshots, two registered renderers, and the frozen sparse quadratic language,
the Fisher sharp null of renderer irrelevance at every one of the 32 slots is
rejected at one-sided $\pv \le \pval{0.05}$, and the unbiased estimate of the
registered finite-experiment estimand is positive and above threshold, for
direct cross-model matched-account distance and for closed-set same-account
identifiability, in both \Hone{} and \Htwo{} finite-difference response
geometry and beyond the registered unsigned, row-free, and broken-alignment
controls.
\end{quote}

Success is not redefined as schema validity, a positive subgroup, account
recognizability, or a post-hoc threshold. The registered conjunction failed.
The verdict is \notsupported{}, the two sentences above are reproduced because
they were registered rather than because they were earned, and the sentence
this paper is permitted to write is the middle row of \Cref{tab:claim-matrix}.
\studyname{} is not rerun; a new attempt requires a new version, a new
preregistration seal, and complete disclosure of the failed run.

\subsection{What the verdict does and does not license}
\label{sec:analysis-plan:licence}

A joint \supported{} verdict would have licensed exactly two statements, both
about this apparatus. First, the sharp null that renderer label is irrelevant
at every slot is rejected: something in the 32 slots responded to the renderer
label. Second, the unbiased estimate of the registered estimand is positive and
above the registered threshold for every primary endpoint in both pipelines.
Both are conditional on the two untestable apparatus assumptions --- no
interference between slots, and one version of each renderer. A reader who
rejects either rejects the randomization argument, and neither statement
survives.

Even then, the verdict licenses no mechanism claim that the \contract{}
renderer \emph{made} the programs more similar: rejecting a sharp null
identifies that renderer label mattered somewhere, not where, why, or through
what. It licenses no unqualified verb of change --- the permitted form is ``the
estimated contrast was positive and above threshold'' --- because the estimand
is an average over \nblocks{} blocks of a within-block randomization average,
individual blocks may run the other way, and the verdict says nothing about
which ones. It licenses no statement about renderers, models, propositions,
probes, or programs outside this fixed set, in either direction, and no claim
that the two renderers carry identical theoretical content: they carry the same
registered propositions, rendered differently, and whether that preserves
theoretical content is exactly what is not tested here. A supported result
would not have shown that a language model thinks like a person, that a
psychiatric condition was simulated, or that any source theory is true.

None of that was established. What the outcome licenses is the registered
not-supported sentence, every failed criterion, and every estimate with its
exact randomization \pv{} value.

\subsection{Deviations from the registered plan}
\label{sec:analysis-plan:deviations}

The plan above was followed. Both registered stages ran, the criteria were
evaluated as written under both pipelines, the verdict was read off the
registered rule, and no analysis was added, dropped, reweighted, or retimed
after any outcome was visible.

One incident touched a registered validity gate and is therefore stated here
rather than only by reference. Three \Htwo{} direction-runs, in three different
samples, timed out on a local subprocess call to \texttt{openssl ts -verify}
while it was checking stored, study-level attestation files. That check runs as
the execution worker's first action, before any sample-specific input is read;
it opens no socket and has no model in the loop, so it was not a provider
failure, not a model failure, not a data failure and not a network failure.
Only the affected direction-runs were repeated. All \ncalls{} \Htwo{} samples
completed, every data-bearing validity check passed, and re-hashing the stage's
tensor corpus returned \num{1280} digests with zero mismatches and zero
missing. What the timeout moved is the registered single-execution-attempt
counter --- one of the 13 validity gates --- and nothing else: not tensor
integrity, and not the scientific result, which is \notsupported{} for
\Htwo{} exactly as it is for \Hone{}. Every other deviation and incident,
including those that preceded the executed run, is enumerated in \annexref{C}.

%% file: sections/04a-results-primary.tex
\section{Results}\label{sec:results}

\begin{keyfinding}
Both registered hypothesis stages completed on all \ncalls{} samples each, with
no invalid program in the corpus and no model call retried, repaired or
substituted anywhere in it. The preregistered
conjunction is \notsupported{}, the registered verdict token for a rule under
which every registered support criterion must hold. 19 of 108 scientific
criterion evaluations passed and 89 failed. The second primary endpoint,
closed-set same-account identifiability, is at chance in all four
stage-by-pipeline cells. The first primary endpoint, matched-distance reduction,
produces exactly one significant result in the whole study --- in \Htwo{} ---
and the two registered pipelines disagree about how large it is.
\end{keyfinding}

\subsection{Execution}\label{sec:results:execution}

The generation stage ran once, as registered. All \ncalls{} preauthorized calls
were issued, all \ncalls{} returned programs that compiled and validated, 0 were
invalid, and no call was retried, repaired, replaced, or substituted. Every
scientific generation call used temperature 1.0. The calls produced 960 provider
exchanges, three per call: the initial request, the retrieval of the completed
response, and the retrieval of the provider's own record of the request's input
items, so that what the provider recorded as having been asked can be compared
byte-for-byte against what was sent rather than trusted. The corpus therefore
holds exactly the \ncalls{} first responses, with no survivorship filtering in
it. The two execution stages then ran the deterministic evaluator over that
corpus, and no language model is in the loop from that point on. Every sample in
both stages was executed in both directions, forward and reverse, inside the
attested analysis environment, which re-verifies the entire bound artifact set
before it runs; all 640 forward-reverse pairs agree byte-for-byte, which is the
design's determinism requirement. Three \Htwo{} direction-runs, in three
different samples, hit a timeout in a local subprocess verification step; only
those direction-runs were repeated, all \ncalls{} \Htwo{} samples completed, and
\annexref{C} documents the incident in full. \Cref{tab:execution} gives the
stage record.

\begin{table}[t]
\caption{Execution of the three registered stages. The wall clock for each
execution stage is a single pass over all \ncalls{} samples; the completion time
is the last of the \ncalls{} per-sample receipts.}
\label{tab:execution}
\centering
\small
\begin{tabular}{llrrrr}
\toprule
Stage & Completed (UTC) & Wall clock & Samples & Valid & Invalid \\
\midrule
Generation        & ---                  & 501.8 s              & 320 & 320 & 0 \\
\Hone{} execution & 2026-08-08T02:53:22Z & 24441.985 s (6.79 h) & 320 & 320 & 0 \\
\Htwo{} execution & 2026-08-08T10:40:05Z & 24382.828 s (6.77 h) & 320 & 320 & 0 \\
\bottomrule
\end{tabular}
\end{table}

\subsection{Validity}\label{sec:results:validity}

Under the registered protocol the validity layer is evaluated before, and
separately from, the scientific criteria. Thirteen validity criteria run per
stage, 26 evaluations in total, and the registered rule asks for all of them.
\Hone{} passes 13/13. \Htwo{} passes 12/13, the single non-passing criterion
being the registered single-execution-attempt counter, which records the
history of the run rather than the content of the data. Every criterion that
bears on the data passes in both stages, all \ncalls{} \Htwo{} samples
completed, and \Htwo{}'s scientific result is therefore reported on the same
footing as \Hone{}'s. The mechanics are set out below rather than left to be
inferred from the counts.

The one criterion \Htwo{} does not pass is an attempt counter, not a data check:
it records whether every sample completed on its first attempt, and three did
not. It is the only validity criterion in either stage whose subject is the
history of the run rather than the content of the data. Its two conjuncts split
--- \texttt{all\_320\_valid} is \emph{true} and
\texttt{single\_execution\_attempt} is \emph{false} --- so the criterion fails
on the counter alone. Three \Htwo{} direction-runs, in three different samples,
hit a timeout in a local subprocess verification of stored study-level
attestation files, a step that runs before any sample-specific input is read
and opens no socket; only the affected direction-runs were repeated, and all
\ncalls{} \Htwo{} samples completed. Every integrity condition
the registered ladder names as verdict-invalidating
(\Cref{tab:outcome-ladder}) holds in both stages, and every validity criterion
that bears on the data passes in both stages: all \ncalls{} panels loaded
exactly once, the manifest verifies, the blind score was committed before
semantic unblinding, the row-free signature is exactly invariant to the
registered shuffles, the fresh-process traversal count is exact, and the tensor
audit found 0 mismatched digests. Transition clamp events and evaluator cap
activations are 0 in both stages. \annexref{C} carries that criterion, the
timeout, the repetition of the affected direction-runs, and the
digest-for-digest re-verification of the whole stage: \num{1280} \Htwo{}
digests re-hashed --- 640 tensor archives and their 640 sidecars, the three
affected samples included --- with zero mismatches and zero missing.

\subsection{The preregistered conjunction}\label{sec:results:conjunction}

\registered{support required every one of the following, with no criterion
weighted, no criterion serving as a tiebreak, and no criterion tradeable against
another.}

\begin{lstlisting}
joint SUPPORTED
  = H1 SUPPORTED and H2 SUPPORTED
  = for each stage:  13/13 validity criteria
                  and 27/27 scientific criteria under LINEAR_TRANSFORM_RULE
                  and 27/27 scientific criteria under RANK_TRANSFORM_RULE
\end{lstlisting}

That is 108 scientific criterion evaluations and 26 validity evaluations.
Requiring both transforms to pass is a deliberate robustness demand: an effect
that survives only one transform is a transform-dependent effect, and the design
declines in advance to call that support.

The outcome is \notsupported{} for \Hone{}, \notsupported{} for \Htwo{}, and
\notsupported{} jointly. \Cref{tab:criterion-counts} gives the counts. Support
needs all 108 evaluations; 89 of them fail. The joint line follows from the
registered rule by inspection rather than by computation. No arithmetic on top
of the two stage verdicts can produce anything but \notsupported{}, and running
the joint-analysis step to restate it would add a record, not a fact. One raw
artifact writes a different token for \Htwo{}, taken from the validity flag
rather than from the criterion tree; \annexref{C} documents that discrepancy.

\begin{table}[t]
\caption{Scientific criterion evaluations passed, of the 27 preregistered
support criteria in each stage-by-pipeline cell. Validity criteria are counted
separately and are not included here.}
\label{tab:criterion-counts}
\centering
\small
\begin{tabular}{lccc}
\toprule
Stage & \linpipe & \rankpipe & Stage total \\
\midrule
\Hone{}      & 4/27 & 3/27 & 7/54 \\
\Htwo{}      & 6/27 & 6/27 & 12/54 \\
\midrule
Conjunction  &      &      & 19/108 \\
\bottomrule
\end{tabular}
\end{table}

The preregistration also fixed the exact sentence to be published under a
\notsupported{} outcome, and \annexref{A} reproduces it verbatim. It is
registered that this sentence is followed by every failed criterion and by
estimates with exact randomization p values, not by a summary count. It is
further registered that success is not redefined as schema validity, a positive
subgroup, account recognizability, or a post-hoc threshold. None of those four
appears here. In \Hone{}, 47 of 54 scientific criterion evaluations failed; in
\Htwo{}, 42 of 54. Every failed criterion is named in \suppref{3} and every
estimate is given with its exact p.

\paragraph{Two hypotheses, not two halves.}
Both stages draw on the same \ncalls{}-program corpus, but they are not a split
of it and they do not test the same thing. \Hone{} poses single-input challenges
and takes central finite differences; \Htwo{} poses interaction challenges and
takes mixed finite differences. They ask whether the intervention geometry
survives translation in two different regimes, and the preregistration treats
them as two hypotheses that must both hold. Two consequences run through
everything below. First, a criterion passing in one stage and failing in the
other is not an inconsistency: it is the design reporting that the two regimes
behave differently. Second, and more restrictive, there is no registered
estimator for the \Hone{}-versus-\Htwo{} contrast. Nothing in the preregistration
seal defines a test statistic for the difference between the stages, no such
estimator was run, and no difference between the stages is tested anywhere in
this paper. Where the text observes that \Htwo{} moved and \Hone{} did not, that
is a description of two separately registered results reported side by side.

\subsection{Primary endpoints}\label{sec:results:primary}

Two primary endpoints were registered. \Mstat{} is the direct matched-distance
reduction between same-account cross-model programs. \Gstat{} is closed-set
same-account identifiability over 40 per-block comparisons. Both are reported
for both stages and both pipelines, with the observed magnitude beside the
criterion outcome in every cell, and with the exact p rather than an inequality.

\registered{the smallest attainable p under this design is
$1/\nperm = \num{1.526e-5}$, because inference enumerates all $2^{16}$
renderer-label assignments. A statement of the form ``$p<0.001$'' would
therefore be a statement about the enumeration rather than about how far into
the tail the observed statistic fell. Exact values are reported throughout.}

\Cref{tab:primary-endpoints} is the whole primary matrix and
\Cref{tab:auc} the identifiability endpoint. Each cell is narrated below.

\begin{table}[t]
\caption{The full primary-endpoint matrix. Five registered test families in each
of the four stage-by-pipeline cells: 20 test evaluations. Each entry gives the
observed mean over the \nblocks{} randomized blocks and the exact one-sided p
from the enumeration of all \nperm{} renderer-label assignments. The registered
magnitude floor is 0.15 and the registered p threshold is 0.05. Per-block,
per-family and tail-count detail is in \suppref{3}.}
\label{tab:primary-endpoints}
\centering
\setlength{\tabcolsep}{4pt}
\resizebox{\textwidth}{!}{%
\begin{tabular}{llrrrrr}
\toprule
Cell & & \Mstat & Raw geometry & Directional & $D_{\mathrm{signature}}$ & $D_{\mathrm{conditional}}$ \\
\midrule
\Hone{}, \linpipe & mean         & $0.031049424$  & $-0.0484375$ & $-0.0078125$ & $0.015625$    & $-0.03848877$    \\
                  & exact \pv{}  & $0.39901733$   & $0.94567871$ & $0.60580444$ & $0.31304932$  & $0.91810608$     \\
\addlinespace
\Hone{}, \rankpipe & mean        & $-0.038265658$ & $-0.0359375$ & $0.0515625$  & $0.0671875$   & $-0.032324219$   \\
                   & exact \pv{} & $0.84689331$   & $0.91552734$ & $0.18435669$ & $0.059265137$ & $0.87815857$     \\
\addlinespace
\Htwo{}, \linpipe & mean         & $0.51739819$   & $0.0109375$  & $0.0296875$  & $-0.00625$    & $-0.0023895264$  \\
                  & exact \pv{}  & $0.0012512207$ & $0.30279541$ & $0.03515625$ & $0.59716797$  & $0.58496094$     \\
\addlinespace
\Htwo{}, \rankpipe & mean        & $0.069331606$  & $-0.00625$   & $0.078125$   & $0.0484375$   & $-0.0097320557$  \\
                   & exact \pv{} & $0.0284729$    & $0.67211914$ & $0.1038208$  & $0.07371521$  & $0.71354675$     \\
\bottomrule
\end{tabular}%
}
\end{table}

\begin{table}[t]
\caption{Closed-set same-account identifiability \Gstat{}, reported as an AUC
over 40 per-block comparisons, against the registered threshold.}
\label{tab:auc}
\centering
\small
\begin{tabular}{lccc}
\toprule
Stage & \linpipe & \rankpipe & Registered threshold \\
\midrule
\Hone{} & \auc{0.515625} & \auc{0.5234375} & 0.80 \\
\Htwo{} & \auc{0.46875}  & \auc{0.503125}  & 0.80 \\
\bottomrule
\end{tabular}
\end{table}

\paragraph{\Hone{}, linear pipeline.}
Four of the 27 criteria pass. The matched-distance reduction is
\effect{0.031049424} with an exact p of \pval{0.39901733}. No test in the cell
meets the registered magnitude floor of 0.15, and no test in the cell has an
exact p at or below 0.05. Identifiability is \auc{0.515625} against the
registered 0.80, and conditional alignment is \effect{0.019885254} against the
registered 0.20.

\paragraph{\Hone{}, rank pipeline.}
Three of the 27 criteria pass, one fewer than under the linear transform. The
matched-distance reduction is negative, \effect{-0.038265658}, with an exact p
of \pval{0.84689331}. The smallest exact p anywhere in \Hone{}, across its ten
tests, is \pval{0.059265137}, on the signature effect in this cell; it does not
cross 0.05. Identifiability is \auc{0.5234375} and conditional alignment is
\effect{0.02628479}.

\paragraph{\Htwo{}, linear pipeline.}
Six of the 27 criteria pass. The matched-distance reduction is
\effect{0.51739819} with an exact p of \pval{0.0012512207}: this is the one cell
in the study where both the magnitude criterion and the p criterion pass on the
same test. The directional-beyond-unsigned effect is \effect{0.0296875} with an
exact p of \pval{0.03515625}, which clears the p threshold and misses the
magnitude floor. The rest of the cell is null: raw geometry \effect{0.0109375}
at \pval{0.30279541}, signature \effect{-0.00625} at \pval{0.59716797},
conditional \effect{-0.0023895264} at \pval{0.58496094}. Identifiability is
\auc{0.46875} and conditional alignment is \effect{0.026071167}.

\paragraph{\Htwo{}, rank pipeline.}
Six of the 27 criteria pass. The matched-distance reduction is
\effect{0.069331606} with an exact p of \pval{0.0284729}: it clears the p
threshold and misses the 0.15 magnitude floor. Nothing else in the cell crosses
either threshold. Identifiability is \auc{0.503125} and conditional alignment is
\effect{\num{2.746582e-05}}.

Across the four cells the identifiability endpoint never leaves chance and never
approaches the registered 0.80 threshold (\Cref{tab:auc}). The conditional
alignment threshold is 0.20; the four observed values are \effect{0.019885254},
\effect{0.02628479}, \effect{0.026071167} and \effect{\num{2.746582e-05}}.

\subsection{The one endpoint that moves}\label{sec:results:moving-endpoint}

Of the two primary endpoints across four cells, exactly one moves, and it moves
in one stage. \Htwo{}'s matched-distance reduction is positive with an exact
one-sided p below 0.05 in both pipelines. Under the linear transform the mean is
\effect{0.5173981945762773} with an exact p of \pval{0.001251220703125}, a tail
of 82 assignments out of \nperm{}. Under the rank transform the mean is
\effect{0.06933160555977207} with an exact p of \pval{0.028472900390625}, a tail
of 1866. The two pipelines agree on sign and both clear 0.05. They part on
magnitude.

The linear p is not merely under 0.05. It is under the Bonferroni threshold of
0.0025 over the study's 20 test evaluations, and it is the only value in the
study that is. Clearing that correction is a statement about the p value alone:
the statistic sits deep in the enumerated null. It is not a statement about
magnitude, and magnitude is where the two pipelines disagree.

The statistic is a ratio, so a denominator near zero would manufacture it. It
does not. The registered primary test aggregates the \nblocks{} block-level
effects, and every block-level denominator in this cell is well conditioned:
they range from 0.0393384 to 0.157305 around a median of 0.0837244, the smallest
being 0.47 times the median. No block-level denominator floor event occurs in
any cell of the study. 14 of the 16 block effects are positive under the linear
transform and 11 of 16 under rank.

Floor events do occur in this cell, at a different level of aggregation. The
linear pipeline fails the registered denominator criterion on 3 of 80
block-by-family denominators, which are zero to within floating-point noise, of
order \num{1e-29}: blocks 1 and 3 in the comparator family, and block 12 in the
source-monitoring family. Those collapsed denominators are a genuine failure and
they legitimately gate the five family-positivity criteria, all of which fail.
They cannot reach the primary statistic, which does not aggregate families. The
rank pipeline reports no floor event and passes that criterion.

Dropping the single largest-magnitude block and re-enumerating the exact null
separates the two pipelines again. Under the linear transform, removing block
11, whose value is 1.18419, leaves a mean of 0.472945 with an exact p of
0.00250244; the result survives. Under the rank transform, removing block 9,
whose value is 0.326828, leaves a mean of 0.0521652 with an exact p of
0.0569458, which does not clear 0.05.

So the linear number is a real statistic on well-conditioned inputs, and it is
large. The rank transform puts the same quantity at \effect{0.069331606}, less
than half the registered 0.15 floor, against \effect{0.51739819} under linear.
Nothing registered in advance says which transform is right about the magnitude.
Two readings are available and this study cannot distinguish them. The effect
may be small with the linear pipeline overstating it: the statistic is bounded
by 2, and four linear blocks exceed 0.94, so a handful of blocks with unusually
large relative separations could carry the mean. Or the two transforms may not
be on a comparable scale for a fixed absolute floor: replacing values by their
ranks bounds how far apart two matched distances can be as a fraction of their
sum, and rank denominators here run near 102.359 against 0.0837244 in linear, so
the same underlying separation yields a smaller ratio under rank. That second
reading was not anticipated in the preregistration; it was noticed after
unblinding.

The support rule requires the magnitude criterion to pass under both transforms,
and it does not. Both transforms were required in advance precisely so that an
effect appearing at one scale and not the other would not count as a finding.
That is why the conjunction fails.

Two further facts sit beside this one. It is one endpoint of two: closed-set
identifiability is at chance in this cell as in every other, \auc{0.46875} under
linear and \auc{0.503125} under rank against the 0.80 threshold, so whatever
moves the matched distance does not make the two renderings mutually
recognizable. And the corresponding \Hone{} cells show nothing at all, which is
two separately registered hypotheses returning different answers and not a
tested contrast between them.

\suppref{3} carries the full per-cell, per-block and per-family tables,
including every tail count, every matched-distance vector, and every failed
criterion by name.

%% file: sections/04b-results-secondary.tex
\subsection{Negative controls}\label{sec:results:controls}

The registered controls exist so that a positive result cannot be produced by
the measurement apparatus alone. The Methods name three --- unsigned, row-free
and broken-alignment. None of them is a separate reported statistic. Each is
built into a registered quantity, as \Cref{tab:negative-controls} sets out, and
each therefore constrains a number that is reported anyway.

\begin{table}[tb]
  \centering
  \caption{The three registered negative controls, the registered quantity that
  implements each one, and what each showed. The controls are not free-standing
  statistics; they are subtractions and invariance requirements built into
  endpoints that are reported in their own right.}
  \label{tab:negative-controls}
  \begin{tabularx}{\linewidth}{lXX}
    \toprule
    control & implemented by & what it showed \\
    \midrule
    unsigned &
    the directional endpoint: raw geometry effect minus unsigned geometry
    effect, so only the directional residue can clear the floor &
    the magnitude criterion
    \texttt{d\_directional\_beyond\_unsigned\_at\_least\_0\_15} fails in all
    four cells; its p criterion passes only for \Htwo{} under the \linpipe \\
    \addlinespace
    row-free &
    the signature endpoint must be exactly invariant to the registered row
    shuffles, at tie tolerance 1e-12 &
    \texttt{row\_free\_signature\_exactly\_invariant\_to\_registered\_shuffles}
    passed in both stages \\
    \addlinespace
    broken-alignment &
    conditional alignment: AUC minus the mean AUC under deliberately shuffled
    alignment, so only performance above broken alignment counts &
    \texttt{invariant\_conditional\_alignment\_at\_least\_0\_20} fails in all
    four cells \\
    \bottomrule
  \end{tabularx}
\end{table}

The third control is easily misread as a plain accuracy number. It is not: the
registered floor of 0.20 is a floor on the margin over broken alignment, not on
the raw AUC. The shuffle controls behind it used 512 permutations per arm per
block, seed 2026083600, tie tolerance 1e-12.

What the three rule out is narrow and worth stating exactly. Directional
structure over and above unsigned geometry never reaches the registered
magnitude floor, so no reported directional effect is the unsigned geometry
wearing a sign. The signature endpoint is exactly invariant to the registered
row shuffles in both stages, so no part of it depends on the ordering of the
rows. Conditional alignment is only ever credited above deliberately broken
alignment, and it fails the 0.20 criterion in all four cells, so nothing in the
identifiability picture is an artifact of scoring against intact alignment.

Two further registered criteria are structural rather than statistical and are
easy to misread as results.
\texttt{all\_16\_invariant\_panels\_at\_least\_4\_reciprocal} requires at least
4 reciprocal matches per block in the invariant arm --- a check that the
matching had enough mutual structure to be meaningful at all, not a check that
the intervention worked.
\texttt{all\_16\_invariant\_assignments\_unique\_identity} requires that each
block's matching be a permutation rather than a many-to-one collapse. Both fail
in all four cells, which is the more informative fact: the matched-distance
statistic is computed over matchings that are, in most blocks, neither
reciprocal enough nor injective. That is a limitation of the measurement, and it
applies equally to \Hone's null and to \Htwo's one positive result.

Denominator floor events were counted in all four cells, zeros included. No
block-level denominator floor event occurs in any cell. \Htwo's \linpipe{}
records three at the family level, one aggregation level below the registered
primary statistic, against 80 family denominators per cell; that is the cell
examined in \Cref{sec:results:moving-endpoint}.

\subsection{Block and family structure}\label{sec:results:structure}

\begin{figure}[tb]
  \centering
  \includegraphics[width=\linewidth]{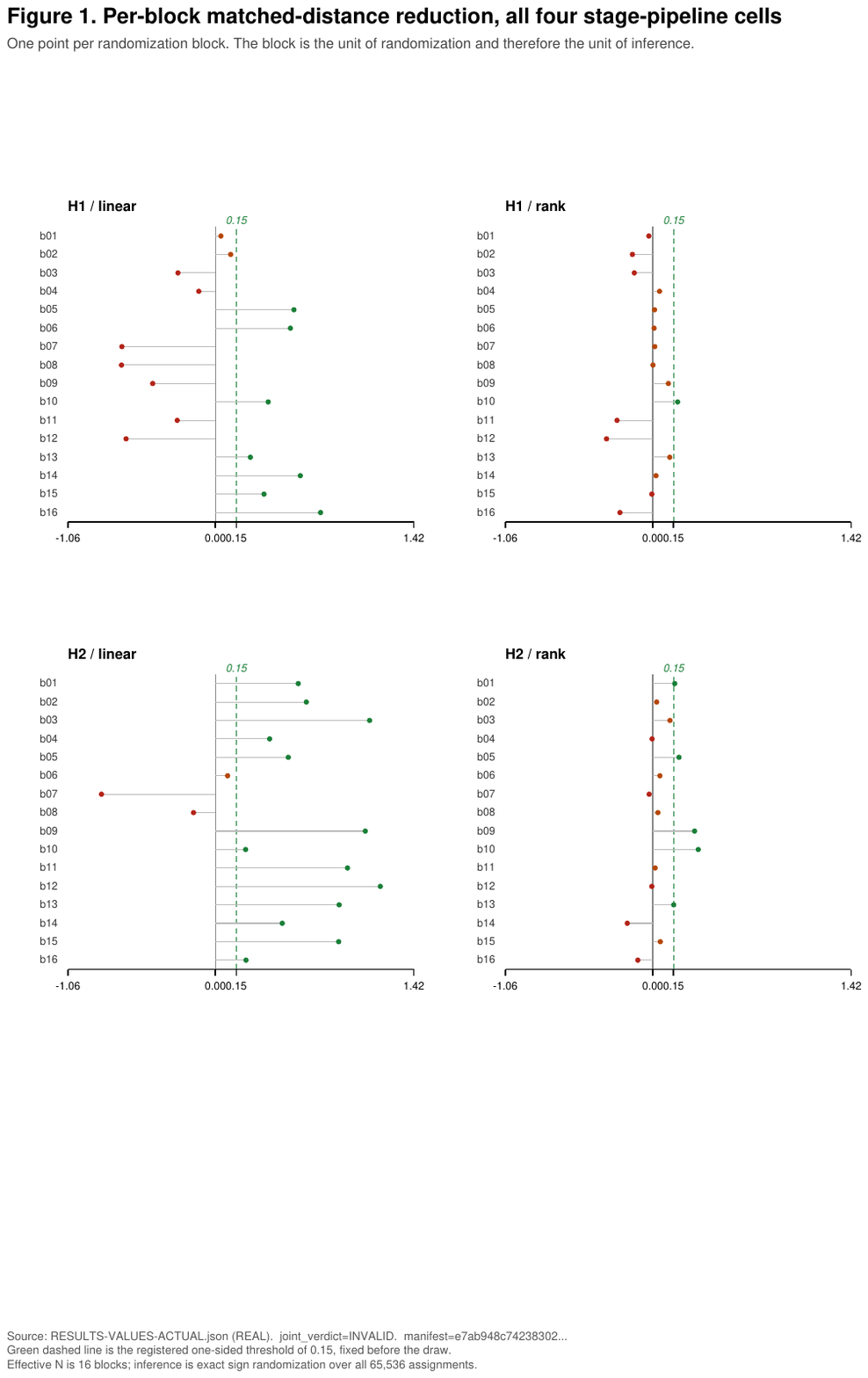}
  \caption{Per-block matched-distance reduction for all four stage-by-pipeline
  cells. Each value is one block's matched-distance reduction; there are 16 per
  cell, and effective N is 16, so these are the units the primary test
  aggregates rather than a breakdown of it. The dashed line is the registered
  0.15 threshold, which lies inside the axis range in every cell. What to
  notice is how thin the majorities are: 9/16 values are positive for \Hone{}
  under the \linpipe{} and 9/16 under the \rankpipe, 11/16 for \Htwo{} under
  the \rankpipe, and 14/16 for \Htwo{} under the \linpipe{} --- the one cell
  whose mean clears the threshold.}
  \label{fig:block-effects}
\end{figure}

Effective N is 16, so the per-block effects are not a supplementary detail. They
are the entire evidential base, and they are reported individually rather than
summarized. \Cref{fig:block-effects} shows them.

\begin{figure}[tb]
  \centering
  \includegraphics[width=\linewidth]{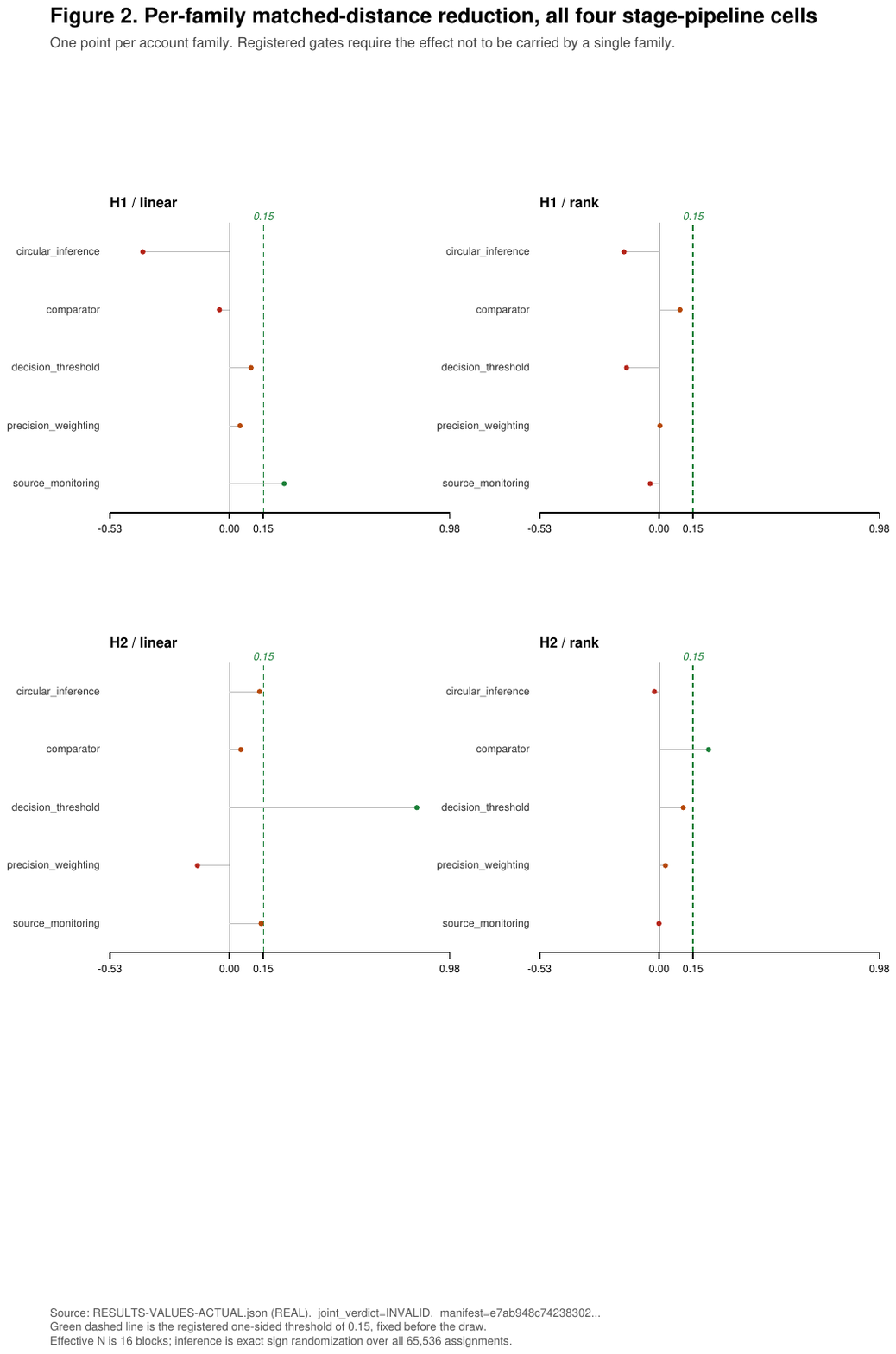}
  \caption{Per-family effects for the same four stage-by-pipeline cells, 5
  values per cell, one for each of \texttt{source\_monitoring},
  \texttt{comparator}, \texttt{precision\_weighting},
  \texttt{circular\_inference} and \texttt{decision\_threshold}. The registered
  family criteria are conjunctions over all five, so what to notice is not the
  average height of a cell but whether any bar falls below zero: one negative
  family fails the criterion however large the other four are. No cell is free
  of negative families.}
  \label{fig:family-effects}
\end{figure}

The five families are \texttt{source\_monitoring}, \texttt{comparator},
\texttt{precision\_weighting}, \texttt{circular\_inference} and
\texttt{decision\_threshold}. The registered family criteria require all five
effects to be positive, which is why a single negative family fails the
criterion regardless of the others. \Cref{fig:family-effects} shows which
families carry the sign in each cell, which the criterion boolean conceals.

Both figures are rendered from the same in-memory values that supply the prose,
in the same generation run, with the threshold line always inside the axis
range; their digests are recorded in \annexref{B}. The full per-block and
per-family tables, including the four component endpoints behind each family
effect, are in \suppref{3}.

\subsection{Registered secondary analyses}\label{sec:results:secondary}

Secondary analyses S0 through S7 were registered in advance and are non-gating
and exploratory. S0 governs how everything else in this section may be read, so
its registered wording is quoted rather than summarised:

\begin{quote}
Every analysis in this section is \textbf{non-gating and descriptive}. None of
them can change the study verdict, in either direction. [\ldots] no secondary
result may be offered as a partial or consolation finding if the conjunction
fails. If the study does not pass, it did not pass; the material below is then a
characterisation of a null, not a rescue of it.
\end{quote}

The headline of each analysis follows. The full treatment of all seven,
including every table, is in \suppref{4}.

\begin{description}[style=unboxed,leftmargin=0pt]

\item[S1. Full exact randomization null distribution.]
The null is exact rather than sampled --- all \nperm{} sign vectors are
enumerated --- so its resolution is a property of the design, fixed before any
data existed. The smallest attainable one-sided \pv{} is one in \nperm{}, or
1.526e-5, and the largest \pv{} that still satisfies the registered maximum is
0.049988, a tail of 3,276 assignments. The last configuration that can pass is
12 of 16 blocks concordant; at 11 of 16 no set of magnitudes can reach 0.05.

\item[S2. Marginal descriptives of the response tensors.]
Registered as reported whatever the outcome. Every median and every
interquartile range over the 11 tensor coordinates is exactly zero, in all 44
coordinate-by-model-by-stage cells, and between 0.6842 and 0.9608 of each
coordinate's values are exactly 0.0, so the 25th, 50th and 75th percentiles all
land inside the block of zeros. The programs the models emit respond to most
interventions on most coordinates by not moving at all. S2 also records the
transition-clamp and cap conditions, which occurred 0 times in both stages, so
no reported estimate, p value or criterion verdict was produced by a clamped or
capped value.

\item[S3. Dispatch order and drift.]
All 320 dispatches fall inside a window of 491.939 seconds, or 0.1366 hours,
with eight requests in flight at a time. The largest absolute correlation
between any per-sample derived quantity and dispatch position is 0.05653, and
between-sample scatter exceeds every fitted trend. No drift correction is
applied, because none is indicated and because applying one would be an
unregistered analysis; an order effect could not bias the primary contrast in
any case, since the registered separation audit reports that schedule position
is invariant to the arm assignment.

\item[S4. The distribution of the 512 shuffle controls.]
Conditional alignment is the observed AUC minus the mean of these shuffles, so
the shuffle null is subtracted into that endpoint and should be visible rather
than implicit. The shuffles were recomputed with the registered functions and
every recomputed vector was required to hash equal to the digest the analyser
recorded: 64/64 panels agree in each stage, 128 in all. Three of the four cells
sit within 0.00136 of 0.5; \Htwo{} under the \linpipe{} does not, with a
shuffled AUC mean of 0.43601532, roughly 0.064 below 0.5, sd 0.0632042, which
mechanically inflates conditional alignment in that one cell. The registered
conditional result there is nonetheless negative ($-0.00238953$, \pv{} =
\pval{0.584961}) and does not clear the registered 0.05 threshold, so the
displacement did not manufacture a result.

\item[S5. Criterion-margin reporting.]
How far each primary test cleared or missed its 0.15 threshold, so that a near
miss is legible as a near miss. Exactly one cell-and-test combination clears the
floor: \Htwo's matched-distance reduction under the \linpipe, by 0.367398
(\Cref{sec:results:moving-endpoint}). Every
other margin is negative, most of them by more than the threshold itself, which
is the shape of a null rather than of a near miss.

\item[S6. Multiplicity.]
Across the study there are $5 \times 2 \times 2 = 20$ test evaluations --- five
tests, two pipelines, two stages --- and a Bonferroni threshold of $0.05/20 =
0.0025$. One observed p value falls below it: \Htwo's matched-distance reduction
under the \linpipe{} at \pval{0.00125122}. Two are nominally significant without
being corrected-significant: \Htwo's directional test under the \linpipe{} at
\pval{0.0351562} and \Htwo's matched-distance reduction under the \rankpipe{} at
\pval{0.0284729}. The correction is taken over all 20 evaluations because the
plan registered it that way, and surviving it is a statement about the p value
only --- that the statistic sits deep in the enumerated null --- not about
magnitude.

\item[S7. What this study cannot answer.]
Effective N is 16 for every inferential statement in this study. The 320
registered calls, 960 provider exchanges and 1,280 tensors increase the
precision with which each block effect is estimated; they do not increase the
number of independent units, and no statement anywhere treats them as if they
did. Two model snapshots is two, and no model-general conclusion is drawn. The
study also cannot say whether the two renderers carry identical theoretical
content: they carry the same registered propositions, rendered differently, and
whether that preserves theoretical content is exactly what is not tested here.

\end{description}

\subsection{Robustness}\label{sec:results:robustness}

Two questions were put to the results after unblinding: whether \Hone's null is
an artifact of the apparatus, and whether the one endpoint that moved can be
killed. The full analyses are in \suppref{5}.

\paragraph{Is \Hone's null an artifact of the apparatus?}
A null is only informative if the instrument could have detected an effect. Two
candidate artifacts were checked. \emph{Denominator floors: none.} \Hone{}
records zero block-level and zero family-level denominator floor events in both
pipelines, over 80 family denominators, so the null is not a floored-divisor
artifact. \emph{Support floors: not global.} The criterion
\texttt{all\_registered\_support\_nonzero\_in\_both\_arms} failed 60 of 64
audits in both pipelines, which is equally consistent with the programs never
moving the designated rows at all and with one family sitting at zero while the
criterion is a conjunction. Opening the audits separates the two: the failures
are heterogeneous and concentrated in \texttt{comparator}, at 51/64 audits below
floor against 5/64 for \texttt{circular\_inference}; no family is below floor in
every audit and none is above floor in every audit. The derived verdict is
\verdict{NO\_FLOOR} under both pipelines.

Two specific explanations for \Hone's null are therefore excluded: a floored
divisor and a global support floor. Neither exclusion is a demonstration that
the apparatus was sensitive enough to detect an effect of the registered size.
No positive control was registered and none was run, so there is no measured
floor on what this instrument could have found. A third limitation is resolved
by neither audit and is carried forward from \Cref{sec:results:controls}: the
matched-distance statistic is computed over matchings that in most blocks are
neither reciprocal enough nor injective. \Hone's null is reported as a null with
those two candidate artifacts excluded and that third one open.

\paragraph{Four attempts to kill the one result that moved.}
\Htwo's matched-distance reduction is the largest statistic observed anywhere in
the study, which is exactly the circumstance in which an artifact is most likely
and least likely to be looked for, so four specific ways it could have been an
artifact were checked before it was written down. It is positive with an exact
one-sided p below 0.05 in both pipelines: \effect{0.5173981945762773},
\pv{} = \pval{0.001251220703125} (tail 82/65536) under the \linpipe, and
\effect{0.06933160555977207}, \pv{} = \pval{0.028472900390625} (tail
1866/65536) under the \rankpipe.

\emph{Challenge 1 --- a denominator floor artifact? No.} This restates, with
the audit behind it, the conditioning check in \Cref{sec:results:moving-endpoint}.
The three floor events
sit at the family aggregation level and are zero to within floating-point noise,
while the registered test aggregates block-level effects; all 16 block-level
denominators are healthy, ranging from 0.0393384 to 0.157305 with none below a
quarter of the median, including the three blocks named in the family-level
failure. Block-level floor events are empty in both pipelines, so the floor
contaminates the family-effect criteria and not the primary test.

\emph{Challenge 2 --- carried by a single block? Largely no, in the linear
pipeline only.} Dropping the single largest-magnitude block and re-enumerating
the exact null leaves the linear result at \effect{0.472945} with
\pv{} = \pval{0.00250244} (block 11, value 1.18419 dropped), while the rank
result falls to \effect{0.0521652} with \pv{} = \pval{0.0569458} (block 9, value
0.326828 dropped) and no longer clears 0.05. With 15 blocks the smallest
attainable p rises to 1/32,768.

\emph{Challenge 3 --- do the two pipelines disagree? On sign and direction, no;
on the registered magnitude criterion, yes, and the criterion is what counts.}
Both pipelines agree on sign and both fall below 0.05. They part on
\texttt{matched\_distance\_reduction\_at\_least\_0\_15}, which linear passes and
rank fails. The rank cell failed that criterion; that is the finding for that
cell. One candidate explanation is that the rank transform compresses the
relative gap between the narrative and invariant matched distances, so the same
underlying separation produces a smaller value of a ratio statistic under rank
than under linear while a single absolute floor of 0.15 is applied to both. That
was not anticipated in the preregistration, it was noticed after unblinding, and
it is carried forward as a design limitation in
\Cref{sec:discussion:limitations}.

\emph{Challenge 4 --- is the \Htwo-versus-\Hone{} difference in this statistic a
divisor artifact? No --- but no test of the difference was run.} Within pipeline
and across stages the denominators are of the same order, with a median
denominator ratio of \Htwo{} to \Hone{} of 1.805 under the linear pipeline and
1.595 under rank, so the larger \Htwo{} value is not produced by dividing by a
smaller number. That excludes the divisor explanation and nothing more. No test
of the \Hone/\Htwo{} difference was registered and none was run: the two stages
are separate hypotheses over different probe batteries and different
finite-difference schemes, not two arms of a comparison, and the study has no
estimator for the contrast between them.

Passing those four rules out those four. It does not make the registered
conjunction pass, and S0 forbids offering any of it as a partial or consolation
finding. What the four licence is one sentence: under \Htwo's interaction
challenges, holding the words fixed and presenting them as explicit
hold-change-direction instructions was associated with a reduction in matched
distance between the two snapshots' executable behaviour --- below the corrected
threshold and robust to dropping the largest block in the linear pipeline,
nominal only and not robust to that deletion in the rank pipeline, where it also
failed the registered magnitude criterion. \Hone{} is reported separately and on
its own terms: under single-input challenges no such association was detected.
The two statements are not joined into a claim that the effect is present in one
setting and absent in the other, because no contrast between them was tested and
a non-detection is not a detection of absence.

What the four do not licence is calling the hypothesis supported. The effect is
a mean shift on one contrast inside a pattern that is otherwise flatly null:
chance-level classification, conditional alignment orders of magnitude below
threshold, and every family-invariance and rendering-relation-invariance
criterion failing. The theory predicted a uniform, family-invariant,
classifiable geometry. That is not what came out. One endpoint moving in one
regime, at a size two of the study's own instruments disagree about, is a result
worth reporting and reproducing --- not a hypothesis confirmed.

%
%
%
%

%% file: sections/05-discussion.tex
\section{Discussion}\label{sec:discussion}

\subsection{What this licenses and what it does not}\label{sec:discussion:licenses}

Both registered stages completed on all \ncalls{} of their samples. \Hone{}
carries the registered verdict \notsupported{}, \Htwo{} carries the registered
verdict \notsupported{}, and the preregistered conjunction of the two is
\notsupported{}. Each stage fails the conjunction on its own registered
criteria, and the two failure patterns are described below side by side and not
as a contrast: no \Hone{}-versus-\Htwo{} comparison was registered, no
estimator for one exists, and none was run.

\textbf{\Hone{} is a null throughout.} Under single-input challenges with
central finite differences, the \contract{} renderer did not produce measurably
more similar executable behaviour between two model snapshots than \prose{}
did. Matched-distance reduction was \effect{0.0310494}
(\pv{}~$=$~\pval{0.399017}) under the \linpipe{} and \effect{-0.0382657}
(\pv{}~$=$~\pval{0.846893}) under the \rankpipe{}, against a registered floor
of \effect{0.15} and a registered $\alpha$ of \pval{0.05}. Classification was
at chance. Two candidate apparatus artifacts were excluded --- a floored
divisor and a global support floor --- so the null is not explained by either.
That is narrower than saying the apparatus was adequate. No positive control
was registered, and the matching-structure limitation is unresolved: the
matched-distance statistic is computed over matchings that in most blocks are
neither reciprocal enough nor injective, and that applies equally to \Hone{}'s
null and to \Htwo{}'s one positive result.

\textbf{\Htwo{} is a null with one endpoint inside it that moved.} Under
interaction challenges with mixed finite differences, matched-distance
reduction was positive in both pipelines and reached nominal significance in
both, and in the \linpipe{} it cleared the corrected threshold, the only test
in the study that did. It nevertheless fails the registered conjunction,
because the registered rule demands the magnitude floor in both transforms and
the \rankpipe{} misses it by a wide margin. It is a result of this study,
reported at the size it came out and with the instrument that disagrees about
that size reported beside it.

The claim boundary registered before the draw, quoted verbatim:

\begin{quote}
This stage tests cross-model agreement in bounded signed finite-difference
response geometry, scored under the registered linear and rank pipelines with
no post-transform centering and no unit normalization, for the complete
registered renderer package. H2 is a numeric interaction holdout and is not
claimed to be entailed by, or a conceptual generalization of, the five prose
accounts. The study does not test people, diagnoses, brains, or cognitive
laws.
\end{quote}

That sentence is the ceiling on what may be said, and it was fixed before the
randomness was drawn precisely so that it could not be loosened afterwards to
fit the result.

What no verdict here licenses, registered in advance:

\begin{itemize}
  \item \textbf{No mechanism claim.} Rejecting a \emph{Fisher sharp null}
    identifies that renderer label mattered somewhere; it does not identify
    where, why, or through what. No sharp null was rejected in \Hone{} in any
    case.
  \item \textbf{No unqualified verb of change.} The estimand is an average over
    \nblocks{} blocks of a within-block randomization average; individual
    blocks may run the other way, and the verdict says nothing about which
    ones.
  \item \textbf{No extrapolation} to renderers, models, propositions, probes,
    or programs outside this fixed set, in either direction. A null here is not
    evidence that no such effect exists anywhere.
  \item \textbf{No claim that the two renderers carry identical theoretical
    content.}
\end{itemize}

Nothing here shows that an LLM thinks like a person, that a psychiatric
condition was simulated, that any source theory is true, or that the generated
programs faithfully preserve those theories. The five account axes carry names
borrowed from a theoretical literature because the stimuli were drawn from it
\citep{adams2013,frith2000,gold2007,jardri2013,jardri2017,johnson1993,ratcliff2008};
they must not be read as claims about the phenomena those names were coined
for.

\subsection{Interpreting the one result that moved}\label{sec:discussion:disagreement}

The two pipelines agree that \Htwo{}'s matched-distance reduction is positive,
and both put its exact one-sided \pv{} below \pval{0.05}. They disagree about
its size by an order of magnitude: \effect{0.517398} under the \linpipe{} and
\effect{0.0693316} under the \rankpipe{}, on the same \nblocks{} blocks, with
the same sign, straddling the registered floor of \effect{0.15}. Two readings
of that disagreement are available and this study cannot distinguish them.

\textbf{Reading one: the effect is small and the \linpipe{} overstates it.}
\Mstat{} is bounded by 2, and four linear blocks exceed \effect{0.94}. If a
handful of blocks with unusually large relative separations drive the mean, the
rank value is the better estimate of a genuinely modest effect.

\textbf{Reading two: the two transforms are not on a comparable scale for a
fixed absolute floor.} The rank transform replaces values by ranks, which
bounds how far apart two matched distances can be as a fraction of their sum:
rank denominators here run near \effect{102.36} against \effect{0.0837244} in
linear, and the same underlying separation yields a smaller ratio under rank. A
single floor of \effect{0.15} applied to both transforms is then a stiffer test
under rank than under linear.

The statistic is a dimensionless ratio, so this is not a units problem, and a
common rescaling of both distances leaves it unchanged. Nothing registered in
advance distinguishes the two readings. The second was noticed after
unblinding, which is exactly when such an observation carries least weight; it
is carried as a design question in \Cref{sec:discussion:limitations}. It
changes nothing here: the criteria stand as registered, and every cell that
failed the magnitude criterion, including \Htwo{} rank, failed it. Both
transforms were required in advance precisely so that an effect appearing at
one scale and not the other would not count as a finding.

Two further facts sit beside the moving endpoint. It is one endpoint of two:
closed-set same-account identifiability \Gstat{} is at chance in this cell as
in every other, \auc{0.46875} under the \linpipe{} and \auc{0.503125} under the
\rankpipe{} against a registered \auc{0.80} threshold, so whatever moves the
matched distance does not make the two renderings mutually recognizable. And
the corresponding \Hone{} cells show nothing at all --- which is two separately
registered hypotheses returning different answers, not a tested contrast
between them.

\subsection{Limitations}\label{sec:discussion:limitations}

These hold regardless of verdict.

\begin{itemize}
  \item \textbf{The verification preamble re-verifies the same study-level
    files on every direction-run.} Each of the \num{1280} direction-runs across
    the two stages re-verifies the same registered attestation files by
    shelling out to OpenSSL under a hard 30-second limit per invocation, with
    up to 8 such runs in flight, and the limit does not scale with contention
    on the host. Verifying those files once per stage, or scaling the limit
    with concurrency, would remove the redundancy without weakening any check.
    The limit fired three times, on three \Htwo{} direction-runs in three
    different samples. In this run the two possibilities were never confusable:
    the timeout was raised by a local \texttt{openssl ts -verify} subprocess
    checking stored study-level attestation files before any sample-specific
    input was read, with no socket opened and no model in the loop, so it was
    not a provider failure, not a model failure, not a data failure and not a
    network failure. Only the affected direction-runs were repeated; all
    \ncalls{} \Htwo{} samples completed; re-hashing the stage gave \num{1280}
    digests with zero mismatches and zero missing; and the only registered
    quantity the episode moved is the single-execution-attempt counter. A future
    version should nonetheless preregister a bounded, disclosed,
    digest-verified retry policy for infrastructure failures, so that the same
    guarantee is mechanical rather than reconstructed after the fact.

  \item \textbf{One absolute floor is applied to two estimators with different
    achievable ranges, and this is where the study's one moving endpoint comes
    apart.} The registered criteria apply a single floor of \effect{0.15} to
    the primary statistic under both the linear and the rank transform. Under
    \Htwo{} that statistic is \effect{0.517398} under linear and
    \effect{0.0693316} under rank: an order of magnitude apart, on the same
    \nblocks{} blocks, with the same sign, both nominally significant, and
    straddling the floor. The statistic is a dimensionless ratio, so this is
    not a units problem, and the denominator scales do not bear on it; the
    issue is that ranking compresses the relative gap the ratio is built from,
    so the two transforms do not have the same headroom against a common floor.
    Nothing registered in advance distinguishes that reading from the competing
    one, that the effect is simply small and the linear figure is the inflated
    one. This was not anticipated when the thresholds were fixed and it was
    noticed after unblinding, which is exactly when such an observation carries
    least weight. It is recorded as a design question for a future
    preregistration, which should either calibrate the two transforms onto a
    comparable scale or register an effect measure that is not a ratio of
    distances. It changes nothing here: the criteria stand as registered, and
    every cell that failed the magnitude criterion, including \Htwo{} rank,
    failed it.

  \item \textbf{Host conditions during execution were not fully controlled.}
    Real-time antivirus scanning of the evidence tree could not be excluded
    without administrative elevation, and CPU thermal telemetry was unavailable
    on this host, so thermal throttling during a multi-hour saturating load
    could not be measured. Both are consistent with the observed contention.

  \item \textbf{The claim is about 32 renderer slots, two model snapshots, and
    this fixed set of programs and probes.} Two snapshots is two.

  \item \textbf{Effective N is \nblocks{}.} The exactness of the enumeration is
    a statement about validity, not about power.

  \item \textbf{Meeting a threshold is not a large effect}, and missing one is
    not evidence of absence at any particular effect size.

  \item \textbf{Temperature was 1.0}, so responses are stochastic.
    Randomization and the block structure handle this for inference, but no
    individual response is reproducible.

  \item \textbf{The programs were executed by the experiment driver.}
    Executability is measured relative to that driver's compiler and validator,
    both of which are archived.
\end{itemize}

\subsection{Provenance}\label{sec:discussion:provenance}

The audit trail is why these numbers can be believed, and it is not an
assurance about the authors' conduct; it is a chain a third party can check
against records the authors do not control. The sealed analysis predates its
own randomness by 1 hour 54 minutes: the preregistration seal names a future
\emph{NIST Randomness Beacon} pulse, that pulse did not exist when the seal was
written, and the assignment is a deterministic function of it, so the analysis
could not have been chosen to suit the assignment. That ordering is enforced
rather than asserted --- the experiment driver parses the \emph{RFC 3161}
signed time out of the seal's timestamp token, compares it against the target
pulse time, and refuses to proceed unless the seal is strictly earlier; the
check is fail-closed, and both times it compares are published, as is the
independently attested digest of the assignment bytes. \Hone{}'s \notsupported{}
verdict was itself sealed with both \emph{OpenTimestamps} and \emph{RFC 3161}
hours before \Htwo{} was analysed, and the registered conjunction requires both
stages supported, so no \Htwo{} outcome could have made this study positive:
the null was not written after seeing the numbers. Three earlier sealing
episodes were each sealed and each abandoned before any result was read, each
carrying its own attested invalidation record written and dated before the next
began, because an audit trail that records only successes is not evidence of
care but evidence of selective recording. Sealing the analysis before the
randomness protects each episode against fitting the analysis to its own draw;
it does not address knowledge carried across episodes --- about timing, about
the instrument, about which parts of the design were fragile --- which could
have shaped the design and thresholds the executed run was sealed with.
\annexref{B} gives the full event timeline, the digests, the \emph{Merkle
root}s and the attestation state of every proof; \annexref{C} documents the
abandoned episodes and every deviation.

\subsection*{Data and code availability}\label{sec:availability}

The two datasets are released as \dsone{} (single-input challenges) and
\dstwo{} (interaction challenges). Together they comprise the model outputs,
the compiled programs, the response tensor archives, the assignment, and the
full provenance chain. They are an explicit deliverable rather than a
by-product, archived off-machine against a documented schema so that secondary
analyses and replications do not depend on this machine, this filesystem
layout, or these authors. \annexref{D} is the data descriptor: it gives the
schema, the variable dictionary, the licence, and the citation to use for each
dataset. \annexref{E} gives the environment and the procedure for re-running
the analysis. The digests that pin every released artifact are in
\annexref{B}, including the digests of the code.

Both are complete registered corpora and both are released on the same terms.
\dstwo{} in particular is not an exploratory dataset and not a damaged one: all
\ncalls{} registered samples completed, 12 of its 13 validity gates passed with
the single failure being the registered single-execution-attempt counter, and
re-hashing its tensor corpus returned \num{1280} digests --- 640 archives and
their 640 sidecars --- with zero mismatches and zero missing. A reuser working
through the provenance chain will meet bound historical artifacts carrying
mechanically-written driver fields that read \texttt{INVALID}, \texttt{h2\_valid
= false} and a post-hoc provenance constant. Those are driver and provenance
fields recording how the run was executed. They are not this publication's
scientific verdicts, which are \notsupported{} for \Hone{}, \notsupported{} for
\Htwo{} and \notsupported{} jointly. \Cref{tab:status-resolution} states the
resolution, \annexref{C} gives the field-by-field reading, and a
publication-level status record ships alongside the data carrying the same
thing in machine-readable form; \annexref{D} names it and documents its
schema. Numerical values come from the released raw dump; scientific claims
and verdicts come from this paper and \annexref{C}.

The per-block effects figure and the per-family effects figure are released as
separate vector PDF files, drawn from the same extraction that supplies the
prose. Their digests are computed when this document is generated and recorded
in \annexref{B}, so a figure that was re-rendered after the text was written
cannot go unnoticed.

\subsection*{Preregistration statement}\label{sec:prereg-statement}

The introduction, the methods, and the preregistered outcome and claim matrix
--- including the support criteria, the thresholds, the two transforms, and the
sentence to be published under each verdict --- were compiled on 2026-08-04 and
sealed with an \emph{RFC 3161} timestamp and three \emph{OpenTimestamps}
calendar commitments, before the assignment randomness existed. That file is
unmodified on disk and its proofs verify against it directly. \annexref{A}
reproduces the sealed text byte for byte, unaltered, including the three
statements execution has since overtaken, together with the digest needed to
check it against the timestamp proofs.

\section{Conclusion}\label{sec:conclusion}

Both registered stages ran to completion on all \ncalls{} samples, and the
preregistered conjunction is \notsupported{}. Of 108 scientific criterion
evaluations, 19 passed.

\Hone{} is a null. Under single-input challenges, presenting the same
registered propositions as explicit hold-change-direction instructions rather
than as \prose{} was not associated with measurably more similar executable
behaviour between two model snapshots. The smallest exact \pv{} anywhere in
that stage, across ten tests, is \pval{0.0593}, and closed-set identifiability
sat at chance. Two explanations that would have written the null off as
instrument failure --- a floored divisor and a global support floor --- were
excluded; a third, the matching structure, was not.

\Htwo{} is a null with one endpoint inside it that moved. Under interaction
challenges, matched-distance reduction was positive and nominally significant
in both pipelines, and in the \linpipe{} it was the one test in the study to
clear the Bonferroni-corrected threshold, surviving deletion of its largest
block. Under the rank transform the same contrast is an order of magnitude
smaller and misses the registered magnitude floor, so this endpoint does not
satisfy the registered rule for \Htwo{}. The joint conjunction is
\notsupported{} because both registered stages independently failed the
registered support rule --- \Hone{} at 7 of 54 evaluations and \Htwo{} at 12 of
54 --- and not because of any single cell. The study cannot say which of the two
transforms is telling the truth about the size;
\Cref{sec:discussion:disagreement} sets out both readings.
Everything else in that stage is flat: identifiability at chance, \emph{conditional
alignment} orders of magnitude below threshold, every family-invariance and
rendering-relation-invariance criterion failing. The theory predicted a
uniform, family-invariant, classifiable geometry, and that is not what came
out.

No test comparing the two stages was registered or run, so nothing here
establishes that the picture differs between single-input and interaction
challenges. \Hone{} and \Htwo{} are two hypotheses reported side by side, not a
contrast.

What the study also establishes is procedural, and it was the harder half of
the work: an experiment of this kind can be run so that its whole course is
checkable afterwards against records the authors do not control. \Hone{}'s null
was fixed and attested hours before anyone had seen a number that might have
made a different answer attractive. The one result that moved is reported at
the size it came out, with the instrument that disagrees about that size
printed beside it.

%% file: supplement/sections/S1-design-detail.tex
\section{Design and materials in full}\label{supp:design}

This section states the design at the length the paper could not carry. It
covers the units and the no-interference conditions, the two pinned model
snapshots, the five account cards and their proposition-by-proposition
formalization audit, the two matched renderers, the two independent wordings,
the public assignment procedure, the frozen program language, the probe
battery, and the quantities the evaluator computes. Nothing here is new
analysis; it is the registered apparatus written out.

Both registered stages completed and the registered hypothesis is
\notsupported{}, the registered verdict token for a valid execution in which one
or more preregistered support criteria fail. This section describes the
apparatus that produced that outcome. It reports none of the outcome.

Two labels recur and are glossed once here for this document:
\Hone{} denotes \emph{single-input challenges} and \Htwo{} denotes
\emph{interaction challenges}. Each is one of the two registered hypothesis
stages.

\subsection{Extended related work}\label{supp:design:related}

\studyLong{} is not the first natural-language-to-program system, the first
LLM autoformalization study, or the first attempt to automate cognitive
modeling. The gap it occupies is narrow, and it is easier to see against the
four literatures it borders.

The motivating problem is older than any of them. Psychological theories are
often stated at a level that does not uniquely determine a computer program.
\citet{guest2021} argue that computational modeling is valuable partly because
implementation forces assumptions into the open. \citet{oberauer2019} similarly
distinguish broad organizing frameworks from theories capable of risky
quantitative prediction. Formalization can therefore reveal ambiguity, but it
does not guarantee that two implementers will resolve that ambiguity in the
same way.

An LLM makes this problem measurable. Treat the model call as a stochastic
translation step: one anonymous verbal account goes in, one executable program
comes out. If two model snapshots receive the same semantic statements and
repeatedly build different response functions, the prose has not constrained
this translation apparatus tightly enough to produce cross-model agreement.

\subsubsection{Autoformalization and cognitive-model generation}

\citet{wu2022} demonstrated LLM autoformalization from informal mathematics to
a formal language. \citet{hahn2022} fine-tuned language models to translate
English into regular expressions, first-order logic, and linear-time temporal
logic, and found that pre-training generalization carried over to unseen
variable names and rephrased operators. \citet{zhang2024} later showed that
syntax feedback, retrieval, denoising, and correction can improve consistency
across models when constructing mathematical libraries. The AutoSpec+ system of
\citet{wen2026} uses symbolic verification to constrain and check
LLM-generated program specifications. SpecFix, from \citet{jia2025}, studies
ambiguous problem descriptions through the distribution of programs they
induce, then repairs wording and tests downstream generation. NL2CA, from
\citet{deng2026}, translates descriptions of cognitive decision rules into
temporal logic and executable production rules. GeCCo, from \citet{rmus2025},
uses LLMs to identify or generate cognitive models from behavioral data.

Those systems address translation, correction, verification, or model recovery.
\studyname{} does something narrower: it randomizes one matched presentation
contrast while holding the semantic proposition strings fixed, then compares the
resulting programs by executing them on unseen numeric interventions. It
neither repairs the prompts nor optimizes a program against behavioral data.

\subsubsection{Prompt sensitivity}

Prompt sensitivity is an immediate alternative explanation.
\citet{hua2025} showed that apparent sensitivity can be inflated by rigid answer
extraction and other evaluation artifacts. \citet{liu2026} analyzed why
meaning-preserving prompt templates can still produce different model behavior.
\citet{xie2026} found that lexical choice alone, holding the task fixed,
measurably shifts output quality.

That last result is the sharpest threat to \studyname{}. The two renderers share
the registered proposition strings but not the surrounding template words, so
any observed contrast is compatible with a purely lexical effect. \studyname{}
therefore cannot treat any output difference as evidence of deeper formalization
quality. Its response is methodological:

\begin{itemize}
  \item the output language is strict and shared across arms;
  \item no regular-expression answer matcher or LLM judge decides correctness;
  \item the same proposition strings appear in both renderers;
  \item programs are executed on frozen numeric probes;
  \item direct same-account distance is reported separately from closed-set
        identifiability; and
  \item unsigned, row-free, and broken-alignment controls test whether a result
        depends on the claimed response structure.
\end{itemize}

This does not eliminate every prompt effect. It makes the exact renderer effect
and its endpoint explicit.

\subsubsection{Interventions and model discrimination}

In causal discovery, interventions can separate structures that observational
data leaves equivalent; \citet{hauser2012} characterize the equivalence classes
that interventional data collapses. In cognitive model comparison,
\citet{myung2009} show that experiments can be chosen to maximally discriminate
rival models. \studyname{} borrows that general logic but not its formal
guarantees: its programs are not established causal models of people, and its
probes are fixed evaluation inputs rather than experiments on human
participants. \citet{liu2025} also found that explicit conditional or
program-like structure can improve LLM causal-reasoning performance;
\studyname{} differs by randomizing two complete renderers and measuring
executed cross-model response similarity rather than answer accuracy.

The contribution, had the conjunction been supported, would have been a
randomized evaluation method and a bounded result about these renderers. It is
not a newly discovered cognitive mechanism.
The phrase \emph{intervention-response similarity} is used for measured
similarity under the frozen probes; \studyname{} does not claim a mathematical
equivalence class, and the novelty claim is deliberately limited to this
prospective AI formalization methodology. It claims no priority over general
prompt-sensitivity experiments, autoformalization, cognitive-model automation,
causal discovery, or optimal model discrimination.

\subsection{Randomized unit, treated unit, and no-interference
conditions}\label{supp:design:unit}

The treatment-bearing unit is an \textbf{anonymous renderer slot}. Each block
contains two slots, so the experiment contains \textbf{32 slots in \nblocks{}
blocks}. A slot is not a single call: it collects the 10 calls
(two model snapshots $\times$ five account axes) that share that block and that
slot label.

The \textbf{independently randomized unit is a block}, not a slot. The design
draws \nblocks{} independent fair bits, one per block. A block's bit fixes a
complete, complementary renderer-to-slot mapping for that block: exactly one of
its two slots carries the \contract{} and the other carries \prose{}. The two
slots inside a block are therefore never independently randomized; they are a
matched pair under one shared coin.

\subsubsection{The two senses of ``block''}

The word \emph{block} is used in two senses in this package and they are always
distinguished by qualification. Unqualified, a \textbf{block} is one of the
\nblocks{} randomized blocks just defined: the unit of randomization, carrying
one fair bit, two renderer slots, and 20 of the \ncalls{} preauthorized
requests. The probe bank separately divides its anchors into \textbf{calibration
blocks} and \textbf{probe blocks}, which are groupings of frozen probe content,
are never randomized, and carry no fair bit; there are eight calibration blocks
and twelve probe blocks for \Hone{}, with twelve more for \Htwo{}. Those two
terms always appear with their qualifier. Every randomized block presents the
entire probe bank, so the two partitions are orthogonal rather than nested.

Calls, accounts, probes, anchors, outputs, and shuffle controls are repeated
measurements inside a slot and are never treated as independent randomized
observations.

\subsubsection{The registered SUTVA-style conditions}

Write $Y(b,s)(\text{contract})$ and $Y(b,s)(\text{prose})$ for the two potential
response tensors of slot $s$ in block $b$. Only one of the two is ever observed
for any slot. The registered analysis requires two conditions on these potential
outcomes, and they are the whole of what was registered.

\begin{description}
  \item[No interference between slots.] A slot's potential response tensor
    depends only on the renderer assigned to that slot, not on the assignment of
    any other slot or block. This is credible here because each of the
    \ncalls{} preauthorized requests is an independent stateless call, issued at
    most once, with no shared conversation state and no cross-request memory.
  \item[One version of each renderer.] Each renderer is a frozen template
    applied to frozen propositions, byte-identical in every block, so
    ``contract'' and ``prose'' name exactly one intervention each.
\end{description}

Both conditions are assumptions about the apparatus, not results. Neither is
testable from the collected data, and both are stated here so that a reader can
reject the causal reading if they reject the assumptions.

\subsubsection{Block composition}

\begin{table}[htbp]
  \centering
  \caption{Composition of one randomized block. The full experiment contains
           exactly $16 \times 2 \times 2 \times 5 = 320$ preauthorized API
           requests.}
  \label{tab:block-composition}
  \begin{tabular}{@{}lS@{}}
    \toprule
    Component & {Levels} \\
    \midrule
    Pinned model snapshot     & 2  \\
    Anonymous renderer slot   & 2  \\
    Anonymous account axis    & 5  \\
    Calls per block           & 20 \\
    \bottomrule
  \end{tabular}
\end{table}

\Cref{tab:block-composition} gives the composition of a single randomized block.
The experiment driver may issue at most one POST for each of the \ncalls{}
requests. Provider-side execution count cannot be independently proven.

\subsection{The two pinned model snapshots}\label{supp:design:snapshots}

Two snapshots translate every account:

\begin{itemize}
  \item \snapshotA{}
  \item \snapshotB{}
\end{itemize}

They are fixed comparison cases, not a random sample of language models. Every
scientific generation call uses temperature \texttt{1.0}, \texttt{top\_p} of
\texttt{1.0}, zero frequency and presence penalties, medium verbosity, and the
remaining explicit response controls bound into the request contract.

\subsubsection{Why dated snapshots rather than model families}

The claim under test is about a \emph{fixed} theory-to-program apparatus: with
the words carrying the theoretical content held fixed, does the presentation
format change the program that comes out? That question is only well posed if
everything other than the renderer is pinned. A model family is not a pinned
object --- it is a moving set of weights behind a moving alias --- so a family
cannot serve as the fixed half of the apparatus. A dated snapshot can, and the
generation settings above pin the sampling distribution on top of it.

The cost is stated plainly rather than argued away. Because the two snapshots
are fixed comparison cases and not a sample, the design supports no inference to
their families, to other providers, or to language models in general. In exactly
the same way, and for exactly the same reason, any supported claim applies only
to the complete pair of renderers described in
\Cref{supp:design:renderers}. A supported result would be a randomized finding
about an AI formalization method. It would not prove that either program is
faithful to the literature, reveal a human cognitive architecture, validate a
psychiatric theory, or establish a universal benefit of structured prompts.

\subsection{The five engineered account cards}\label{supp:design:cards}

The cards are inspired by (1) source monitoring; (2) comparator processing;
(3) precision weighting; (4) circular inference; and (5) decision thresholds.
Each card contains five proposition objects. The generation model never sees the
account names.

This is not a label-recovery task for the generation model. Each API call sees
one anonymous account only. It receives no account name, competing account,
target program, example solution, probe, scoring rule, or output from another
call. Labels are used only by the frozen evaluator after a blinded score
artifact has been committed.

\subsubsection{The three-way formalization audit}

Every one of the 25 semantic propositions is classified in a separate
formalization audit as:

\begin{description}
  \item[L --- literature core:] a broad directional relationship central to a
    cited account;
  \item[O --- operational bridge:] the mapping from a broad idea to the public
    variables and outputs in this experiment; or
  \item[D --- designed discriminator:] a dominance or interaction challenge
    created to separate the five engineered cards.
\end{description}

Only \texttt{L} is attributed directly to the cited authors. The exact
variables, source categories, output groups, sparse quadratic language, finite
differences, support masks, interaction pairs, and thresholds are \studyname{}
design choices. The executable inventory contains \textbf{2 literature-core
propositions, 14 operational bridges, and 9 designed discriminators}. That
conservative count prevents broad theoretical inspiration from being presented
as 25 literature-derived predictions. The exact 25-position class vector, not
only these three totals, is registered in the frozen design source and asserted
by the executable design audit, so a reclassification that preserves the totals
still fails.

\subsubsection{Card provenance}

\paragraph{Source monitoring.} \citet{johnson1993} describe source monitoring as
the processes used to attribute memories, knowledge, and beliefs to their
origins. \studyname{} maps that broad framework to four source-output
coordinates and public cues for origin category, cue strength, recurrence,
parent reference, confidence, and reliability. The exact four-category output,
the cue-relative coordinate frame, and every dominance or interaction
requirement are engineering choices; none of this card's five propositions is
labeled literature core.

\paragraph{Comparator processing.} \citet{frith2000} discuss internal
representations of intended, predicted, and actual states in the awareness and
control of action. \studyname{} abstracts this into a bounded forecast-evidence
transaction. The signed reversal of a prediction-error response is treated as
literature core; forecast, support, memory, and continuation outputs are
operationalizations or discriminators. This is not a motor-control simulation.

\paragraph{Precision weighting.} \citet{adams2013} describe predictive-coding
accounts in which prediction errors are weighted by expected precision.
\studyname{} has no direct measure of neural or expected precision; it uses
textual reliability, parse confidence, dependency, and source uncertainty as
observable proxies. The card is a precision-inspired engineering abstraction,
not an implementation of the computational anatomy of psychosis.

\paragraph{Circular inference.} \citet{jardri2013} model circular message
passing in which information can reverberate through a hierarchy and be counted
repeatedly, and \citet{jardri2017} later reported behavioral evidence consistent
with circular inference in schizophrenia. \studyname{} treats excessive reuse
through recurrent dependence as literature core, and the registered change
channel for that proposition is an increase in memory activation or dependency
load; content similarity, previous signed update, parent lineage, and the
registered superadditive interaction are engineering bridges or discriminators.
The program is not a neural simulation of excitation-inhibition balance,
hallucination, delusion, or schizophrenia.

\paragraph{Decision threshold.} Sequential-sampling models describe decisions as
evidence accumulation toward a boundary; \citet{ratcliff2008} separate evidence
quality, boundary setting, starting point, and nondecision components, and
\citet{gold2007} discuss deliberation and commitment. None of this card's five
propositions is labeled literature core: the cited accounts do not state a
directional relationship between a continuation-value feature and a
stopping-pressure output in a bounded sparse quadratic DSL. No choice or
reaction-time data are fitted, so this is not a diffusion-model implementation.

\subsubsection{Proposition-by-proposition audit}

The IDs in \Cref{tab:proposition-audit} refer to semantic positions shared by the
independently written \texttt{R1} and \texttt{R2} versions. The exact prompt
strings are frozen in the registered design source and reproduced in
\annexref{A}; the summaries state the same registered relationships in shorter
form.

\begin{longtable}{@{}l >{\raggedright\arraybackslash}p{\dimexpr\linewidth-9em\relax} c@{}}
  \caption{The 25 registered propositions and their formalization class.
           \texttt{L} = literature core, \texttt{O} = operational bridge,
           \texttt{D} = designed discriminator. Totals: 2 \texttt{L},
           14 \texttt{O}, 9 \texttt{D}.}
  \label{tab:proposition-audit} \\
  \toprule
  ID & Registered relationship & Class \\
  \midrule
  \endfirsthead
  \multicolumn{3}{@{}l}{\itshape (continued)} \\
  \toprule
  ID & Registered relationship & Class \\
  \midrule
  \endhead
  \midrule
  \multicolumn{3}{r@{}}{\itshape continued overleaf} \\
  \endfoot
  \bottomrule
  \endlastfoot
  SM-P1 & Clearer definite origin cue increases movement toward the cued source & O \\
  SM-P2 & Weak or absent origin cues increase mixed or unknown attribution & O \\
  SM-P3 & Recurrence increases ambiguity and memory carry more without a parent reference & D \\
  SM-P4 & Confidence or reliability strengthens the clear-origin response & O \\
  SM-P5 & Under attribution-support conflict, source outputs dominate support carry & D \\
  \addlinespace
  CP-P1 & Reversing discrepancy sign reverses the signed forecast-error response & L \\
  CP-P2 & A larger forecast gap increases forecast and support correction & O \\
  CP-P3 & Greater clarity or reliability strengthens mismatch correction & O \\
  CP-P4 & Replacing forecast conflict with alignment reduces carried error, memory, and continuation pressure & D \\
  CP-P5 & Forecast mismatch effects dominate a change in origin identity alone & D \\
  \addlinespace
  PW-P1 & Stronger explicit reliability increases evidence-driven updating & O \\
  PW-P2 & Greater parse confidence increases evidence influence & O \\
  PW-P3 & Reliability and parse confidence together exceed either isolated increase & D \\
  PW-P4 & Dependence or source uncertainty reduces effective precision and update influence & O \\
  PW-P5 & Precision-cue effects dominate recurrence alone & D \\
  \addlinespace
  CI-P1 & Greater recurrence increases carry from the previous update & O \\
  CI-P2 & Greater memory activation or dependency increases reuse of prior support & L \\
  CI-P3 & Recurrence combined with memory or dependency exceeds either isolated response & D \\
  CI-P4 & Removing lineage or reducing recurrence decreases memory and dependency carry & O \\
  CI-P5 & Reversing the prior signed update reverses signed carried support & O \\
  \addlinespace
  DT-P1 & Greater confidence increases stopping pressure & O \\
  DT-P2 & Greater value of another sample decreases stopping pressure & O \\
  DT-P3 & Confidence combined with lower continuation value exceeds either isolated commitment effect & D \\
  DT-P4 & Dependence in repeated evidence reduces the value of another repeated sample & O \\
  DT-P5 & Policy and boundary outputs dominate support and forecast outputs & D \\
\end{longtable}

An earlier draft classified DT-P2 as literature core. That was inconsistent with
DT-P1, which states the structurally symmetric claim about the same
stopping-pressure output and was already an operational bridge. DT-P2 was
reclassified to \texttt{O} before the preregistration seal. No citation was
added in its place, because no cited source states the directional claim in this
form.

\subsection{The two matched renderers}\label{supp:design:renderers}

Within a block, model side, and account, both prompts reuse the exact same
proposition strings and the exact same feature, monomial, output, and account
adapter. This is the guarantee that makes the contrast a presentation contrast:
the bytes carrying the theoretical content are shared, and only the frame around
them changes.

\Prose{} renders each proposition as:

\begin{lstlisting}
With [hold], if we [change], then [response].
\end{lstlisting}

The \contract{} renders the same strings with the labels \holdfixed{},
\intervene{}, and \reqdir{}:

\begin{lstlisting}
HOLD FIXED: [hold]
INTERVENE: [change]
REQUIRED DIRECTION: [response]
\end{lstlisting}

The prompt-word imbalance must remain at or below 3\%.

The manipulation is the complete renderer package: it changes labels,
capitalization, bullets, and layout together. \studyname{} cannot identify which
ingredient caused an effect, and any supported claim applies only to this
complete pair of renderers.

\subsection{Two independent wordings per card}\label{supp:design:wordings}

Each account has two independently written versions, \texttt{R1} and
\texttt{R2}. Four blocks use each cross-model wording relation:
\texttt{R1}--\texttt{R1}, \texttt{R2}--\texttt{R2}, \texttt{R1}--\texttt{R2},
and \texttt{R2}--\texttt{R1}. The registered matched-distance and
relative-geometry effects must be positive in all four cells.

Identical wording across the two model sides is therefore unnecessary, and a
pooled effect cannot hide a reversed wording relation.

\subsection{Public random assignment}\label{supp:design:assignment}

Renderer assignment is unknowable when the study is committed. The procedure is
fixed in advance and has no branch that the experimenter can take.

\begin{enumerate}
  \item The complete apparatus, runtime, documents, tests, criteria, and failure
        rules are written into the preregistration seal.
  \item The seal is timestamped before one exact future \emph{NIST Randomness
        Beacon} 2.0 pulse, fixed 120 one-minute pulses after the observed
        anchor.
  \item One durable assignment-attempt marker starts a six-hour availability
        window. The monitor may send only bodyless \texttt{HEAD} requests to the
        exact target URI. Redirects are rejected.
  \item After the first \texttt{HEAD 200}, one durable target-fetch marker is
        written. The target is then fetched exactly once with \texttt{GET}, and
        its raw bytes are persisted before parsing.
  \item The target pulse, its predecessor, and the Beacon certificate are
        verified against the pinned certificate chain, the RSA-4096/SHA-512
        signature, the output hash, the predecessor link, and the predecessor
        commitment.
  \item The assignment digest is the SHA-256 of a fixed domain separator
        concatenated with the sealed preregistration digest bytes and the target
        pulse output-value bytes.
  \item The first 16 digest bits, most-significant bit first, assign the
        \contract{} to neutral slot \texttt{U} or \texttt{V} in each block.
  \item The remaining 240 disjoint bits generate opaque IDs, adapters,
        anonymous axis order, and execution order.
\end{enumerate}

A timeout, failure of the sole target \texttt{GET}, or a cryptographic failure
stops the study. There is no alternate pulse, seed, target retry, balancing
rule, or reroll.

\subsubsection{The complete complementary within-block mapping}

Step 7 does not assign one slot and leave the other free. Each block has exactly
two neutral slots, \texttt{U} and \texttt{V}, and that block's single bit
selects which of the two carries the \contract{}; the other necessarily carries
\prose{}. The mapping is therefore complete --- every slot in the block is
assigned --- and complementary --- the two slots always carry different
renderers. A block never has two contract slots or two prose slots, and there
are exactly two possible assignments per block rather than four. This is what
makes the two slots of a block a matched pair under one coin, and it is why the
block, not the slot, is the independently randomized unit.

\subsubsection{The inference distribution}

The inference distribution contains all $2^{16} = \nperm{}$ block-level
renderer-label assignments while holding the 240 nuisance bits fixed. An
all-zero-versus-all-one audit must show that changing the 16 assignment bits
changes only the renderer-to-slot mapping and nothing else.

The randomness source, the pulse identifiers, the digests, the timestamps, and
the chain of custody are in \annexref{B}. The four sealing episodes, three of
which were abandoned, are in \annexref{C}.

\subsection{The frozen program language}\label{supp:design:language}

Each call returns strict JSON describing 11 equations. It cannot return Python
or free prose.

\subsubsection{Inputs}

The language exposes 40 bounded features, each finite and inside $[-1,\,1]$,
drawn from current state, current evidence, and proposed-action context.
\Cref{tab:feature-inventory} lists them by group.

\begin{table}[htbp]
  \centering
  \caption{The 40 bounded input features, by group. Confirmatory \Hone{} and
           \Htwo{} interventions alter state or evidence only; action features
           remain fixed context.}
  \label{tab:feature-inventory}
  \begin{tabularx}{\linewidth}{@{}lX@{}}
    \toprule
    Group & Features \\
    \midrule
    Current state &
      centered belief; centered forecast; four centered source probabilities;
      centered memory activation; centered dependency load; centered
      continuation value; centered confidence; previous signed update; and
      bounded turn index \\
    \addlinespace
    Current evidence &
      signed stance; signed reliability direction; centered reliability
      strength; four origin indicators; centered origin strength; three
      parent-reference indicators; centered content similarity; three
      forecast-relation indicators; and centered parse confidence \\
    \addlinespace
    Proposed action context &
      centered proposed belief; four centered proposed source probabilities;
      centered proposed forecast; centered proposed continuation probability;
      and five retrieval-choice indicators \\
    \bottomrule
  \end{tabularx}
\end{table}

``Centered'' has a fixed executable meaning. An ordinary probability or strength
$p$ becomes $2p - 1$. Each coordinate $p$ of a four-way source distribution
becomes $4p - 1$, so the uniform value $0.25$ becomes $0$. Signed evidence
direction and previous signed update are used unchanged. Turn index becomes
$\min(1, \text{turn index} / 10)$. A categorical indicator is $1$ for the
observed category and $0$ for every other category in that group. The
always-false stopped flag and the always-zero citation count are excluded.

\subsubsection{The 823 legal monomials}

The compiler pre-enumerates every linear feature and every legal unordered
quadratic product, including squares where legal. Factor order is canonical, so
$x_i \cdot x_j$ and $x_j \cdot x_i$ are the same term. The compiler removes
indicator squares and products between indicators that are mutually exclusive
within the origin, parent-reference, forecast-relation, or retrieval-choice
groups. The resulting language contains exactly \textbf{823 legal monomials}.

\subsubsection{Equations and the compiler}

Each equation has one integer bias and exactly six term records. A record
selects one opaque monomial, or the reserved neutral monomial \texttt{zero}, and
an integer coefficient from $-20$ to $20$, including zero. The six records are
an unordered additive multiset: repeated monomials add, opposite coefficients
cancel, record order has no meaning, and \texttt{zero} contributes nothing.

The deterministic compiler maps opaque IDs to registered monomials, removes
neutral contributions, groups duplicates as exact integers, removes exact
cancellation, and sorts only to create a stable artifact. Raw provider output
and compiled artifact are separately preserved and hash-bound. Compilation never
repairs, substitutes, clips, or asks for another response.

The fixed divide-by-seven normalization proves that every generated output is
bounded, exactly neutral at strength zero, and exactly proportional to strength
without runtime clipping. Invalid JSON, an incorrect request echo, a wrong
account key, missing or extra outputs, unknown terms, malformed coefficients, or
a seventh term invalidates the call.

\subsubsection{Outputs and adapters}

Each program defines 11 outputs: belief shift; forecast shift; raw self-,
external-, mixed-, and unknown-source shifts; memory shift; dependency shift;
continuation shift; departure of update gain from one; and the stopping-boundary
term.

Every feature, monomial, output, and anonymous account receives an opaque ID; an
\textbf{adapter} is the bijective lookup table mapping semantic names to opaque
IDs and back. There are 160 pair-specific adapters, one for every block
$\times$ model side $\times$ anonymous account pair. The two renderer slots
share the adapter within a pair, which is what allows the two arms of a block to
reuse byte-identical proposition strings and byte-identical symbol names. Opaque
IDs and registry order depend only on neutral positions and the 240 nuisance
bits.

\subsection{The probe battery}\label{supp:design:probes}

All probes are deterministic and frozen before generation. Confirmatory
interventions alter state or evidence only; proposed actions remain fixed
context.

\subsubsection{Common calibration reference}

Eight calibration blocks use three fields: reliability direction; current
belief; and previous signed update. For each randomized block, both renderer
slots and both model sides are pooled into one four-panel, arm-blind reference.
A \textbf{calibration panel} is one of those four cells: one renderer slot on
one model side. Each randomized block has exactly four, being its two slots
crossed with the two model sides, and pooling all four is what makes the
reference arm-blind.

\subsubsection{The five senses of ``panel''}\label{supp:design:panel-senses}

\emph{Panel} is registered vocabulary in five different places, with five
different referents and five different counts. The names below are the
registered ones and cannot be changed without renaming criteria, so they are
disambiguated here instead. Every use of the word in this package is one of
these five, and each is qualified at the point of use.

\begin{description}
  \item[Calibration panel --- four per block.] The cell just defined: one
    renderer slot on one model side. These are the pooling cells of the
    arm-blind scale reference, and the unit counted by the registered criterion
    \texttt{all\_blocks\_use\_one\_common\_four\_panel\_calibration}.
  \item[Probe panel --- three in the frozen battery.] One of the three named
    groups the frozen probe bank is divided into: the eight calibration blocks,
    the twelve \Hone{} probe blocks, and the twelve \Htwo{} probe blocks, in
    the qualified senses of \Cref{supp:design:unit}. The three groups are
    disjoint. This is the sense in which the
    released response tensors are named, shaped and indexed \seeannex{D}, and
    the sense used in the marginal descriptives of
    \Cref{supp:secondary:s2}.
  \item[Invariant panel --- \nblocks{} per stage.] The \contract{} arm's set of
    same-account comparisons within one randomized block, which is the unit of
    the registered criterion
    \texttt{all\_16\_invariant\_panels\_at\_least\_4\_reciprocal}. There are
    \nblocks{} and not $2 \times \nblocks{}$ of these because the criterion is a
    floor on the \contract{} arm alone: \emph{invariant} in a registered key
    name is that arm's name in the analysis code, not a claim that the quantity
    is arm-blind (\Cref{supp:cells}).
  \item[Shuffle panel --- 64 per stage.] One arm of one randomized block under
    one transform family, which is the unit over which the 512
    broken-alignment shuffles are drawn, summarised and hash-bound
    (\Cref{supp:secondary:s4}).
  \item[Loaded panel --- \ncalls{} per stage.] One registered call's response
    tensor set, which is the unit counted by the validity criterion requiring
    that every one of them is loaded exactly once.
\end{description}

\noindent
The first and the last are the pair most easily confused, because both are
counted per block: there are four calibration panels and 20 loaded panels in
each of the \nblocks{} randomized blocks, and the four are a coarsening of the
20 that discards the account axis.

The reference serves two registered purposes:

\begin{itemize}
  \item the \linpipe{} leaves every bounded signed finite difference unchanged;
        and
  \item the \rankpipe{}, a sign-preserving magnitude-rank transform, maps each
        value $v$ to
        \[
          \operatorname{sign}(v) \times
          \text{pooled percentile}(|v|),
          \qquad
          \text{pooled percentile} =
          \frac{\text{lower rank} + \text{upper rank} + 1}
               {2 \times \text{reference count} + 1},
        \]
        with exact zero remaining zero.
\end{itemize}

The rank percentile is strictly positive for every magnitude, so the
transformation is sign-preserving rather than merely odd: a nonzero finite
difference smaller than every calibration magnitude keeps its direction instead
of collapsing to zero. The percentile is nondecreasing in magnitude, assigns
tied magnitudes the same score, and reaches exactly $1$ only strictly above the
largest calibration magnitude. Exactly at the largest calibration magnitude,
held with multiplicity $k$ in a reference of $N$ values, the percentile is
$(2N - k + 1) / (2N + 1)$, which is strictly below $1$. No registered criterion
depends on the boundary value, since the transform is monotone and
sign-preserving either way, but a replicator implementing from this description
would otherwise compute a different number there.

There is no post-transform centering and no unit-length normalization. The
reference also audits structural-zero coordinates and must contain variation
above the frozen floor. Both pipelines must pass every criterion.

\subsubsection{What the calibration reference is not}

The calibration panel's registered role is to be that arm-blind scale reference.
It is not a specificity test or a null control, and a zero response on it is not
evidence of anything. This matters because the reference is deliberately pooled
rather than per-account. Evaluating the five hand-authored canonical witnesses
against the calibration panel gives identically zero tensors for three of them
--- source monitoring, comparator, and decision threshold --- because those
accounts do not use reliability direction, current belief, or previous signed
update at all; only precision weighting and circular inference respond, while
all five respond on the atomic holdout. The canonical witnesses are reference
programs, not predictions of what the models will write, so this is a statement
about which accounts the calibration fields can engage, not a forecast of the
observed reference. It is disclosed because a reader who inspects one account's
calibration tensor will find zeros and should know that this is the account's
structure rather than a failed measurement.

Pooling is what keeps the reference usable. The
\texttt{calibration\_reference\_below\_floor} check of \gate{27} is a global
check --- it fires only when \emph{no} pooled magnitude exceeds the scale floor
--- and the pool spans all five accounts across both renderer slots and both
model sides. The separate per-coordinate audit,
\texttt{structural\_zero\_coordinates}, flags any output coordinate whose spread
across the pooled reference is at or below the floor. Both are reported rather
than absorbed.

\subsubsection{\Honefull{}}

Twelve probe blocks measure central finite differences for ten fields.
\Cref{tab:h1-fields} gives the two fields designated to each account for the
support criterion.

\begin{table}[htbp]
  \centering
  \caption{The ten \Hone{} single-input challenge fields, and the account whose
           support criterion each pair serves. Every generated account is
           evaluated on all ten fields.}
  \label{tab:h1-fields}
  \begin{tabularx}{\linewidth}{@{}lX@{}}
    \toprule
    Account used for the support criterion & Two designated \Hone{} fields \\
    \midrule
    Source monitoring    & origin-cue strength; content similarity \\
    Comparator           & stored forecast; evidence stance \\
    Precision weighting  & reliability-cue strength; parse confidence \\
    Circular inference   & memory activation; dependency load \\
    Decision threshold   & confidence; state continuation value \\
    \bottomrule
  \end{tabularx}
\end{table}

Every generated account is evaluated on all ten fields. For support, each
account's two designated rows must have a raw nonzero response in at least one
defining output group, in both renderers and on both model sides. Activity on
another row cannot satisfy this criterion. The rows are not target vectors and
no literature-specified sign is graded.

\subsubsection{\Htwofull{}}

After \Hone{} execution, analysis, sealing, and independent timestamp
verification, 12 separate blocks measure mixed finite differences for ten pairs.
\Cref{tab:h2-pairs} lists them.

\begin{table}[htbp]
  \centering
  \caption{The ten \Htwo{} interaction challenge pairs, and the account whose
           support criterion each pair serves.}
  \label{tab:h2-pairs}
  \begin{tabularx}{\linewidth}{@{}lX@{}}
    \toprule
    Account used for the support criterion & Two designated \Htwo{} pairs \\
    \midrule
    Source monitoring &
      origin strength $\times$ parse confidence;
      origin strength $\times$ content similarity \\
    Comparator &
      stored forecast $\times$ parse confidence;
      stored forecast $\times$ evidence stance \\
    Precision weighting &
      reliability strength $\times$ parse confidence;
      reliability strength $\times$ dependency \\
    Circular inference &
      content similarity $\times$ memory;
      dependency $\times$ previous update \\
    Decision threshold &
      confidence $\times$ continuation value;
      continuation value $\times$ dependency \\
    \bottomrule
  \end{tabularx}
\end{table}

These are study-created numeric challenges. They are not claimed to be entailed
by the prose cards or by the cited literature, and they are not evidence of
conceptual generalization. Positive and negative mirror orientations remain
separate so a real interaction cannot disappear by cancellation.

\subsubsection{Balanced anchors and the source frame}

Every probe block contains 24 interior anchors and their exact orientation
mirrors, for 48 anchors. The 24 positive anchors form four contiguous
source-origin strata (\texttt{self}, \texttt{external}, \texttt{mixed}, and
\texttt{absent}), with six repeats per stratum. Parent and forecast categories
retain balanced margins and complete pair coverage.

Before averaging origin strata, the four source-output coordinates are permuted
so the category named by the current origin cue is first. For an absent cue, the
unknown-source coordinate is first. This cue-relative frame preserves a shift
toward the currently cued category; averaging fixed absolute source labels would
cancel that effect across the four balanced strata.

\subsection{What the evaluator measures}\label{supp:design:evaluator}

Each generated program is treated as a small machine. The evaluator changes one
field for \Hone{} or two fields for \Htwo{} and records how 11 outputs change.
No LLM grades the program.

The four source outputs are projected to sum to zero, so they contain three
independent coordinates. The evaluator compares ten independent coordinates in
eight equally weighted groups: belief; forecast; source attribution; memory;
dependency; continuation; update gain; and stopping boundary.

For any program from model A and any program from model B, the signed response
distance is the average, over the eight output groups, of $0.5$ times the sum of
squared row-by-row response differences. No account-wise centering and no unit
normalization are applied. Smaller distance means more similar responses to the
same controlled inputs.

\subsubsection{Direct matched-distance reduction, \Mstat{}}

The five same-account comparisons are the diagonal of a $5 \times 5$ distance
matrix. For each block,
\[
  \Mstat{} = \frac{2\,(\text{prose matched distance} -
                       \text{contract matched distance})}
                  {\text{prose matched distance} +
                   \text{contract matched distance}}.
\]
\Mstat{} ranges from $-2$ to $2$. Positive values mean the \contract{} produced
lower direct same-account distance.

The denominator must exceed the frozen floor in every block and, separately, in
each of the 80 block-by-account cells that feed the account-specific reductions.
A floored cell contributes a substituted zero; every such substitution is
counted and disclosed in the analysis artifact, and any substitution fails the
denominator criterion.

The threshold $\Mstat{} \geq \effect{0.15}$ is equivalent to about a 13.95\%
ordinary reduction when prose distance is treated as the baseline. That
equivalence is exact per block and is an upper bound in aggregate. \gate{1} of
the 27 preregistered support criteria is placed on the block \emph{mean} of
\Mstat{}, and the map from \Mstat{} to the ordinary reduction,
$r(\Mstat{}) = 2\Mstat{} / (2 + \Mstat{})$, is strictly concave, so by Jensen's
inequality a mean \Mstat{} of $\effect{0.15}$ implies a mean per-block ordinary
reduction of at most 13.95\%, with equality only when every block carries the
same \Mstat{}. Heterogeneous blocks clear the criterion at a lower mean
reduction: a set of block effects of $\effect{0.5}$ in five blocks and
$\effect{0}$ in the other eleven clears \gate{1} with a mean \Mstat{} of
$\effect{0.15625}$ while the mean per-block ordinary reduction is 12.5\%.

\subsubsection{Relative same-account identifiability, \Gstat{}}

Each of the five diagonal cells is compared with four foils in its row and four
in its column, yielding 40 strict comparisons per block. \Gstat{} is the
proportion won by the same-account match: $\auc{1.00}$ means all 40 comparisons
are won; about $\auc{0.50}$ is chance-like only when distances are exchangeable;
$\auc{0.00}$ means none are won. A tie is not a win. The smallest step is
$1/40 = \auc{0.025}$; an arm effect of $\auc{0.15}$ is six net comparisons.

\Gstat{} is a closed-set relative-identifiability score. It is not direct
agreement, population accuracy, or evidence of literature fidelity. This is why
the two endpoints are kept apart: a renderer could improve relative
identifiability simply by moving all five accounts farther apart while making
the same-account pairs no closer. That cannot support \studyname{}, because both
measures, their exact tests, and their heterogeneity criteria must pass.

\subsubsection{Required negative controls}

The registered verdict also asks what information produced any \Gstat{}
advantage. Three controls are registered.

\begin{enumerate}
  \item \textbf{Unsigned magnitude, $U$.} Replace every response with its
    absolute value. Then
    $D_{\mathrm{directional}} =
      (G_{\mathrm{contract}} - G_{\mathrm{prose}}) -
      (U_{\mathrm{contract}} - U_{\mathrm{prose}})$.
    This requires a signed advantage beyond output support or magnitude.
  \item \textbf{Row-free signature, $S$.} Find the best reassignment of complete
    response rows before computing identifiability. Then
    $D_{\mathrm{signature}} =
      (G_{\mathrm{contract}} - G_{\mathrm{prose}}) -
      (S_{\mathrm{contract}} - S_{\mathrm{prose}})$.
    This requires an advantage beyond an unordered bag of response rows.
  \item \textbf{Broken alignment.} Independently permute complete rows on both
    model sides 512 times, separately for each arm within each randomized block.
    For each arm, that arm's \textbf{conditional alignment} is its signed
    \Gstat{} minus the mean of its own shuffled \Gstat{} values.
    $D_{\mathrm{conditional}}$ is the contract arm's conditional alignment minus
    the prose arm's, and so compares each renderer's aligned \Gstat{} with its
    mean shuffled \Gstat{}. This requires an advantage tied to correct
    cross-model row correspondence.
\end{enumerate}

Conditional alignment is a per-arm level, not a contrast. It is the quantity
\gate{22} places an absolute floor on, and it is distinct from
$D_{\mathrm{conditional}}$, which is the difference between the two arms'
conditional alignments and is what \gate{16} places a floor on. A study can
clear one and fail the other. Because the subtrahend is a mean over 512
permutations, conditional alignment does not lie on the $1/40$ comparison grid
that \Gstat{} itself lies on; the registered floor of $\auc{0.20}$ corresponds
to eight of 40 comparison wins by arithmetic scale rather than as an attainable
grid point.

The \nperm{} shuffle-seed tuples are collision-free. The row-free signature must
be exactly invariant to every registered shuffle within \texttt{1e-12}. The
evaluator also reports reciprocal nearest matches and whether identity is the
unique minimum-cost one-to-one assignment.

The estimand, the estimator, and the exact randomization test that consume
\Mstat{}, \Gstat{}, and these controls are stated in the paper's Methods.

%% file: supplement/sections/S2-support-criteria.tex
\section{The 27 preregistered support criteria}\label{supp:criteria}

\subsection{What the criteria are and how many evaluations they generate}

The 27 preregistered support criteria are the decision rule of \studylong. They
were frozen before data collection and they carry the whole burden of the
verdict: no criterion is weighted, no criterion serves as a tiebreak, and no
criterion is tradeable against another. Each is a boolean on a registered
quantity, and the registered thresholds are part of the sealed text rather than
choices made after unblinding.

Support required all 27 criteria to pass in both registered transforms --- the
\linpipe{} and the \rankpipe{} --- in both hypothesis stages, where a stage is
one of the two registered hypothesis stages: \Honefull{} and \Htwofull{}.
That is 108 scientific evaluations. Requiring both transforms is a deliberate
robustness demand: an effect that survives only one transform is a
transform-dependent effect, and the design declines in advance to call that
support.

A separate validity layer of 13 criteria runs per stage, 26 validity
evaluations in total. Validity is evaluated before, and separately from, the
scientific criteria. Under the registered rule an execution-integrity failure
and a scientific-criterion failure are different outcomes, and it is the second
that yields the registered verdict token \notsupported{} for that stage.

Both stages are reported at that second outcome. \Hone{} passes 13 of its 13
validity criteria and \Htwo{} passes 12 of 13. The single criterion \Htwo{}
does not pass is the registered single-execution-attempt counter, which is a
record of how many dispatches a sample took rather than a check on what the
data contain; its two conjuncts split, with all \ncalls{} samples valid and the
single-attempt flag false. Every criterion in either stage whose subject is the
content of the data passes. Both stages therefore carry the registered verdict
\notsupported{} on the strength of their scientific criteria, and both are
reported on the same footing throughout this supplement.

19 of the 108 scientific evaluations passed. \Hone{} passed 7/54 and \Htwo{}
passed 12/54. The registered conjunction failed, and it failed across the
board rather than at one margin: 89 of the 108 evaluations fail.

\subsection{What the record states per criterion, and what it states in aggregate}

For each of the four cells --- \Hone{} under each pipeline, \Htwo{} under each
pipeline --- the record names \emph{every} criterion that failed, gives the
number that failed, and gives the number that passed. The failing lists are
exhaustive: 23 of 27 failed in \Hone{} under the \linpipe{}, 24 of 27 in
\Hone{} under the \rankpipe{}, 21 of 27 in \Htwo{} under each pipeline. The
passing set in each cell is therefore the complement of that cell's failing
list, and the complement reproduces the reported per-cell totals exactly: 4/27,
3/27, 6/27 and 6/27, summing to the 19 passes above.
\Cref{supp:tab:criteria-results} is built that way.

Eight of the 27 are additionally reported as an explicit four-cell pass/fail
grid in the registered negative-control table: \gates{5, 10, 11, 23, 24, 25, 26}
and~27. Those eight entries agree with the complement in all 32 of their cells.
No cell of \Cref{supp:tab:criteria-results} is inferred from anything other than
the exhaustive per-cell failing lists and that grid.

\subsection{The criteria as registered}

\Cref{supp:tab:criteria-definitions} reproduces the 27 criteria in their
registered order and grouping. \Mstat{} is the matched-distance reduction
between same-account cross-model programs; \Gstat{} is closed-set same-account
identifiability over 40 per-block comparisons. The signed contract-minus-prose
\Gstat{} effect is recorded under the identifier \texttt{raw\_geometry} in the
deposited analysis, and appears under that name in the paper's
primary-endpoint matrix and under the name \emph{signed \Gstat{}} in the
criterion tables here; the two names denote one quantity. Exact one-sided
\pv{} values come from enumerating all \nperm{} sign assignments. The \contract{} renderer and the
\prose{} renderer are the two arms; the criteria below are written on the
contract-minus-prose direction. \Htwo{} applies the same 27 criteria to mixed
finite differences rather than central ones.

\begin{longtable}{@{}r>{\raggedright\arraybackslash}p{0.24\linewidth}>{\raggedright\arraybackslash}p{0.60\linewidth}@{}}
\caption{The 27 preregistered support criteria, in registered order and
grouping. Every criterion had to pass in both pipelines in both stages for the
registered conjunction to be supported.}\label{supp:tab:criteria-definitions}\\
\toprule
No. & Criterion & What it requires \\
\midrule
\endfirsthead
\caption[]{The 27 preregistered support criteria \emph{(continued)}.}\\
\toprule
No. & Criterion & What it requires \\
\midrule
\endhead
\midrule
\multicolumn{3}{r@{}}{\footnotesize\itshape continued on next page}\\
\endfoot
\bottomrule
\endlastfoot

\multicolumn{3}{@{}l}{\textit{Direct same-account closeness}}\\
\addlinespace[2pt]
1 & Mean \Mstat{}, magnitude & Mean \Mstat{} is at least 0.15. \\
2 & Mean \Mstat{}, exact \pv{} & Its exact one-sided \pv{} value is at most 0.05. \\
3 & Account-level \Mstat{} & All five account-specific matched-distance reductions are positive. \\
4 & Wording-relation \Mstat{} & All four wording-relation matched-distance reductions are positive. \\
5 & Denominator floor & Every matched-distance denominator is above its floor, both the 16 pooled block denominators and the 80 block-by-account denominators. \\
\addlinespace[4pt]

\multicolumn{3}{@{}l}{\textit{Relative signed geometry}}\\
\addlinespace[2pt]
6 & Mean signed \Gstat{}, magnitude & Mean contract-minus-prose \Gstat{} is at least 0.15. \\
7 & Mean signed \Gstat{}, exact \pv{} & Its exact one-sided \pv{} value is at most 0.05. \\
8 & Account-level signed \Gstat{} & All five account effects are positive. \\
9 & Wording-relation signed \Gstat{} & All four wording-relation effects are positive. \\
\addlinespace[4pt]

\multicolumn{3}{@{}l}{\textit{Direction beyond magnitude}}\\
\addlinespace[2pt]
10 & Mean $D_{\mathrm{directional}}$, magnitude & Mean $D_{\mathrm{directional}}$ is at least 0.15. \\
11 & Mean $D_{\mathrm{directional}}$, exact \pv{} & Its exact one-sided \pv{} value is at most 0.05. \\
12 & Account-level directional & All five account effects are positive. \\
13 & Wording-relation directional & All four wording-relation effects are positive. \\
\addlinespace[4pt]

\multicolumn{3}{@{}l}{\textit{Row identity and correct alignment}}\\
\addlinespace[2pt]
14 & Mean $D_{\mathrm{signature}}$, magnitude & Mean $D_{\mathrm{signature}}$ is at least 0.15. \\
15 & Mean $D_{\mathrm{signature}}$, exact \pv{} & Its exact one-sided \pv{} value is at most 0.05. \\
16 & Mean $D_{\mathrm{conditional}}$, magnitude & Mean $D_{\mathrm{conditional}}$ is at least 0.15. \\
17 & Mean $D_{\mathrm{conditional}}$, exact \pv{} & Its exact one-sided \pv{} value is at most 0.05. \\
18 & Account-level signature & All five $D_{\mathrm{signature}}$ account effects are positive. \\
19 & Account-level conditional & All five $D_{\mathrm{conditional}}$ account effects are positive. \\
20 & Wording-relation signature & $D_{\mathrm{signature}}$ is positive in all four wording-relation cells. \\
\addlinespace[4pt]

\multicolumn{3}{@{}l}{\textit{Absolute level and auditability}}\\
\addlinespace[2pt]
21 & Absolute contract \Gstat{} & Mean contract-renderer \Gstat{} is at least 0.80, or 32 of 40 comparisons on average. \\
22 & Absolute conditional alignment & Mean contract-arm conditional alignment, averaged over the 16 randomized blocks, is at least 0.20. This is the per-arm level, not the between-arm difference $D_{\mathrm{conditional}}$ of \gate{16}. \\
23 & Unique identity assignment & In each of the 16 randomized blocks, the contract arm's five-by-five cross-model account comparison recovers identity as the unique minimum-cost one-to-one assignment. \\
24 & Reciprocal matching & In each of the 16 randomized blocks, the contract arm's five-by-five cross-model account comparison has at least four of its five reciprocal nearest matches. \\
25 & Symmetric output support & Registered defining-output support is nonzero in both renderers and on both model sides. \\
26 & One calibration reference & Each of the 16 randomized blocks uses exactly one calibration reference, pooled from all four of that block's calibration panels. \\
27 & Calibration variation & Every common calibration reference contains variation above its floor. \\
\end{longtable}

\subsection{Results, criterion by criterion}

\Cref{supp:tab:criteria-results} gives the outcome of all 108 evaluations. The
registered identifier is the literal key the deposited analysis records for each
criterion, so every row is checkable against the data.

Three patterns are worth reading off the table before the individual rows. The
whole of the relative-geometry group --- \gates{6--9} --- fails in every cell,
as does every account-level positivity criterion, and \gates{21} and~22 fail in
every cell, so closed-set identifiability never approaches its registered level.
All of \gates{23--25} fail in all four cells; the first two of those are
structural checks on whether the matching had enough mutual structure to be
meaningful at all, not checks on whether the intervention worked, and their
failure applies equally to \Hone{}'s null and to \Htwo{}'s one positive result.
The calibration criteria, \gates{26} and~27, pass in all four cells.

\begin{longtable}{@{}r>{\raggedright\arraybackslash\ttfamily\footnotesize}p{0.45\linewidth}cccc@{}}
\caption{Outcome of each of the 27 preregistered support criteria in each of the
four cells. Passing evaluations are set in bold; there are 19 of them out of
108. Column headings: \Hone{} and \Htwo{} under the \linpipe{} (lin.) and the
\rankpipe{} (rank).}\label{supp:tab:criteria-results}\\
\toprule
No. & \normalfont\normalsize Registered identifier & \Hone{} lin. & \Hone{} rank & \Htwo{} lin. & \Htwo{} rank \\
\midrule
\endfirsthead
\caption[]{Outcome of each of the 27 preregistered support criteria
\emph{(continued)}.}\\
\toprule
No. & \normalfont\normalsize Registered identifier & \Hone{} lin. & \Hone{} rank & \Htwo{} lin. & \Htwo{} rank \\
\midrule
\endhead
\midrule
\multicolumn{6}{r@{}}{\footnotesize\itshape continued on next page}\\
\endfoot
\bottomrule
\endlastfoot

1 & matched\_distance\_reduction\_\allowbreak at\_least\_0\_15 & fail & fail & \textbf{pass} & fail \\
2 & matched\_distance\_reduction\_\allowbreak exact\_p\_at\_most\_0\_05 & fail & fail & \textbf{pass} & \textbf{pass} \\
3 & all\_five\_matched\_distance\_\allowbreak reductions\_positive & fail & fail & fail & fail \\
4 & all\_four\_rendering\_relation\_\allowbreak matched\_distance\_reductions\_\allowbreak positive & fail & fail & fail & fail \\
5 & all\_matched\_distance\_\allowbreak denominators\_above\_floor & \textbf{pass} & \textbf{pass} & fail & \textbf{pass} \\
\addlinespace[3pt]
6 & raw\_geometry\_effect\_\allowbreak at\_least\_0\_15 & fail & fail & fail & fail \\
7 & raw\_geometry\_exact\_p\_\allowbreak at\_most\_0\_05 & fail & fail & fail & fail \\
8 & all\_five\_raw\_geometry\_\allowbreak family\_effects\_positive & fail & fail & fail & fail \\
9 & all\_four\_rendering\_relation\_\allowbreak raw\_geometry\_effects\_positive & fail & fail & fail & fail \\
\addlinespace[3pt]
10 & d\_directional\_beyond\_unsigned\_\allowbreak at\_least\_0\_15 & fail & fail & fail & fail \\
11 & d\_directional\_beyond\_unsigned\_\allowbreak exact\_p\_at\_most\_0\_05 & fail & fail & \textbf{pass} & fail \\
12 & all\_five\_directional\_\allowbreak family\_effects\_positive & fail & fail & fail & fail \\
13 & all\_four\_rendering\_relation\_\allowbreak directional\_effects\_positive & fail & fail & \textbf{pass} & \textbf{pass} \\
\addlinespace[3pt]
14 & d\_signature\_at\_least\_0\_15 & fail & fail & fail & fail \\
15 & d\_signature\_exact\_p\_\allowbreak at\_most\_0\_05 & fail & fail & fail & fail \\
16 & d\_conditional\_at\_least\_0\_15 & fail & fail & fail & fail \\
17 & d\_conditional\_exact\_p\_\allowbreak at\_most\_0\_05 & fail & fail & fail & fail \\
18 & all\_five\_signature\_\allowbreak family\_effects\_positive & fail & fail & fail & fail \\
19 & all\_five\_conditional\_\allowbreak family\_effects\_positive & fail & fail & fail & fail \\
20 & all\_four\_rendering\_relation\_\allowbreak signature\_effects\_positive & \textbf{pass} & fail & fail & \textbf{pass} \\
\addlinespace[3pt]
21 & invariant\_geometry\_auc\_\allowbreak at\_least\_0\_80 & fail & fail & fail & fail \\
22 & invariant\_conditional\_alignment\_\allowbreak at\_least\_0\_20 & fail & fail & fail & fail \\
23 & all\_16\_invariant\_assignments\_\allowbreak unique\_identity & fail & fail & fail & fail \\
24 & all\_16\_invariant\_panels\_\allowbreak at\_least\_4\_reciprocal & fail & fail & fail & fail \\
25 & all\_registered\_support\_\allowbreak nonzero\_in\_both\_arms & fail & fail & fail & fail \\
26 & all\_blocks\_use\_one\_common\_\allowbreak four\_panel\_calibration & \textbf{pass} & \textbf{pass} & \textbf{pass} & \textbf{pass} \\
27 & common\_calibration\_reference\_\allowbreak above\_floor & \textbf{pass} & \textbf{pass} & \textbf{pass} & \textbf{pass} \\
\addlinespace[3pt]
\midrule
& \normalfont\normalsize Criteria passed, of 27 & \textbf{4} & \textbf{3} & \textbf{6} & \textbf{6} \\
\end{longtable}

On the primary endpoint the criterion with the narrowest pass footprint is
\gate{1}, the magnitude floor on \Mstat{}: it passes in exactly one cell of the
four. The exact \pv{} criterion on the same statistic, \gate{2}, passes in both
\Htwo{} cells: the two pipelines agree on sign and both clear 0.05, and they
part on magnitude. Magnitude is a registered criterion, so \Htwo{} is
\notsupported{} on its own criteria, exactly as \Hone{} is on its own. Each cell
is read inside its own stage. No cross-stage estimator was registered, so this
footprint describes where each stage's criteria fell and is not a measured
contrast. Only \gate{5} passes in three cells and fails in the fourth,
and the failure is in \Htwo{} under the \linpipe{}, on 3 of the 80
block-by-account denominators --- an aggregation level below the one the
registered primary statistic uses.

\subsection{The 13 validity criteria}

Thirteen validity criteria run per stage, 26 evaluations in total, and the
registered rule asks for all of them. Twenty-five passed. The one that did not
is an attempt counter rather than a data check, so both stages are complete
registered stages whose scientific results stand on the same footing; the
paragraph after \Cref{supp:tab:validity-counts} gives the mechanics.
They check execution integrity rather than the science: whether the
assignment, the timestamps, the manifest, the dispatch accounting, the
provider attestations, the tensor replay, the stage ordering and traversal, the
finite-value and clamp rules, the blind scoring commitment and the phase
inventory came out as registered.

\begin{table}[htbp]
\caption{Validity criteria passed, by stage. \Htwo{}'s one non-passing
criterion is the registered single-execution-attempt counter, which is an
attempt counter and not a data check; every criterion bearing on the data
passes in both stages, and both stages carry the registered scientific verdict
\notsupported{}.}\label{supp:tab:validity-counts}
\centering
\begin{tabular}{@{}lc@{}}
\toprule
Stage & Validity criteria passed \\
\midrule
\Hone{} & 13/13 \\
\Htwo{} & 12/13 \\
\bottomrule
\end{tabular}
\end{table}

\Hone{} passed all thirteen. \Htwo{} passed twelve. The one that did not pass is
the single-execution-attempt criterion, which is an attempt counter rather than
a data check: it records whether every sample completed on its first attempt,
and three did not. Its two conjuncts split --- \texttt{all\_320\_valid} is
\emph{true} and \texttt{single\_execution\_attempt} is \emph{false} --- so the
criterion fails on the counter alone. What moved it was a timeout in a local
\texttt{openssl ts -verify} subprocess checking stored study-level attestation
files. That check runs as the execution worker's first action, before any
sample-specific input is read; it opens no socket and has no model in the loop,
so it was not a provider failure, not a model failure, not a data failure and
not a network failure. It affected three \Htwo{} direction-runs, one in each of
three different samples, and only those direction-runs were repeated: 3 of the
640 direction-runs in the stage, not 3 whole samples. All \ncalls{} \Htwo{}
samples completed. The first-attempt record was left on disk unedited rather
than overwritten.

That unedited first-attempt record is why the released provenance chain carries
mechanically-written driver fields reading \texttt{INVALID} and
\texttt{h2\_valid = false}. Those are driver and provenance fields describing
how the stage executed, not this study's scientific verdict, which is
\notsupported{} for \Htwo{} exactly as it is for \Hone{}. The repetition, the
digest-for-digest re-verification run against it, and the field-by-field
reconciliation are in \annexref{C}; the same reconciliation ships in
machine-readable form as a publication-level status record beside the data,
documented in \annexref{D}.

Every validity criterion that bears on the data passes in both stages. All 320
panels loaded exactly once, the manifest verifies, the blind score was committed
before semantic labels were used, the row-free signature is exactly invariant to
the registered shuffles, and the tensor audit found 0 mismatched digests.

The record names the criterion that did not pass and characterises the
data-bearing criteria that did; it does not print all thirteen individually, so
no thirteen-row roster is given here. The registered diagnostics that are
reported whether or not they gate are complete, and all three are clean in both
stages: 320 metadata records read in each stage, 0 transition-clamp events, and
0 evaluator-cap activations. Nothing produced by either analyser --- which is
every criterion, estimate and \pv{} value reported in this supplement --- came
from a clamped or capped value.

\subsection{How far the failures were from passing}

One of the registered secondary analyses, S5 in the sealed numbering, reports
how far each primary test cleared or missed the registered 0.15 floor, so that a
near miss is legible as a near miss. It covers the five magnitude criteria ---
\gates{1, 6, 10, 14} and~16 --- in each of the four cells, 20 combinations in
all.

\begin{longtable}{@{}llS[table-format=-1.8]S[table-format=-1.8]@{}}
\caption{Criterion margins: observed value and signed distance from the
registered 0.15 magnitude floor, for the five magnitude criteria in each cell.
One of the 20 combinations clears the floor.}\label{supp:tab:criteria-margins}\\
\toprule
Cell & Test & {Observed} & {Margin vs.\ 0.15} \\
\midrule
\endfirsthead
\caption[]{Criterion margins \emph{(continued)}.}\\
\toprule
Cell & Test & {Observed} & {Margin vs.\ 0.15} \\
\midrule
\endhead
\midrule
\multicolumn{4}{r@{}}{\footnotesize\itshape continued on next page}\\
\endfoot
\bottomrule
\endlastfoot

\Hone{} \linpipe{} & \Mstat{} & 0.0310494 & -0.118951 \\
\Hone{} \linpipe{} & signed \Gstat{} & -0.0484375 & -0.198437 \\
\Hone{} \linpipe{} & $D_{\mathrm{directional}}$ & -0.0078125 & -0.157812 \\
\Hone{} \linpipe{} & $D_{\mathrm{signature}}$ & 0.015625 & -0.134375 \\
\Hone{} \linpipe{} & $D_{\mathrm{conditional}}$ & -0.0384888 & -0.188489 \\
\addlinespace[3pt]
\Hone{} \rankpipe{} & \Mstat{} & -0.0382657 & -0.188266 \\
\Hone{} \rankpipe{} & signed \Gstat{} & -0.0359375 & -0.185937 \\
\Hone{} \rankpipe{} & $D_{\mathrm{directional}}$ & 0.0515625 & -0.0984375 \\
\Hone{} \rankpipe{} & $D_{\mathrm{signature}}$ & 0.0671875 & -0.0828125 \\
\Hone{} \rankpipe{} & $D_{\mathrm{conditional}}$ & -0.0323242 & -0.182324 \\
\addlinespace[3pt]
\Htwo{} \linpipe{} & \Mstat{} & 0.517398 & 0.367398 \\
\Htwo{} \linpipe{} & signed \Gstat{} & 0.0109375 & -0.139062 \\
\Htwo{} \linpipe{} & $D_{\mathrm{directional}}$ & 0.0296875 & -0.120313 \\
\Htwo{} \linpipe{} & $D_{\mathrm{signature}}$ & -0.00625 & -0.15625 \\
\Htwo{} \linpipe{} & $D_{\mathrm{conditional}}$ & -0.00238953 & -0.15239 \\
\addlinespace[3pt]
\Htwo{} \rankpipe{} & \Mstat{} & 0.0693316 & -0.0806684 \\
\Htwo{} \rankpipe{} & signed \Gstat{} & -0.00625 & -0.15625 \\
\Htwo{} \rankpipe{} & $D_{\mathrm{directional}}$ & 0.078125 & -0.071875 \\
\Htwo{} \rankpipe{} & $D_{\mathrm{signature}}$ & 0.0484375 & -0.101562 \\
\Htwo{} \rankpipe{} & $D_{\mathrm{conditional}}$ & -0.00973206 & -0.159732 \\
\end{longtable}

One combination clears the floor: \Mstat{} in \Htwo{} under the \linpipe{},
which clears it by \effect{0.367398}. Every other margin is negative, most of
them by more than the threshold itself, which is the shape of a null rather than
of a near miss. The closest of the misses is \Mstat{} in \Htwo{} under the
\rankpipe{}, short by \effect{0.0806684}; the next is
$D_{\mathrm{directional}}$ in \Htwo{} under the \rankpipe{}, short by
\effect{0.071875}.

The margin secondary does not cover the two absolute-level criteria, whose
floors are not 0.15. For \gate{21} the floor is \Gstat{} of at least 0.80, and
the observed values run from \auc{0.46875} to \auc{0.5234375}, which is chance
in every cell. For \gate{22} the floor is 0.20, and no cell exceeds
\effect{0.02628479}. Neither is a near miss.

Multiplicity was registered as well. Over the 5 tests, 2 pipelines and 2 stages
there are 20 test evaluations, and the Bonferroni threshold is
$0.05/20 = 0.0025$.
Exactly one observed \pv{} value falls below it: \Mstat{} in \Htwo{} under the
\linpipe{}, at \pval{0.0012512207}. Two more are nominally under 0.05 without
clearing the corrected threshold: $D_{\mathrm{directional}}$ in \Htwo{} under
the \linpipe{} at \pval{0.0351562}, and \Mstat{} in \Htwo{} under the
\rankpipe{} at \pval{0.0284729}. Clearing a corrected threshold is a statement
about the \pv{} value alone; it says nothing about magnitude, and magnitude is
where the two pipelines disagree.

%% file: supplement/sections/S3-endpoint-tables.tex
\section{Complete endpoint results}\label{supp:endpoints}

Two primary endpoints were registered. \Mstat{} is the direct matched-distance
reduction between same-account cross-model programs. \Gstat{} is closed-set
same-account identifiability over 40 per-block comparisons. Both are reported
here for both stages --- \Honefull{} and \Htwofull{} --- and both pipelines, with
the observed magnitude beside the criterion boolean in every cell, and with the
exact \pv{} rather than an inequality.

The reporting rule was fixed in advance. The smallest attainable \pv{} under
this design is $1/65536 = \num{1.526e-5}$, because inference enumerates all $2^{16}$
assignments. Any statement of the form ``\pv{} $<$ 0.001'' would therefore be a
statement about the enumeration rather than about how far into the tail the
observed statistic fell. Exact values are reported throughout.

The registered conjunction failed. No cell of the design passed the 27
preregistered support criteria: 4 of 27 passed in \Hone{} \linpipe{}, 3 of 27 in
\Hone{} \rankpipe{}, 6 of 27 in \Htwo{} \linpipe{}, and 6 of 27 in \Htwo{}
\rankpipe{}.

\subsection{The full test matrix}\label{supp:full-matrix}

Five endpoint families were tested by exact randomization in each of the four
cells, for 20 tests in total. \Cref{supp:tab:full-matrix} gives every one, with
the observed mean, the exact one-sided \pv{}, and the position of the observed
statistic in the enumerated null.

\begin{small}
\begin{longtable}{@{}lrrrcc@{}}
\caption{All 20 registered exact tests: observed mean contrast, exact one-sided
\pv{}, and tail count out of \nperm{} enumerated assignments. All five families
share the registered magnitude threshold 0.15 and the registered \pv{} threshold
0.05. ``Mag.'' is the magnitude criterion, ``\pv{}'' the \pv{} criterion.}
\label{supp:tab:full-matrix}\\
\toprule
Endpoint family & Observed & Exact \pv{} & Tail & Mag. & \pv{} \\
\midrule
\endfirsthead
\toprule
Endpoint family & Observed & Exact \pv{} & Tail & Mag. & \pv{} \\
\midrule
\endhead
\midrule
\multicolumn{6}{r@{}}{\emph{continued}}\\
\endfoot
\bottomrule
\endlastfoot
\multicolumn{6}{@{}l}{\textbf{\Hone{}, \linpipe{}}}\\
matched-distance reduction \Mstat{} & \effect{0.031049424} & \pval{0.39901733} & \num{26150} & FAIL & FAIL \\
raw geometry effect                 & \effect{-0.0484375}  & \pval{0.94567871} & \num{61976} & FAIL & FAIL \\
directional beyond unsigned         & \effect{-0.0078125}  & \pval{0.60580444} & \num{39702} & FAIL & FAIL \\
\texttt{D\_signature}               & \effect{0.015625}    & \pval{0.31304932} & \num{20516} & FAIL & FAIL \\
\texttt{D\_conditional}             & \effect{-0.03848877} & \pval{0.91810608} & \num{60169} & FAIL & FAIL \\
\addlinespace
\multicolumn{6}{@{}l}{\textbf{\Hone{}, \rankpipe{}}}\\
matched-distance reduction \Mstat{} & \effect{-0.038265658} & \pval{0.84689331}  & \num{55502} & FAIL & FAIL \\
raw geometry effect                 & \effect{-0.0359375}   & \pval{0.91552734}  & \num{60000} & FAIL & FAIL \\
directional beyond unsigned         & \effect{0.0515625}    & \pval{0.18435669}  & \num{12082} & FAIL & FAIL \\
\texttt{D\_signature}               & \effect{0.0671875}    & \pval{0.059265137} & \num{3884}  & FAIL & FAIL \\
\texttt{D\_conditional}             & \effect{-0.032324219} & \pval{0.87815857}  & \num{57551} & FAIL & FAIL \\
\addlinespace
\multicolumn{6}{@{}l}{\textbf{\Htwo{}, \linpipe{}}}\\
matched-distance reduction \Mstat{} & \effect{0.51739819}    & \pval{0.0012512207} & \num{82}    & PASS & PASS \\
raw geometry effect                 & \effect{0.0109375}     & \pval{0.30279541}   & \num{19844} & FAIL & FAIL \\
directional beyond unsigned         & \effect{0.0296875}     & \pval{0.03515625}   & \num{2304}  & FAIL & PASS \\
\texttt{D\_signature}               & \effect{-0.00625}      & \pval{0.59716797}   & \num{39136} & FAIL & FAIL \\
\texttt{D\_conditional}             & \effect{-0.0023895264} & \pval{0.58496094}   & \num{38336} & FAIL & FAIL \\
\addlinespace
\multicolumn{6}{@{}l}{\textbf{\Htwo{}, \rankpipe{}}}\\
matched-distance reduction \Mstat{} & \effect{0.069331606}   & \pval{0.0284729}  & \num{1866}  & FAIL & PASS \\
raw geometry effect                 & \effect{-0.00625}      & \pval{0.67211914} & \num{44048} & FAIL & FAIL \\
directional beyond unsigned         & \effect{0.078125}      & \pval{0.1038208}  & \num{6804}  & FAIL & FAIL \\
\texttt{D\_signature}               & \effect{0.0484375}     & \pval{0.07371521} & \num{4831}  & FAIL & FAIL \\
\texttt{D\_conditional}             & \effect{-0.0097320557} & \pval{0.71354675} & \num{46763} & FAIL & FAIL \\
\end{longtable}
\end{small}

The registered test keys, in the order of the table rows, are
\texttt{matched\_distance\_reduction\_test}, \texttt{raw\_geometry\_test},
\texttt{d\_directional\_test}, \texttt{d\_signature\_test} and
\texttt{d\_conditional\_test}.

Three entries in \Cref{supp:tab:full-matrix} are given at reduced precision in
the table for width. Their exact values are \effect{0.5173981945762773} with
\pval{0.001251220703125} for \Htwo{} \linpipe{} matched-distance reduction,
\effect{0.06933160555977207} with \pval{0.028472900390625} for \Htwo{}
\rankpipe{} matched-distance reduction, and \pval{0.05926513671875} for the
\texttt{D\_signature} test in \Hone{} \rankpipe{} --- which is the smallest exact
\pv{} anywhere in \Hone{}, across all ten of its tests, and does not cross 0.05.
Every \pv{} in this design is a multiple of $1/65536$, so the exact value is the
tail count divided by \nperm{}; the table's tail column is therefore the exact
\pv{} in integer form for every row.

Exactly one endpoint moves, and it moves in one stage. \Htwo{}'s
matched-distance reduction \Mstat{} is positive with an exact one-sided \pv{}
below 0.05 in both pipelines. The two pipelines disagree about how large it is:
the \linpipe{} puts \Mstat{} at \effect{0.51739819}, above the registered 0.15
floor; the \rankpipe{} puts it at \effect{0.069331606}, less than half that
floor. Both pipelines agree on sign and both clear 0.05; they part on magnitude,
and magnitude is a registered criterion. The support rule requires the magnitude
criterion to pass in both, and it does not.

The \linpipe{} \pv{} of \pval{0.0012512207} is not merely under 0.05; it is under
the 0.0025 threshold that a Bonferroni correction over all 20 test evaluations
would impose. It is the only value in the study that is.

\subsection{Per-cell detail}\label{supp:cells}

Each cell is reported at full length: the seven registered quantities, the tail
counts, the two matched-distance vectors that \Mstat{} differences, the
denominators, and the complete list of failing criteria. \Mstat{} is a
reduction, so the two quantities it differences are reported rather than only
their difference. A reader can then see whether a reduction came from the
intervention arm moving or from the prose arm moving.

The registered definition and denominator rule are quoted below rather than
paraphrased, which means they arrive in the vocabulary of the frozen analysis
code rather than the vocabulary of this paper. The two are the same two arms
under different names, and the mapping is fixed for the whole study: the
registered name \texttt{causal\_invariants}, abbreviated to \emph{invariant} in
key names and in the quoted rule, is the \contract{}; the registered name
\texttt{narrative} is \prose{}. The released per-call records carry the
registered names \seeannex{D}, so the mapping is what a reader needs in order to
join these tables to the data. It is used without further comment in the
quantity names below --- \texttt{invariant\_matched\_distances} is the
\contract{} arm and \texttt{narrative\_matched\_distances} is the \prose{} arm
--- and in the identifiability quantities and criteria, where \emph{invariant
geometry AUC}, \emph{invariant conditional alignment}, \emph{invariant
assignments} and \emph{invariant panels} are all the \contract{} arm's values of
those quantities and not contrasts between the arms.

\begin{quote}
2 * (narrative matched distance - invariant matched distance) / (narrative
matched distance + invariant matched distance)

every pooled block denominator and every block-by-family denominator must exceed
the frozen scale floor; a floored family denominator contributes a substituted
zero that is disclosed here and fails the registered denominator criterion
\end{quote}

\subsubsection{\texorpdfstring{\Hone{}, \linpipe{}}{H1, linear pipeline}}

Criteria passed: 4 of 27.

\begin{table}[htbp]
\caption{\Hone{}, \linpipe{}: the seven registered quantities against their
registered thresholds.}
\label{supp:tab:h1-linear}
\centering
\begin{tabularx}{\linewidth}{@{}XXrrcc@{}}
\toprule
Quantity & Key & Threshold & Observed & Mag. & \pv{} \\
\midrule
matched-distance reduction \Mstat{} & \texttt{matched\_distance\_reduction\_test} & $\geq 0.15$ & \effect{0.031049424} & FAIL & FAIL \\
raw geometry effect & \texttt{raw\_geometry\_test} & $\geq 0.15$ & \effect{-0.0484375} & FAIL & FAIL \\
directional beyond unsigned & \texttt{d\_directional\_test} & $\geq 0.15$ & \effect{-0.0078125} & FAIL & FAIL \\
\texttt{D\_signature} & \texttt{d\_signature\_test} & $\geq 0.15$ & \effect{0.015625} & FAIL & FAIL \\
\texttt{D\_conditional} & \texttt{d\_conditional\_test} & $\geq 0.15$ & \effect{-0.03848877} & FAIL & FAIL \\
invariant geometry AUC & \texttt{invariant\_geometry\_auc} & $\geq 0.80$ & \auc{0.515625} & FAIL & --- \\
invariant conditional alignment & \texttt{invariant\_conditional\_alignment} & $\geq 0.20$ & \effect{0.019885254} & FAIL & --- \\
\bottomrule
\end{tabularx}
\end{table}

Tail counts, so the exact \pv{} is legible as a position in the enumerated null:
\num{26150} of \nperm{} for \texttt{matched\_distance\_reduction\_test},
\num{61976} for \texttt{raw\_geometry\_test},
\num{39702} for \texttt{d\_directional\_test},
\num{20516} for \texttt{d\_signature\_test}, and
\num{60169} for \texttt{d\_conditional\_test}.

Matched distances and denominators, 16 values each, in block order:

\begin{itemize}
\item \texttt{invariant\_matched\_distances}: 0.022803, 0.0213518, 0.019497, 0.0244889, 0.00937379, 0.0170578, 0.0342236, 0.0504268, 0.0302576, 0.012872, 0.0289896, 0.0223302, 0.018496, 0.0133857, 0.0195453, 0.0148489
\item \texttt{narrative\_matched\_distances}: 0.0237235, 0.0238256, 0.0148777, 0.021737, 0.0167355, 0.0296298, 0.0170248, 0.0250014, 0.0191391, 0.0188918, 0.0220024, 0.011486, 0.0238105, 0.0251294, 0.0278083, 0.0328546
\item \texttt{matched\_distance\_denominators}: 0.0465264, 0.0451775, 0.0343747, 0.0462259, 0.0261093, 0.0466876, 0.0512484, 0.0754283, 0.0493968, 0.0317638, 0.0509921, 0.0338161, 0.0423065, 0.0385151, 0.0473535, 0.0477035
\end{itemize}

Failing criteria in this cell, 23 of 27:

\begin{itemize}\small
\item \texttt{all\_16\_invariant\_assignments\_unique\_identity}
\item \texttt{all\_16\_invariant\_panels\_at\_least\_4\_reciprocal}
\item \texttt{all\_five\_conditional\_family\_effects\_positive}
\item \texttt{all\_five\_directional\_family\_effects\_positive}
\item \texttt{all\_five\_matched\_distance\_reductions\_positive}
\item \texttt{all\_five\_raw\_geometry\_family\_effects\_positive}
\item \texttt{all\_five\_signature\_family\_effects\_positive}
\item \texttt{all\_four\_rendering\_relation\_directional\_effects\_positive}
\item \texttt{all\_four\_rendering\_relation\_matched\_distance\_reductions\_positive}
\item \texttt{all\_four\_rendering\_relation\_raw\_geometry\_effects\_positive}
\item \texttt{all\_registered\_support\_nonzero\_in\_both\_arms}
\item \texttt{d\_conditional\_at\_least\_0\_15}
\item \texttt{d\_conditional\_exact\_p\_at\_most\_0\_05}
\item \texttt{d\_directional\_beyond\_unsigned\_at\_least\_0\_15}
\item \texttt{d\_directional\_beyond\_unsigned\_exact\_p\_at\_most\_0\_05}
\item \texttt{d\_signature\_at\_least\_0\_15}
\item \texttt{d\_signature\_exact\_p\_at\_most\_0\_05}
\item \texttt{invariant\_conditional\_alignment\_at\_least\_0\_20}
\item \texttt{invariant\_geometry\_auc\_at\_least\_0\_80}
\item \texttt{matched\_distance\_reduction\_at\_least\_0\_15}
\item \texttt{matched\_distance\_reduction\_exact\_p\_at\_most\_0\_05}
\item \texttt{raw\_geometry\_effect\_at\_least\_0\_15}
\item \texttt{raw\_geometry\_exact\_p\_at\_most\_0\_05}
\end{itemize}

\subsubsection{\texorpdfstring{\Hone{}, \rankpipe{}}{H1, rank pipeline}}

Criteria passed: 3 of 27.

\begin{table}[htbp]
\caption{\Hone{}, \rankpipe{}: the seven registered quantities against their
registered thresholds.}
\label{supp:tab:h1-rank}
\centering
\begin{tabularx}{\linewidth}{@{}XXrrcc@{}}
\toprule
Quantity & Key & Threshold & Observed & Mag. & \pv{} \\
\midrule
matched-distance reduction \Mstat{} & \texttt{matched\_distance\_reduction\_test} & $\geq 0.15$ & \effect{-0.038265658} & FAIL & FAIL \\
raw geometry effect & \texttt{raw\_geometry\_test} & $\geq 0.15$ & \effect{-0.0359375} & FAIL & FAIL \\
directional beyond unsigned & \texttt{d\_directional\_test} & $\geq 0.15$ & \effect{0.0515625} & FAIL & FAIL \\
\texttt{D\_signature} & \texttt{d\_signature\_test} & $\geq 0.15$ & \effect{0.0671875} & FAIL & FAIL \\
\texttt{D\_conditional} & \texttt{d\_conditional\_test} & $\geq 0.15$ & \effect{-0.032324219} & FAIL & FAIL \\
invariant geometry AUC & \texttt{invariant\_geometry\_auc} & $\geq 0.80$ & \auc{0.5234375} & FAIL & --- \\
invariant conditional alignment & \texttt{invariant\_conditional\_alignment} & $\geq 0.20$ & \effect{0.02628479} & FAIL & --- \\
\bottomrule
\end{tabularx}
\end{table}

Tail counts: \num{55502} of \nperm{} for
\texttt{matched\_distance\_reduction\_test}, \num{60000} for
\texttt{raw\_geometry\_test}, \num{12082} for \texttt{d\_directional\_test},
\num{3884} for \texttt{d\_signature\_test}, and \num{57551} for
\texttt{d\_conditional\_test}.

\begin{itemize}
\item \texttt{invariant\_matched\_distances}: 32.9119, 36.5455, 29.7234, 25.71, 34.1278, 31.85, 31.6121, 32.8905, 32.3754, 29.9771, 34.3254, 35.8013, 30.2028, 29.8501, 33.2363, 35.3669
\item \texttt{narrative\_matched\_distances}: 32.0042, 31.5794, 26.0358, 26.9971, 34.5721, 32.1607, 32.0723, 32.9435, 36.2218, 35.8557, 26.5237, 25.6153, 34.1347, 30.5546, 33.0227, 27.9159
\item \texttt{matched\_distance\_denominators}: 64.9161, 68.1249, 55.7592, 52.7072, 68.6999, 64.0107, 63.6844, 65.834, 68.5972, 65.8327, 60.8491, 61.4166, 64.3375, 60.4047, 66.259, 63.2828
\end{itemize}

Failing criteria in this cell, 24 of 27:

\begin{itemize}\small
\item \texttt{all\_16\_invariant\_assignments\_unique\_identity}
\item \texttt{all\_16\_invariant\_panels\_at\_least\_4\_reciprocal}
\item \texttt{all\_five\_conditional\_family\_effects\_positive}
\item \texttt{all\_five\_directional\_family\_effects\_positive}
\item \texttt{all\_five\_matched\_distance\_reductions\_positive}
\item \texttt{all\_five\_raw\_geometry\_family\_effects\_positive}
\item \texttt{all\_five\_signature\_family\_effects\_positive}
\item \texttt{all\_four\_rendering\_relation\_directional\_effects\_positive}
\item \texttt{all\_four\_rendering\_relation\_matched\_distance\_reductions\_positive}
\item \texttt{all\_four\_rendering\_relation\_raw\_geometry\_effects\_positive}
\item \texttt{all\_four\_rendering\_relation\_signature\_effects\_positive}
\item \texttt{all\_registered\_support\_nonzero\_in\_both\_arms}
\item \texttt{d\_conditional\_at\_least\_0\_15}
\item \texttt{d\_conditional\_exact\_p\_at\_most\_0\_05}
\item \texttt{d\_directional\_beyond\_unsigned\_at\_least\_0\_15}
\item \texttt{d\_directional\_beyond\_unsigned\_exact\_p\_at\_most\_0\_05}
\item \texttt{d\_signature\_at\_least\_0\_15}
\item \texttt{d\_signature\_exact\_p\_at\_most\_0\_05}
\item \texttt{invariant\_conditional\_alignment\_at\_least\_0\_20}
\item \texttt{invariant\_geometry\_auc\_at\_least\_0\_80}
\item \texttt{matched\_distance\_reduction\_at\_least\_0\_15}
\item \texttt{matched\_distance\_reduction\_exact\_p\_at\_most\_0\_05}
\item \texttt{raw\_geometry\_effect\_at\_least\_0\_15}
\item \texttt{raw\_geometry\_exact\_p\_at\_most\_0\_05}
\end{itemize}

\subsubsection{\texorpdfstring{\Htwo{}, \linpipe{}}{H2, linear pipeline}}

Criteria passed: 6 of 27.

\begin{table}[htbp]
\caption{\Htwo{}, \linpipe{}: the seven registered quantities against their
registered thresholds. This is the only cell in which the matched-distance
reduction clears both of its registered criteria.}
\label{supp:tab:h2-linear}
\centering
\begin{tabularx}{\linewidth}{@{}XXrrcc@{}}
\toprule
Quantity & Key & Threshold & Observed & Mag. & \pv{} \\
\midrule
matched-distance reduction \Mstat{} & \texttt{matched\_distance\_reduction\_test} & $\geq 0.15$ & \effect{0.51739819} & PASS & PASS \\
raw geometry effect & \texttt{raw\_geometry\_test} & $\geq 0.15$ & \effect{0.0109375} & FAIL & FAIL \\
directional beyond unsigned & \texttt{d\_directional\_test} & $\geq 0.15$ & \effect{0.0296875} & FAIL & PASS \\
\texttt{D\_signature} & \texttt{d\_signature\_test} & $\geq 0.15$ & \effect{-0.00625} & FAIL & FAIL \\
\texttt{D\_conditional} & \texttt{d\_conditional\_test} & $\geq 0.15$ & \effect{-0.0023895264} & FAIL & FAIL \\
invariant geometry AUC & \texttt{invariant\_geometry\_auc} & $\geq 0.80$ & \auc{0.46875} & FAIL & --- \\
invariant conditional alignment & \texttt{invariant\_conditional\_alignment} & $\geq 0.20$ & \effect{0.026071167} & FAIL & --- \\
\bottomrule
\end{tabularx}
\end{table}

Tail counts: \num{82} of \nperm{} for
\texttt{matched\_distance\_reduction\_test}, \num{19844} for
\texttt{raw\_geometry\_test}, \num{2304} for \texttt{d\_directional\_test},
\num{39136} for \texttt{d\_signature\_test}, and \num{38336} for
\texttt{d\_conditional\_test}.

\begin{itemize}
\item \texttt{invariant\_matched\_distances}: 0.031455, 0.0203025, 0.0146213, 0.0633412, 0.0156075, 0.0188109, 0.0605437, 0.06075, 0.0193884, 0.04578, 0.0280078, 0.0285084, 0.017175, 0.0261159, 0.0201, 0.03717
\item \texttt{narrative\_matched\_distances}: 0.05808, 0.03999, 0.0508732, 0.0939637, 0.02667, 0.0205275, 0.0253957, 0.0519, 0.0645412, 0.0569203, 0.07851, 0.111272, 0.04464, 0.0426112, 0.0520012, 0.0463491
\item \texttt{matched\_distance\_denominators}: 0.089535, 0.0602925, 0.0654945, 0.157305, 0.0422775, 0.0393384, 0.0859395, 0.11265, 0.0839297, 0.1027, 0.106518, 0.13978, 0.061815, 0.0687272, 0.0721012, 0.0835191
\end{itemize}

Failing criteria in this cell, 21 of 27:

\begin{itemize}\small
\item \texttt{all\_16\_invariant\_assignments\_unique\_identity}
\item \texttt{all\_16\_invariant\_panels\_at\_least\_4\_reciprocal}
\item \texttt{all\_five\_conditional\_family\_effects\_positive}
\item \texttt{all\_five\_directional\_family\_effects\_positive}
\item \texttt{all\_five\_matched\_distance\_reductions\_positive}
\item \texttt{all\_five\_raw\_geometry\_family\_effects\_positive}
\item \texttt{all\_five\_signature\_family\_effects\_positive}
\item \texttt{all\_four\_rendering\_relation\_matched\_distance\_reductions\_positive}
\item \texttt{all\_four\_rendering\_relation\_raw\_geometry\_effects\_positive}
\item \texttt{all\_four\_rendering\_relation\_signature\_effects\_positive}
\item \texttt{all\_matched\_distance\_denominators\_above\_floor}
\item \texttt{all\_registered\_support\_nonzero\_in\_both\_arms}
\item \texttt{d\_conditional\_at\_least\_0\_15}
\item \texttt{d\_conditional\_exact\_p\_at\_most\_0\_05}
\item \texttt{d\_directional\_beyond\_unsigned\_at\_least\_0\_15}
\item \texttt{d\_signature\_at\_least\_0\_15}
\item \texttt{d\_signature\_exact\_p\_at\_most\_0\_05}
\item \texttt{invariant\_conditional\_alignment\_at\_least\_0\_20}
\item \texttt{invariant\_geometry\_auc\_at\_least\_0\_80}
\item \texttt{raw\_geometry\_effect\_at\_least\_0\_15}
\item \texttt{raw\_geometry\_exact\_p\_at\_most\_0\_05}
\end{itemize}

\subsubsection{\texorpdfstring{\Htwo{}, \rankpipe{}}{H2, rank pipeline}}

Criteria passed: 6 of 27.

\begin{table}[htbp]
\caption{\Htwo{}, \rankpipe{}: the seven registered quantities against their
registered thresholds.}
\label{supp:tab:h2-rank}
\centering
\begin{tabularx}{\linewidth}{@{}XXrrcc@{}}
\toprule
Quantity & Key & Threshold & Observed & Mag. & \pv{} \\
\midrule
matched-distance reduction \Mstat{} & \texttt{matched\_distance\_reduction\_test} & $\geq 0.15$ & \effect{0.069331606} & FAIL & PASS \\
raw geometry effect & \texttt{raw\_geometry\_test} & $\geq 0.15$ & \effect{-0.00625} & FAIL & FAIL \\
directional beyond unsigned & \texttt{d\_directional\_test} & $\geq 0.15$ & \effect{0.078125} & FAIL & FAIL \\
\texttt{D\_signature} & \texttt{d\_signature\_test} & $\geq 0.15$ & \effect{0.0484375} & FAIL & FAIL \\
\texttt{D\_conditional} & \texttt{d\_conditional\_test} & $\geq 0.15$ & \effect{-0.0097320557} & FAIL & FAIL \\
invariant geometry AUC & \texttt{invariant\_geometry\_auc} & $\geq 0.80$ & \auc{0.503125} & FAIL & --- \\
invariant conditional alignment & \texttt{invariant\_conditional\_alignment} & $\geq 0.20$ & \num{2.746582e-05} & FAIL & --- \\
\bottomrule
\end{tabularx}
\end{table}

Tail counts: \num{1866} of \nperm{} for
\texttt{matched\_distance\_reduction\_test}, \num{44048} for
\texttt{raw\_geometry\_test}, \num{6804} for \texttt{d\_directional\_test},
\num{4831} for \texttt{d\_signature\_test}, and \num{46763} for
\texttt{d\_conditional\_test}.

\begin{itemize}
\item \texttt{invariant\_matched\_distances}: 52.4561, 55.9003, 45.5109, 51.6395, 46.0477, 51.0464, 50.1671, 45.5587, 46.5823, 51.1895, 51.8233, 55.782, 44.7092, 49.1818, 45.2094, 50.2759
\item \texttt{narrative\_matched\_distances}: 61.4555, 57.4968, 51.5143, 51.402, 55.6291, 53.7277, 48.8995, 47.2746, 63.0567, 71.1876, 52.7256, 55.4215, 51.9864, 40.9408, 47.7253, 45.1729
\item \texttt{matched\_distance\_denominators}: 113.912, 113.397, 97.0252, 103.041, 101.677, 104.774, 99.0666, 92.8333, 109.639, 122.377, 104.549, 111.204, 96.6956, 90.1226, 92.9347, 95.4488
\end{itemize}

Failing criteria in this cell, 21 of 27:

\begin{itemize}\small
\item \texttt{all\_16\_invariant\_assignments\_unique\_identity}
\item \texttt{all\_16\_invariant\_panels\_at\_least\_4\_reciprocal}
\item \texttt{all\_five\_conditional\_family\_effects\_positive}
\item \texttt{all\_five\_directional\_family\_effects\_positive}
\item \texttt{all\_five\_matched\_distance\_reductions\_positive}
\item \texttt{all\_five\_raw\_geometry\_family\_effects\_positive}
\item \texttt{all\_five\_signature\_family\_effects\_positive}
\item \texttt{all\_four\_rendering\_relation\_matched\_distance\_reductions\_positive}
\item \texttt{all\_four\_rendering\_relation\_raw\_geometry\_effects\_positive}
\item \texttt{all\_registered\_support\_nonzero\_in\_both\_arms}
\item \texttt{d\_conditional\_at\_least\_0\_15}
\item \texttt{d\_conditional\_exact\_p\_at\_most\_0\_05}
\item \texttt{d\_directional\_beyond\_unsigned\_at\_least\_0\_15}
\item \texttt{d\_directional\_beyond\_unsigned\_exact\_p\_at\_most\_0\_05}
\item \texttt{d\_signature\_at\_least\_0\_15}
\item \texttt{d\_signature\_exact\_p\_at\_most\_0\_05}
\item \texttt{invariant\_conditional\_alignment\_at\_least\_0\_20}
\item \texttt{invariant\_geometry\_auc\_at\_least\_0\_80}
\item \texttt{matched\_distance\_reduction\_at\_least\_0\_15}
\item \texttt{raw\_geometry\_effect\_at\_least\_0\_15}
\item \texttt{raw\_geometry\_exact\_p\_at\_most\_0\_05}
\end{itemize}

\subsection{Closed-set identifiability}\label{supp:auc}

The second primary endpoint, \Gstat{}, is at chance in every cell.
\Cref{supp:tab:auc-full} gives the invariant geometry AUC in all four cells
against the registered 0.80 threshold, and the invariant conditional alignment
against its registered 0.20 threshold.

\begin{table}[htbp]
\caption{Closed-set same-account identifiability and conditional alignment in
all four cells, against their registered thresholds. Neither threshold is
cleared anywhere.}
\label{supp:tab:auc-full}
\centering
\begin{tabular}{@{}llrcrc@{}}
\toprule
Stage & Pipeline & AUC & $\geq 0.80$ & Cond.\ alignment & $\geq 0.20$ \\
\midrule
\Hone{} & \linpipe{} & \auc{0.515625}  & FAIL & \effect{0.019885254}  & FAIL \\
\Hone{} & \rankpipe{} & \auc{0.5234375} & FAIL & \effect{0.02628479}   & FAIL \\
\Htwo{} & \linpipe{} & \auc{0.46875}   & FAIL & \effect{0.026071167}  & FAIL \\
\Htwo{} & \rankpipe{} & \auc{0.503125}  & FAIL & \num{2.746582e-05} & FAIL \\
\bottomrule
\end{tabular}
\end{table}

Conditional alignment is easily misread as a plain accuracy number. It is not:
the registered floor of 0.20 is a floor on the margin over broken alignment ---
the AUC minus the mean AUC under deliberately shuffled alignment --- not on the
raw AUC.

The \Htwo{} \linpipe{} cell, where the matched-distance reduction clears both of
its criteria, records an AUC of \auc{0.46875} and the \rankpipe{} an AUC of
\auc{0.503125}, against the 0.80 threshold. Whatever moves the matched distance
does not make the two renderings mutually recognizable.

\subsection{Per-block effects}\label{supp:block-effects}

Effective $N$ is \nblocks{}, so the per-block effects are not a supplementary
detail; they are the entire evidential base, and they are reported individually
rather than summarized. Blocks are indexed 0--15, the same indices used in
\Cref{supp:tab:floor-magnitudes} and \Cref{supp:tab:lobo}.

\begin{small}
\begin{longtable}{@{}rrrrr@{}}
\caption{Per-block matched-distance reductions, \nblocks{} values per cell. The
registered magnitude floor is 0.15. Positive-block counts: \Hone{} \linpipe{}
9/16, \Hone{} \rankpipe{} 9/16, \Htwo{} \linpipe{} 14/16, \Htwo{} \rankpipe{}
11/16.}
\label{supp:tab:block-effects}\\
\toprule
Block & \Hone{} linear & \Hone{} rank & \Htwo{} linear & \Htwo{} rank \\
\midrule
\endfirsthead
\toprule
Block & \Hone{} linear & \Hone{} rank & \Htwo{} linear & \Htwo{} rank \\
\midrule
\endhead
\bottomrule
\endlastfoot
0  & \effect{0.0395676} & \effect{-0.0279626}  & \effect{0.594739}  & \effect{0.158006} \\
1  & \effect{0.109515}  & \effect{-0.145795}   & \effect{0.653066}  & \effect{0.0281566} \\
2  & \effect{-0.268759} & \effect{-0.132267}   & \effect{1.10702}   & \effect{0.123748} \\
3  & \effect{-0.119061} & \effect{0.0488396}   & \effect{0.389339}  & \effect{-0.0046101} \\
4  & \effect{0.563912}  & \effect{0.0129352}   & \effect{0.523328}  & \effect{0.188468} \\
5  & \effect{0.538561}  & \effect{0.0097063}   & \effect{0.0872715} & \effect{0.0511818} \\
6  & \effect{-0.671192} & \effect{0.0144518}   & \effect{-0.817972} & \effect{-0.0255908} \\
7  & \effect{-0.674161} & \effect{0.00161165}  & \effect{-0.157124} & \effect{0.0369677} \\
8  & \effect{-0.450172} & \effect{0.112146}    & \effect{1.07597}   & \effect{0.300521} \\
9  & \effect{0.37904}   & \effect{0.178592}    & \effect{0.216948}  & \effect{0.326828} \\
10 & \effect{-0.27405}  & \effect{-0.256429}   & \effect{0.948239}  & \effect{0.0172611} \\
11 & \effect{-0.641363} & \effect{-0.331701}   & \effect{1.18419}   & \effect{-0.00648361} \\
12 & \effect{0.251235}  & \effect{0.122229}    & \effect{0.888619}  & \effect{0.150519} \\
13 & \effect{0.609823}  & \effect{0.0233247}   & \effect{0.480023}  & \effect{-0.182883} \\
14 & \effect{0.348992}  & \effect{-0.00644858} & \effect{0.884901}  & \effect{0.0541429} \\
15 & \effect{0.754902}  & \effect{-0.235483}   & \effect{0.219808}  & \effect{-0.106927} \\
\end{longtable}
\end{small}

\Mstat{} is bounded by 2, and four \Htwo{} \linpipe{} blocks exceed 0.94.

The two geometry endpoints are also reported per block, because the directional
control is defined as their difference.

\begin{small}
\begin{longtable}{@{}rrrrr@{}}
\caption{Per-block unsigned geometry effects,
\texttt{unsigned\_geometry\_block\_effects}, \nblocks{} values per cell.}
\label{supp:tab:block-unsigned}\\
\toprule
Block & \Hone{} linear & \Hone{} rank & \Htwo{} linear & \Htwo{} rank \\
\midrule
\endfirsthead
\toprule
Block & \Hone{} linear & \Hone{} rank & \Htwo{} linear & \Htwo{} rank \\
\midrule
\endhead
\bottomrule
\endlastfoot
0  & \effect{0.075}  & \effect{0}      & \effect{-0.2}   & \effect{-0.175} \\
1  & \effect{0.075}  & \effect{0.05}   & \effect{0.025}  & \effect{0.2} \\
2  & \effect{0.2}    & \effect{-0.125} & \effect{-0.025} & \effect{-0.325} \\
3  & \effect{-0.15}  & \effect{-0.075} & \effect{-0.175} & \effect{-0.175} \\
4  & \effect{-0.075} & \effect{-0.075} & \effect{-0.05}  & \effect{0.15} \\
5  & \effect{0.025}  & \effect{-0.15}  & \effect{0}      & \effect{-0.125} \\
6  & \effect{-0.225} & \effect{-0.3}   & \effect{-0.05}  & \effect{-0.325} \\
7  & \effect{-0.075} & \effect{-0.275} & \effect{-0.05}  & \effect{-0.075} \\
8  & \effect{0.075}  & \effect{0.175}  & \effect{0}      & \effect{0.35} \\
9  & \effect{-0.275} & \effect{-0.075} & \effect{0.05}   & \effect{-0.225} \\
10 & \effect{-0.05}  & \effect{-0.075} & \effect{0.025}  & \effect{-0.225} \\
11 & \effect{-0.15}  & \effect{0.075}  & \effect{-0.025} & \effect{0.075} \\
12 & \effect{-0.05}  & \effect{-0.025} & \effect{0.15}   & \effect{-0.225} \\
13 & \effect{0.1}    & \effect{0}      & \effect{0.125}  & \effect{0.225} \\
14 & \effect{-0.15}  & \effect{-0.475} & \effect{0.05}   & \effect{-0.425} \\
15 & \effect{0}      & \effect{-0.05}  & \effect{-0.15}  & \effect{-0.05} \\
\end{longtable}
\end{small}

\begin{small}
\begin{longtable}{@{}rrrrr@{}}
\caption{Per-block directional-beyond-unsigned effects,
\texttt{d\_directional\_block\_effects}, \nblocks{} values per cell. Two entries
are zero to within floating-point noise.}
\label{supp:tab:block-directional}\\
\toprule
Block & \Hone{} linear & \Hone{} rank & \Htwo{} linear & \Htwo{} rank \\
\midrule
\endfirsthead
\toprule
Block & \Hone{} linear & \Hone{} rank & \Htwo{} linear & \Htwo{} rank \\
\midrule
\endhead
\bottomrule
\endlastfoot
0  & \effect{-0.25}         & \effect{-0.25}  & \effect{0.125}  & \effect{0.05} \\
1  & \effect{0.05}          & \effect{0.05}   & \effect{0.075}  & \effect{-0.175} \\
2  & \effect{-0.2}          & \effect{0.1}    & \effect{-0.025} & \effect{0.475} \\
3  & \num{-5.55112e-17}     & \effect{-0.025} & \effect{0.1}    & \effect{0.05} \\
4  & \effect{-0.025}        & \effect{0.1}    & \effect{0}      & \effect{-0.15} \\
5  & \effect{-0.025}        & \effect{0.15}   & \effect{0}      & \effect{0.15} \\
6  & \effect{0.25}          & \effect{0.375}  & \effect{0.025}  & \effect{0.3} \\
7  & \effect{-0.15}         & \effect{0.125}  & \effect{0}      & \num{-5.55112e-17} \\
8  & \effect{0.1}           & \effect{-0.15}  & \effect{-0.05}  & \effect{-0.35} \\
9  & \effect{0.25}          & \effect{0.075}  & \effect{0}      & \effect{0.175} \\
10 & \effect{-0.075}        & \effect{-0.075} & \effect{0}      & \effect{0.225} \\
11 & \effect{-0.025}        & \effect{-0.2}   & \effect{0.075}  & \effect{-0.025} \\
12 & \effect{-0.075}        & \effect{0.1}    & \effect{0}      & \effect{0.2} \\
13 & \effect{0.025}         & \effect{-0.05}  & \effect{0}      & \effect{-0.2} \\
14 & \effect{0.075}         & \effect{0.575}  & \effect{0}      & \effect{0.475} \\
15 & \effect{-0.05}         & \effect{-0.075} & \effect{0.15}   & \effect{0.05} \\
\end{longtable}
\end{small}

\subsection{Per-family effects}\label{supp:family-effects}

The five families are \texttt{source\_monitoring}, \texttt{comparator},
\texttt{precision\_weighting}, \texttt{circular\_inference} and
\texttt{decision\_threshold}. The family criteria require all five effects
positive, which is why a single negative family fails the criterion regardless
of the others. \Cref{supp:tab:family-effects} shows which families carry the
sign in each cell, which the criterion boolean conceals.

\begin{small}
\setlength{\tabcolsep}{5pt}
\begin{longtable}{@{}lrrrrr@{}}
\caption{Per-family effects, 5 values per metric per cell. Column headings
abbreviate the five registered families.}
\label{supp:tab:family-effects}\\
\toprule
Metric & source\_mon. & comparator & precision\_w. & circular\_inf. & decision\_thr. \\
\midrule
\endfirsthead
\toprule
Metric & source\_mon. & comparator & precision\_w. & circular\_inf. & decision\_thr. \\
\midrule
\endhead
\bottomrule
\endlastfoot
\multicolumn{6}{@{}l}{\textbf{\Hone{}, \linpipe{}}}\\
matched-distance reduction & \effect{0.242816}   & \effect{-0.0448955} & \effect{0.0467324}  & \effect{-0.384884} & \effect{0.0955778} \\
raw geometry               & \effect{0}          & \effect{-0.0703125} & \effect{-0.0078125} & \effect{-0.1875}   & \effect{0.0234375} \\
directional                & \effect{-0.015625}  & \effect{-0.03125}   & \effect{0.03125}    & \effect{-0.0859375}& \effect{0.0625} \\
signature                  & \effect{0.046875}   & \effect{-0.015625}  & \effect{0.0546875}  & \effect{-0.0703125}& \effect{0.0625} \\
conditional                & \effect{-0.0140533} & \effect{-0.0597382} & \effect{0.0501099}  & \effect{-0.093689} & \effect{-0.0750732} \\
\addlinespace
\multicolumn{6}{@{}l}{\textbf{\Hone{}, \rankpipe{}}}\\
matched-distance reduction & \effect{-0.0402882} & \effect{0.092386}  & \effect{0.0034235}  & \effect{-0.156346} & \effect{-0.144815} \\
raw geometry               & \effect{0.0234375}  & \effect{0.03125}   & \effect{-0.0234375} & \effect{-0.101562} & \effect{-0.109375} \\
directional                & \effect{0.046875}   & \effect{0.0625}    & \effect{0.0625}     & \effect{0.09375}   & \effect{-0.0078125} \\
signature                  & \effect{0.148438}   & \effect{0.0234375} & \effect{-0.0078125} & \effect{0.101562}  & \effect{0.0703125} \\
conditional                & \effect{0.0535278}  & \effect{-0.077179} & \effect{-0.0810394} & \effect{0.00521851}& \effect{-0.062149} \\
\addlinespace
\multicolumn{6}{@{}l}{\textbf{\Htwo{}, \linpipe{}}}\\
matched-distance reduction & \effect{0.140299}   & \effect{0.0503252}  & \effect{-0.142283} & \effect{0.133538}  & \effect{0.831999} \\
raw geometry               & \effect{0.046875}   & \effect{0.0390625}  & \effect{-0.015625} & \effect{-0.140625} & \effect{0.125} \\
directional                & \effect{-0.0078125} & \effect{0}          & \effect{0.0234375} & \effect{0.015625}  & \effect{0.117188} \\
signature                  & \effect{0}          & \effect{0.0078125}  & \effect{-0.0234375}& \effect{-0.0546875}& \effect{0.0390625} \\
conditional                & \effect{-0.0367584} & \effect{-0.0377197} & \effect{0.0399475} & \effect{-0.050354} & \effect{0.072937} \\
\addlinespace
\multicolumn{6}{@{}l}{\textbf{\Htwo{}, \rankpipe{}}}\\
matched-distance reduction & \effect{-0.000904729} & \effect{0.219208}   & \effect{0.0277271}  & \effect{-0.0213243} & \effect{0.106797} \\
raw geometry               & \effect{-0.015625}    & \effect{0.0859375}  & \effect{-0.0625}    & \effect{-0.109375}  & \effect{0.0703125} \\
directional                & \effect{-0.015625}    & \effect{0.101562}   & \effect{0.078125}   & \effect{0.046875}   & \effect{0.179688} \\
signature                  & \effect{-0.078125}    & \effect{0.164062}   & \effect{0.101562}   & \effect{-0.046875}  & \effect{0.101562} \\
conditional                & \effect{0.00340271}   & \effect{-0.0108795} & \effect{-0.0408325} & \effect{-0.0245972} & \effect{0.0242462} \\
\end{longtable}
\end{small}

\subsubsection{Family denominator floor events}

A floored family denominator contributes a substituted zero. Floor events were
counted in all four cells, zeros included.

\begin{table}[htbp]
\caption{Denominator floor events at both aggregation levels, in all four cells.
No block-level floor event occurs anywhere in the study.}
\label{supp:tab:floor-events}
\centering
\begin{tabular}{@{}lrrrr@{}}
\toprule
Record & \Hone{} linear & \Hone{} rank & \Htwo{} linear & \Htwo{} rank \\
\midrule
\texttt{block\_matched\_denominator\_floor\_events} & 0 & 0 & 0 & 0 \\
\texttt{family\_matched\_denominator\_floor\_events} & 0 & 0 & 3 & 0 \\
\texttt{family\_matched\_denominator\_count} & 80 & 80 & 80 & 80 \\
\bottomrule
\end{tabular}
\end{table}

\Htwo{}'s \linpipe{} fails
\texttt{all\_matched\_distance\_denominators\_above\_floor} on 3 of 80
denominators. \Cref{supp:tab:floor-magnitudes} names them.

\begin{table}[htbp]
\caption{The three family denominator floor events, all in \Htwo{} \linpipe{},
with the block and family each occurred in. They are at the block-by-family
aggregation level and are zero to within floating-point noise.}
\label{supp:tab:floor-magnitudes}
\centering
\begin{tabular}{@{}rlr@{}}
\toprule
Block & Family & Denominator \\
\midrule
1  & \texttt{comparator}         & \num{8.680516e-30} \\
3  & \texttt{comparator}         & \num{1.177663e-29} \\
12 & \texttt{source\_monitoring} & \num{1.080759e-29} \\
\bottomrule
\end{tabular}
\end{table}

These are of order \num{1e-29}. They are a genuine failure and they legitimately
gate the five family-positivity criteria, all of which fail.

\subsection{Denominators}\label{supp:denominators}

The registered primary test does not aggregate families. It aggregates the
\nblocks{} block-level effects, and every block-level denominator in the \Htwo{}
cells is healthy. This is the evidence against reading the \Htwo{} \linpipe{}
result as a division by something near zero.

In \Htwo{} \linpipe{} the \nblocks{} block denominators range from 0.0393384 to
0.157305, with median 0.0837244. The smallest is $0.47\times$ the median, and
none is below a quarter of it. In \Htwo{} \rankpipe{} the median is 102.359. The
full \nblocks{}-value denominator vectors for both cells are listed in
\Cref{supp:cells}.

\begin{table}[htbp]
\caption{Median matched-distance denominators. The two transforms operate on
numbers of very different absolute size; that difference does not by itself make
the primary statistics incomparable, because \Mstat{} is a dimensionless ratio
and \Gstat{} is an AUC.}
\label{supp:tab:median-denominators}
\centering
\begin{tabular}{@{}lrrr@{}}
\toprule
Stage & Linear & Rank & Rank/linear \\
\midrule
\Hone{} & 0.0463762 & 64.1741 & $1384\times$ \\
\Htwo{} & 0.0837244 & 102.359 & $1223\times$ \\
\bottomrule
\end{tabular}
\end{table}

\Mstat{} is registered as a dimensionless ratio: multiply both matched distances
by any common factor and \Mstat{} is unchanged. Neither primary endpoint carries
the units \Cref{supp:tab:median-denominators} is measured in, so that table
cannot explain a difference between the pipelines' endpoint values. What can
differ is the relative gap between the narrative and invariant distances, and
the rank transform compresses it: replacing values by their ranks bounds how far
apart two matched distances can be as a fraction of their sum, so the same
underlying separation yields a smaller \Mstat{} under rank than under linear.
Rank denominators here run near 102.36 against 0.0837244 in linear. A single
0.15 floor applied to both transforms is then a stiffer test under rank than
under linear. This was not anticipated in the preregistration; it was noticed
after unblinding, and it is carried as a design limitation.

The collapsed family denominators cannot reach the primary statistic, and
\texttt{block\_matched\_denominator\_floor\_events} is empty in every cell of the
study.

\subsection{Sensitivity}\label{supp:sensitivity}

The \Htwo{} matched-distance result was re-enumerated after dropping the single
largest-magnitude block, separately in each pipeline.
\Cref{supp:tab:lobo} gives the block dropped, the value dropped, the mean over
the remaining blocks, and the exact \pv{} from a fresh enumeration on 15 blocks.

\begin{table}[htbp]
\caption{Leave-one-block-out re-enumeration of the \Htwo{} matched-distance
reduction, both pipelines. Each row drops the single largest-magnitude block and
re-enumerates the exact null on the remaining 15.}
\label{supp:tab:lobo}
\centering
\begin{tabular}{@{}lrrrrr@{}}
\toprule
Pipeline & Block dropped & Value dropped & Mean after & Exact \pv{} after & Blocks \\
\midrule
\linpipe{} & 11 & \effect{1.18419}  & \effect{0.472945}  & \pval{0.00250244} & 15 \\
\rankpipe{} & 9  & \effect{0.326828} & \effect{0.0521652} & \pval{0.0569458}  & 15 \\
\bottomrule
\end{tabular}
\end{table}

The linear result survives; the rank result does not clear 0.05 once its largest
block is removed. With 15 blocks the smallest attainable \pv{} rises to
1/32,768.

One further consequence follows from arithmetic on the two numbers already
given, and is stated here so that it is not left for the reader to notice. The
\linpipe{} \pv{} after the drop is \pval{0.00250244}, which clears 0.05 but no
longer clears the 0.0025 a Bonferroni correction over all 20 test evaluations
would impose. The single value in the study that crossed that corrected
threshold therefore stops crossing it when one block of sixteen is withheld.
This is a restatement of the sensitivity result, not a further test: no
correction was applied to the re-enumeration, and the registered support rule
was already not met.

%% file: supplement/sections/S4-secondary-analyses.tex
\section{Registered secondary analyses}\label{supp:secondary}

Secondary analyses S0 through S7 were registered in advance and are non-gating
and exploratory. Where a description below sets \Honefull{} beside \Htwofull{},
that comparison is unregistered: nothing in the preregistration seal defines a
test statistic for the difference between the two stages, so no difference
between them is tested anywhere in this study.

The registered wording of S0 is quoted rather than summarised, because it
governs how everything else in this section may be read:

\begin{quote}
Every analysis in this section is \textbf{non-gating and descriptive}. None of
them can change the study verdict, in either direction. [\ldots] no secondary
result may be offered as a partial or consolation finding if the conjunction
fails. If the study does not pass, it did not pass; the material below is then a
characterisation of a null, not a rescue of it.
\end{quote}

The registered conjunction failed. Nothing below alters that, and nothing below
is offered as a partial finding.

\subsection{S1. Full exact randomization null distribution}
\label{supp:secondary:s1}

The null is exact, not sampled --- all \nperm{} sign vectors are enumerated ---
so the quantities in \Cref{supp:tab:s1-resolution} are properties of the design,
fixed before any data existed. They are resolution limits, not results: they say
what the enumeration is capable of reporting, whatever the programs turned out
to do.

\begin{table}[tb]
\caption{Resolution limits of the exact randomization null. All \nperm{}
renderer-label sign vectors over the \nblocks{} randomized blocks are
enumerated, so the attainable one-sided \pv{} values form a fixed discrete grid.
The last two rows are concordance configurations: how many of the \nblocks{}
blocks must agree in sign before the design can return a \pv{} at or below the
registered 0.05 threshold at all.}
\label{supp:tab:s1-resolution}
\centering
\small
\begin{tabular}{ll}
\toprule
registered resolution limit & value \\
\midrule
smallest attainable one-sided \pv{}
  & $1/\nperm = \num{1.526e-5}$ \\
largest \pv{} still satisfying \texttt{MAXIMUM\_ONE\_SIDED\_P}
  & $\num{3276}/\nperm = 0.049988$ \\
last passing concordance configuration
  & 12 of 16 blocks concordant \\
cannot pass at 0.05 whatever the magnitudes
  & 11 of 16 blocks concordant \\
\bottomrule
\end{tabular}
\end{table}

Observed tail positions are given per test with the primary endpoints in
\suppref{3}.

\subsection{S2. Marginal descriptives of the response tensors}
\label{supp:secondary:s2}

Registered as reported whatever the outcome, not only under a null: the median,
the interquartile range and the fraction exactly zero over the 11 tensor
coordinates, separately for each model snapshot and each stage, pooled across
blocks, anchors and interventions.

Forward-traversal archives only. The reverse traversal is a fresh-process
replicate whose response tensor archives are byte-identical to forward, so
pooling both would double-count each value rather than add information.

Column labels in \Cref{supp:tab:s2-h1} and \Cref{supp:tab:s2-h2} are abbreviated
to \texttt{gpt-4.1} and \texttt{gpt-5.4}; the full snapshot identifiers are
\snapshotB{} and \snapshotA.

\begin{table}[tb]
\caption{\Hone{} marginal descriptives over the 11 response tensor coordinates,
pooling the \texttt{atomic\_holdout} and \texttt{calibration} panels:
\num{1105920} values per coordinate per model, from 160 samples each. IQR is the
interquartile range; \emph{zero} is the fraction of values exactly equal to
0.0.}
\label{supp:tab:s2-h1}
\centering
\small
\begin{tabular}{lrrrrrr}
\toprule
& \multicolumn{3}{c}{\texttt{gpt-4.1}} & \multicolumn{3}{c}{\texttt{gpt-5.4}} \\
\cmidrule(lr){2-4}\cmidrule(lr){5-7}
coordinate & median & IQR & zero & median & IQR & zero \\
\midrule
0  & 0 & 0 & 0.79515   & 0 & 0 & 0.742123 \\
1  & 0 & 0 & 0.795711  & 0 & 0 & 0.803096 \\
2  & 0 & 0 & 0.684589  & 0 & 0 & 0.890341 \\
3  & 0 & 0 & 0.684861  & 0 & 0 & 0.890928 \\
4  & 0 & 0 & 0.684159  & 0 & 0 & 0.891269 \\
5  & 0 & 0 & 0.684441  & 0 & 0 & 0.890271 \\
6  & 0 & 0 & 0.743862  & 0 & 0 & 0.787234 \\
7  & 0 & 0 & 0.760012  & 0 & 0 & 0.758087 \\
8  & 0 & 0 & 0.782968  & 0 & 0 & 0.825392 \\
9  & 0 & 0 & 0.768553  & 0 & 0 & 0.726047 \\
10 & 0 & 0 & 0.790591  & 0 & 0 & 0.799257 \\
\bottomrule
\end{tabular}
\end{table}

\begin{table}[tb]
\caption{\Htwo{} marginal descriptives over the 11 response tensor coordinates,
from the \texttt{interaction\_holdout} panel: \num{921600} values per coordinate
per model, from 160 samples each. Same conventions as
\Cref{supp:tab:s2-h1}.}
\label{supp:tab:s2-h2}
\centering
\small
\begin{tabular}{lrrrrrr}
\toprule
& \multicolumn{3}{c}{\texttt{gpt-4.1}} & \multicolumn{3}{c}{\texttt{gpt-5.4}} \\
\cmidrule(lr){2-4}\cmidrule(lr){5-7}
coordinate & median & IQR & zero & median & IQR & zero \\
\midrule
0  & 0 & 0 & 0.923174 & 0 & 0 & 0.874132 \\
1  & 0 & 0 & 0.920087 & 0 & 0 & 0.901827 \\
2  & 0 & 0 & 0.907688 & 0 & 0 & 0.960418 \\
3  & 0 & 0 & 0.908937 & 0 & 0 & 0.960328 \\
4  & 0 & 0 & 0.909489 & 0 & 0 & 0.960818 \\
5  & 0 & 0 & 0.909803 & 0 & 0 & 0.959993 \\
6  & 0 & 0 & 0.88212  & 0 & 0 & 0.86452  \\
7  & 0 & 0 & 0.876485 & 0 & 0 & 0.820693 \\
8  & 0 & 0 & 0.905187 & 0 & 0 & 0.885769 \\
9  & 0 & 0 & 0.955985 & 0 & 0 & 0.893446 \\
10 & 0 & 0 & 0.906548 & 0 & 0 & 0.869488 \\
\bottomrule
\end{tabular}
\end{table}

Every median is exactly zero and every interquartile range is exactly zero, in
all 44 coordinate-by-model-by-stage cells. That is not a formatting artifact and
it is not a failure: it is the dominant feature of this corpus. Between 0.6842
and 0.9608 of each coordinate's values are exactly 0.0 --- that is the range
across all 44 cells, not a per-model average --- so the 25th, 50th and 75th
percentiles all land inside the block of zeros. The programs the models emit
respond to most interventions on most coordinates by not moving at all.

Because a table of nothing but zeros says little about the scale of what does
move, \Cref{supp:tab:s2-nonzero} repeats the same three statistics conditional
on the value being non-zero. These are supplementary and unregistered.

\begin{table}[tb]
\caption{Descriptives conditional on the value being non-zero, pooled over the
11 coordinates within each stage-by-model cell. \emph{Supplementary and
unregistered.} The sixth column is the median across the 11 coordinates of each
coordinate's own non-zero median absolute value, not a pooled median over all
values at once; the two differ.}
\label{supp:tab:s2-nonzero}
\centering
\small
\setlength{\tabcolsep}{4pt}
\resizebox{\textwidth}{!}{%
\begin{tabular}{llrrrrrr}
\toprule
stage & model & fraction & fraction & fraction & median non-zero
& pooled & pooled \\
      &       & zero     & negative & positive & median abs
& min & max \\
\midrule
\Hone{} & \snapshotB & 0.743172 & 0.117641  & 0.139187  & 0.00913938
  & $-0.225$    & 0.40175 \\
\Hone{} & \snapshotA & 0.81855  & 0.088708  & 0.0927424 & 0.00439489
  & $-0.163106$ & 0.22 \\
\Htwo{} & \snapshotB & 0.909591 & 0.0429311 & 0.0474778 & \num{2.1684e-17}
  & $-0.28$     & 0.28 \\
\Htwo{} & \snapshotA & 0.904676 & 0.0432971 & 0.0520271 & \num{8.13152e-18}
  & $-0.16$     & 0.2 \\
\bottomrule
\end{tabular}%
}
\end{table}

The sign split is close to symmetric everywhere. That is consistent with, but
does not establish, an absence of systematic directional bias: no test of
asymmetry was registered and none is run here. The non-zero magnitudes are small
in absolute terms. The per-coordinate figures behind the sixth column are in the
deposited secondary-analysis values \seeannex{D}.

S2 also registers the observed frequency of the transition-clamp and cap
conditions, read from the validity diagnostics in each stage's analysis.
\Cref{supp:tab:s2-diagnostics} gives them.

\begin{table}[tb]
\caption{Registered guard-rail conditions, by stage. Both counts are zero in
both stages, so neither guard rail was touched.}
\label{supp:tab:s2-diagnostics}
\centering
\small
\begin{tabular}{lrr}
\toprule
stage & \texttt{transition\_clamp\_events} & \texttt{cap\_activations} \\
\midrule
\Hone{} & 0 & 0 \\
\Htwo{} & 0 & 0 \\
\bottomrule
\end{tabular}
\end{table}

Both conditions occurred zero times in both stages. Neither guard rail was
touched, so no quantity produced by either analyser --- which is every
criterion, estimate and \pv{} value reported in this package --- was produced by
a clamped or capped value.

\subsection{S3. Dispatch order and drift}\label{supp:secondary:s3}

A descriptive check for drift across the generation window: provider-side
changes, load-dependent behaviour, or any other time-varying influence on the
corpus. The dispatch timestamp is read from the \ncalls{} per-sample
provider-attempt records, which is where the experiment driver writes it; it is
not in the generation summary record.

Order is treated as a continuous covariate, not a rank with meaningful
adjacent-pair structure, because eight requests are in flight concurrently. An
order effect does not bias the primary contrast: the registered separation audit
reports that schedule position is invariant to the arm assignment. This is a
check on the corpus, not on the inference.

All \ncalls{} dispatches fall between \code{2026-08-07T19:46:03.043541+00:00}
and \code{2026-08-07T19:54:14.982125+00:00} --- a window of 491.939 seconds,
or 0.1366 hours. Eight requests were in flight at a time, so the window is short
by design; there is correspondingly little room for provider-side drift to
develop, and \Cref{supp:tab:s3-drift} is what that looks like when it is
measured rather than assumed.

\begin{table}[tb]
\caption{Dispatch-order descriptives for eleven per-sample derived quantities
over the \ncalls{} dispatched samples. \emph{first 160} and \emph{last 160} are
the split-half means by dispatch position. $r$ is the Pearson correlation with
dispatch position $1$--$320$, which is treated as a continuous covariate rather
than a rank; the slope column is the ordinary-least-squares slope expressed per
100 positions, in the units of the quantity itself. The first row is constant
across the run, so no correlation or slope is defined for it. No \pv{} value is
attached to any entry: the inferential unit in this study is the block, not the
sample.}
\label{supp:tab:s3-drift}
\centering
\small
\setlength{\tabcolsep}{4pt}
\resizebox{\textwidth}{!}{%
\begin{tabular}{lrrrrrrr}
\toprule
per-sample derived quantity & mean & sd & first 160 & last 160
& $r$ vs position & slope per 100 & $r$ vs elapsed s \\
\midrule
\texttt{generation\_attempt\_count}
  & 1 & 0 & 1 & 1
  & constant & constant & constant \\
\texttt{generation\_elapsed\_seconds}
  & 12.3502 & 6.49484 & 13.0402 & 11.6602
  & $-0.05278$ & $-0.3705$ & $-0.0476$ \\
\texttt{h1\_tensor\_fraction\_zero}
  & 0.780861 & 0.0828299 & 0.775812 & 0.78591
  & 0.05272 & 0.004719 & 0.05574 \\
\texttt{h1\_tensor\_max\_abs}
  & 0.100895 & 0.0492897 & 0.102058 & 0.0997317
  & 0.00443 & 0.000236 & 0.006931 \\
\texttt{h1\_tensor\_mean\_abs}
  & 0.00350507 & 0.00197582 & 0.00340858 & 0.00360157
  & 0.05456 & 0.0001165 & 0.05426 \\
\texttt{h2\_tensor\_fraction\_zero}
  & 0.907133 & 0.0507167 & 0.907838 & 0.906429
  & $-0.02871$ & $-0.001574$ & $-0.0267$ \\
\texttt{h2\_tensor\_max\_abs}
  & 0.0624258 & 0.0586984 & 0.064375 & 0.0604766
  & $-0.01542$ & $-0.0009781$ & $-0.01378$ \\
\texttt{h2\_tensor\_mean\_abs}
  & 0.00149151 & 0.00197617 & 0.00143331 & 0.00154972
  & 0.04981 & 0.0001064 & 0.05007 \\
\texttt{usage\_input\_tokens}
  & 15649.3 & 41.8649 & 15651.8 & 15646.7
  & $-0.02641$ & $-1.195$ & $-0.02456$ \\
\texttt{usage\_output\_tokens}
  & 946.197 & 39.2286 & 949.5 & 942.894
  & $-0.05406$ & $-2.292$ & $-0.05332$ \\
\texttt{usage\_total\_tokens}
  & 16595.5 & 57.0649 & 16601.3 & 16589.6
  & $-0.05653$ & $-3.487$ & $-0.05468$ \\
\bottomrule
\end{tabular}%
}
\end{table}

The largest absolute correlation anywhere in the table is 0.05653, for
\texttt{usage\_total\_tokens}. These are descriptive correlations over the
\ncalls{} dispatched samples, computed to see whether the corpus looks
time-varying. Read as description, the fitted line for that quantity moves it by
11.16 across the whole run of \ncalls{} dispatches, against a between-sample
standard deviation of 57.0649 --- the trend is smaller than the scatter it is
fitted through.

The split-half means are not uniformly close. Relative gaps here are the
difference between the two half-means expressed as a percentage of the
quantity's own overall mean, which is the second column of
\Cref{supp:tab:s3-drift}. The largest is
\texttt{generation\_elapsed\_seconds}: 13.0402 over the first 160 dispatches
against 11.6602 over the last 160, a change of 11.2 percent of its mean of
12.3502. That quantity is wall clock rather than corpus content, and its
correlation with position is still only $-0.05278$. Queue depth across eight concurrent workers is the
obvious candidate; nothing in this study measures it.

Among quantities derived from the response tensors themselves, the largest
relative split-half gap is \texttt{h2\_tensor\_mean\_abs} at 7.8 percent, on a
quantity whose standard deviation across samples is 0.00197617 against a mean of
0.00149151 --- the between-sample spread is larger than the between-half
difference, which is the shape of noise rather than of drift.

No drift correction is applied, because none is indicated and because applying
one would be an unregistered analysis.

\subsection{S4. The distribution of the 512 shuffle controls}
\label{supp:secondary:s4}

The 512 permutations are not independent of the primary inference. Conditional
alignment --- the observed AUC minus the mean AUC under deliberately shuffled
alignment --- is defined by subtracting the mean of these shuffles, so the
shuffle null is subtracted into the conditional endpoint.

That is why this section exists: the quantity being subtracted should be visible
rather than implicit. The registered analyser persists the mean, minimum and
maximum of each panel's 512 shuffled AUCs, but neither the standard deviation
nor the raw vector, so the shuffles were recomputed with the registered
functions. The recomputation is not taken on trust. Every recomputed vector was
hashed with the analyser's own array digest function and required to equal the
AUC digest the analyser recorded for that panel.
\Cref{supp:tab:s4-hash} reports the outcome.

\begin{table}[tb]
\caption{Agreement between the recomputed shuffle vectors and the digests the
registered analyser recorded, by stage. All 128 panels agree, so the
distributions in \Cref{supp:tab:s4-distribution} are the registered ones and not
a reimplementation of them.}
\label{supp:tab:s4-hash}
\centering
\small
\begin{tabular}{lrr}
\toprule
stage & panels & vectors hash-equal to the registered analyser \\
\midrule
\Hone{} & 64 & 64/64 \\
\Htwo{} & 64 & 64/64 \\
\bottomrule
\end{tabular}
\end{table}

\begin{table}[tb]
\caption{Distribution of the shuffled AUCs, by stage and transform family. A
panel is one arm of one block under one transform family, so the 64 panels per
stage counted in \Cref{supp:tab:s4-hash} are $\nblocks \times 2$ arms under each
of the two families. Each row here therefore pools the 512 shuffles from 32
panels, and summarises \num{16384} values. A shuffle null over exchangeable rows
belongs at 0.5.}
\label{supp:tab:s4-distribution}
\centering
\small
\begin{tabular}{llrrrrrr}
\toprule
stage & transform family & values & mean & sd & min & max & range \\
\midrule
\Hone{} & \linpipe  & \num{16384} & 0.500714111 & 0.0288693 & 0.375 & 0.65  & 0.275 \\
\Hone{} & \rankpipe & \num{16384} & 0.498959351 & 0.0390537 & 0.325 & 0.675 & 0.35  \\
\Htwo{} & \linpipe  & \num{16384} & 0.43601532  & 0.0632042 & 0.175 & 0.55  & 0.375 \\
\Htwo{} & \rankpipe & \num{16384} & 0.501356506 & 0.0252956 & 0.375 & 0.675 & 0.3   \\
\bottomrule
\end{tabular}
\end{table}

Three of the four cells sit within 0.00136 of 0.5, which is where a shuffle null
over exchangeable rows belongs. The fourth does not: \Htwo{} under the
\linpipe{} has a shuffled AUC mean of \auc{0.43601532}, roughly 0.064 below 0.5,
with sd 0.0632042. The widest spread of the four cells is \Htwo{} under the
\linpipe{} and the lowest minimum is \Htwo{} under the \linpipe.

This matters for reading one number and only one number. Conditional alignment
is the observed AUC minus this mean, so in that cell the subtraction is not
removing a 0.5 baseline but a baseline that is already displaced downward, which
mechanically inflates \texttt{d\_conditional} relative to the other three cells.
The registered \texttt{d\_conditional} result in that cell is nonetheless
negative, at \effect{-0.00238953} with \pv{} = \pval{0.584961}, and it does not
clear the registered 0.05 threshold, so the displacement did not manufacture a
result.

A per-canonical-family breakdown is also deposited, but it covers only the
panels where the anonymous axis to family map could be recovered by an
exact-value join against the analyser's persisted per-family AUC means, which is
62 of 64 for \Hone{} and 47 of 64 for \Htwo. That shortfall is a property of the
label-recovery join --- an exact-value match that fails when two families share
a persisted mean --- and not a gap in the released \Htwo{} data, which holds all
\ncalls{} completed samples with every tensor matching its sidecar digest. It is
marked partial-coverage there and is not reproduced here. \emph{Transform family} in the registered wording of
the outcome matrix means the \linpipe{} and the \rankpipe, which is
\Cref{supp:tab:s4-distribution} and is complete.

\subsection{S5. Criterion-margin reporting}\label{supp:secondary:s5}

How far each primary test cleared or missed its 0.15 threshold, so a near miss
is legible as a near miss. \Cref{supp:tab:s5-margins} gives all 20 evaluations.

\begin{table}[tb]
\caption{Observed value and margin against the registered 0.15 magnitude
threshold, for all five registered tests in each of the four
stage-by-pipeline cells. A positive margin means the test cleared the
threshold.}
\label{supp:tab:s5-margins}
\centering
\small
\begin{tabular}{llrr}
\toprule
cell & test & observed & margin vs 0.15 \\
\midrule
\Hone{} \linpipe  & \texttt{matched\_distance\_reduction\_test} & 0.0310494    & $-0.118951$  \\
\Hone{} \linpipe  & \texttt{raw\_geometry\_test}                & $-0.0484375$ & $-0.198437$  \\
\Hone{} \linpipe  & \texttt{d\_directional\_test}               & $-0.0078125$ & $-0.157812$  \\
\Hone{} \linpipe  & \texttt{d\_signature\_test}                 & 0.015625     & $-0.134375$  \\
\Hone{} \linpipe  & \texttt{d\_conditional\_test}               & $-0.0384888$ & $-0.188489$  \\
\addlinespace
\Hone{} \rankpipe & \texttt{matched\_distance\_reduction\_test} & $-0.0382657$ & $-0.188266$  \\
\Hone{} \rankpipe & \texttt{raw\_geometry\_test}                & $-0.0359375$ & $-0.185937$  \\
\Hone{} \rankpipe & \texttt{d\_directional\_test}               & 0.0515625    & $-0.0984375$ \\
\Hone{} \rankpipe & \texttt{d\_signature\_test}                 & 0.0671875    & $-0.0828125$ \\
\Hone{} \rankpipe & \texttt{d\_conditional\_test}               & $-0.0323242$ & $-0.182324$  \\
\addlinespace
\Htwo{} \linpipe  & \texttt{matched\_distance\_reduction\_test} & 0.517398     & 0.367398     \\
\Htwo{} \linpipe  & \texttt{raw\_geometry\_test}                & 0.0109375    & $-0.139062$  \\
\Htwo{} \linpipe  & \texttt{d\_directional\_test}               & 0.0296875    & $-0.120313$  \\
\Htwo{} \linpipe  & \texttt{d\_signature\_test}                 & $-0.00625$   & $-0.15625$   \\
\Htwo{} \linpipe  & \texttt{d\_conditional\_test}               & $-0.00238953$& $-0.15239$   \\
\addlinespace
\Htwo{} \rankpipe & \texttt{matched\_distance\_reduction\_test} & 0.0693316    & $-0.0806684$ \\
\Htwo{} \rankpipe & \texttt{raw\_geometry\_test}                & $-0.00625$   & $-0.15625$   \\
\Htwo{} \rankpipe & \texttt{d\_directional\_test}               & 0.078125     & $-0.071875$  \\
\Htwo{} \rankpipe & \texttt{d\_signature\_test}                 & 0.0484375    & $-0.101562$  \\
\Htwo{} \rankpipe & \texttt{d\_conditional\_test}               & $-0.00973206$& $-0.159732$  \\
\bottomrule
\end{tabular}
\end{table}

Only one cell-and-test combination clears the floor: \Htwo's matched-distance
reduction under the \linpipe, characterised with the primary endpoints in the
main text and attacked in \suppref{5}. Every other margin is negative, most of
them by more than the threshold itself, which is the shape of a null rather than
of a near miss.

\subsection{S6. Multiplicity}\label{supp:secondary:s6}

Across the study there are $5 \times 2 \times 2 = 20$ test evaluations --- five
tests, two pipelines, two stages. A Bonferroni threshold is
$0.05/20 = 0.0025$. Applied to every observed \pv{}:

\begin{itemize}
\item \Htwo{} \linpipe{} \texttt{matched\_distance\_reduction\_test}
  (\pv{} = \pval{0.00125122}) --- below the corrected threshold.
\end{itemize}

Nominally significant but not below the corrected threshold:

\begin{itemize}
\item \Htwo{} \linpipe{} \texttt{d\_directional\_test}
  (\pv{} = \pval{0.0351562}).
\item \Htwo{} \rankpipe{} \texttt{matched\_distance\_reduction\_test}
  (\pv{} = \pval{0.0284729}).
\end{itemize}

Surviving Bonferroni is a statement about the \pv{} value only: the statistic
sits deep in the enumerated null. It is not a statement about magnitude, and
magnitude is where the two pipelines disagree; that disagreement is recorded
with the primary endpoints in the main text and examined in \suppref{5}. The
correction is reported over all 20 evaluations because S6 registers it that way.

\subsection{S7. What this study cannot answer}\label{supp:secondary:s7}

Effective N is \nblocks{} for every inferential statement in this study. The
\ncalls{} registered calls, 960 provider exchanges and \num{1280} tensors
increase the precision with which each block effect is estimated; they do not
increase the number of independent units, and no statement anywhere in this
document treats them as if they did.

Two model snapshots is two. It is not a sample from which any model-general
conclusion could be drawn, and none is drawn. The study also cannot say whether
the two renderers carry identical theoretical content --- they carry the same
registered propositions, rendered differently, and whether that preserves
theoretical content is exactly what is not tested here.

%% file: supplement/sections/S5-robustness.tex
\section{Robustness and adversarial checks}\label{supp:robustness}

Two questions were put to the results after unblinding: whether the null under
\Honefull{} is an artifact of the apparatus, and whether the one endpoint that
moved, under \Htwofull{}, can be killed. This section is the full version of
both analyses.

Neither analysis is a registered test. An adversarial check that passes rules
out the artifact it was aimed at and nothing else. None of what follows makes
the preregistered conjunction pass --- it failed, and it failed at \Hone{}
before \Htwo{} was analysed --- and the registered wording of secondary S0
forbids offering any of this as a partial or consolation finding.

\subsection{Is \Hone's null an artifact of the apparatus?}
\label{supp:robustness:h1-null}

A null is only informative if the instrument could have detected an effect.
Two candidate artifacts were checked.

\paragraph{Denominator floors: none.}
The matched-distance reduction \Mstat{} is a ratio, so a divisor pinned at the
registered floor would distort it. \Hone{} records zero block-level and zero
family-level denominator floor events in both the \linpipe{} and the
\rankpipe, over 80 family denominators. \Hone's null is not a floored-divisor
artifact.

\paragraph{Support floors: not global.}
The registered criterion
\texttt{all\_registered\_support\_nonzero\_in\_both\_arms} failed 60 of 64
audits in both pipelines. That single count is equally consistent with two
opposite sentences: that the programs never moved the designated rows at all,
and that one family sits at zero while the criterion is a conjunction over all
five families. The two demand opposite readings of the null, so the audits were
opened family by family. \Cref{tab:supp-support-audits} gives the result.

\begin{table}[tb]
  \centering
  \caption{Registered-support audits below floor, by family, out of 64 audits
  per family per pipeline. The two pipelines agree exactly. The five counts are
  not a partition of the 60 failing audits and do not sum to it: the criterion
  is a conjunction over all five families, so an audit fails if any family is
  below floor, and the five below-floor sets overlap. The failures are
  heterogeneous across families and concentrated in \texttt{comparator}, which
  is the signature of a family-specific effect rather than of a global
  measurement floor.}
  \label{tab:supp-support-audits}
  \begin{tabularx}{\linewidth}{@{}l
      >{\raggedleft\arraybackslash}X
      >{\raggedleft\arraybackslash}X@{}}
    \toprule
    family & audits below floor (\linpipe) & audits below floor (\rankpipe) \\
    \midrule
    \texttt{circular\_inference}  & 5/64  & 5/64  \\
    \texttt{comparator}           & 51/64 & 51/64 \\
    \texttt{decision\_threshold}  & 8/64  & 8/64  \\
    \texttt{precision\_weighting} & 32/64 & 32/64 \\
    \texttt{source\_monitoring}   & 37/64 & 37/64 \\
    \bottomrule
  \end{tabularx}
\end{table}

No family is below floor in every audit, and none is above floor in every
audit. The derived verdict is \verdict{NO\_FLOOR} under the \linpipe{} and
\verdict{NO\_FLOOR} under the \rankpipe.

\paragraph{What that establishes, and what it does not.}
Two specific explanations for \Hone's null have been excluded: a floored
divisor and a global support floor. Neither exclusion is a demonstration that
the apparatus was sensitive enough to detect an effect of the registered size.
No positive control was registered and none was run, so there is no measured
floor on what this instrument could have found.

A third candidate is resolved by neither audit and is carried forward from the
negative-controls analysis: the matched-distance statistic is computed over
matchings that in most blocks are neither reciprocal enough nor injective. The
structural criteria
\texttt{all\_16\_invariant\_panels\_at\_least\_4\_reciprocal} and
\texttt{all\_16\_invariant\_assignments\_unique\_identity} fail in all four
cells of the study. That limitation applies equally to \Hone's null and to
\Htwo's one positive result, and it remains open. \Hone's null is reported as a
null with two candidate artifacts excluded and that third one open.

\subsection{Four attempts to kill the one result that moved}
\label{supp:robustness:four-challenges}

\Htwo's matched-distance reduction is the largest statistic observed anywhere
in the study, which is exactly the circumstance in which an artifact is most
likely and least likely to be looked for. Four specific ways it could have been
an artifact were checked before it was written down. Two of the four
re-enumerate the exact null on modified data.

The statistic is positive with an exact one-sided \pv{} below 0.05 in both
pipelines: \effect{0.5173981945762773}, \pv{} = \pval{0.001251220703125}
(tail 82/65536) under the \linpipe, and \effect{0.06933160555977207},
\pv{} = \pval{0.028472900390625} (tail 1866/65536) under the \rankpipe. The
linear \pv{} is not merely under 0.05; it is under the corrected threshold of
0.0025 taken over all 20 test evaluations in the study, and it is the only
value in the study that is.

\paragraph{Challenge 1 --- a denominator floor artifact? No.}
The statistic is a ratio, and \Htwo's \linpipe{} reports three
\texttt{family\_matched\_denominator\_floor\_events} with
\texttt{all\_matched\_distance\_denominators\_above\_floor} failing, while the
\rankpipe{} reports zero and passes. A floored denominator inflates a ratio
with no change in numerator, so the two observations were not independent. The
floor events are given in \Cref{tab:supp-floor-events}.

\begin{table}[tb]
  \centering
  \caption{The three denominator floor events in \Htwo{} under the \linpipe.
  All three are at the block-by-family aggregation level, one level below the
  registered primary statistic, and all three are zero to within
  floating-point noise.}
  \label{tab:supp-floor-events}
  \begin{tabular}{llr}
    \toprule
    block & family & denominator \\
    \midrule
    12 & \texttt{source\_monitoring} & \num{1.08076e-29} \\
    1  & \texttt{comparator}         & \num{8.68052e-30} \\
    3  & \texttt{comparator}         & \num{1.17766e-29} \\
    \bottomrule
  \end{tabular}
\end{table}

Those three sit at the family aggregation level and are effectively exactly
zero, of order 1e-29. They are a genuine failure and they legitimately carry
the five family-positivity criteria, all of which fail. The registered primary
test does not aggregate families. It aggregates the \nblocks{} block-level
effects, and every block-level denominator is healthy: they range from
0.0393384 to 0.157305, the smallest at 0.47 times the median and none below a
quarter of it, including the three blocks named above.
\texttt{block\_matched\_denominator\_floor\_events} is empty in every cell of
the study. The collapsed denominators cannot reach the primary statistic. The
floor contaminates the family-effect criteria, not the primary test.

\paragraph{Challenge 2 --- carried by a single block? Largely no, in the
\linpipe{} only.}
The single largest-magnitude block was dropped and the exact null re-enumerated
over the remaining blocks. \Cref{tab:supp-loo} gives both pipelines.

\begin{table}[tb]
  \centering
  \caption{Leave-one-block-out sensitivity. The largest-magnitude block is
  removed and the exact randomization null is re-enumerated on the 15
  remaining blocks. The linear result survives; the rank result no longer
  clears 0.05.}
  \label{tab:supp-loo}
  \begin{tabular}{lrrrr}
    \toprule
    pipeline & block dropped & value dropped & mean after & exact \pv{} after \\
    \midrule
    \linpipe  & 11 & 1.18419  & \effect{0.472945}  & \pval{0.00250244} \\
    \rankpipe & 9  & 0.326828 & \effect{0.0521652} & \pval{0.0569458}  \\
    \bottomrule
  \end{tabular}
\end{table}

The linear result survives the deletion. The rank result does not clear 0.05
once its largest block is removed. With 15 blocks the smallest attainable \pv{}
rises to 1/32,768.

\paragraph{Challenge 3 --- do the two pipelines disagree? On sign and
direction, no; on the registered magnitude criterion, yes, and the criterion is
what counts.}
Both pipelines agree on sign and both fall below 0.05. They part on
\texttt{matched\_distance\_reduction\_at\_least\_0\_15}, which the \linpipe{}
passes at \effect{0.517398} and the \rankpipe{} fails at \effect{0.0693316},
less than half the registered floor of 0.15.

\textbf{The rank cell failed that criterion; that is the finding for that
cell.} The registered support rule requires the magnitude criterion to pass in
both transforms, and it does not. Both transforms were required in advance
precisely so that an effect appearing at one scale and not the other would not
count as a finding.

Two readings of the disagreement are available and this study cannot
distinguish them. On the first, the effect is small and the \linpipe{}
overstates it: \Mstat{} is bounded by 2, four linear blocks exceed 0.94, and if
a handful of blocks with unusually large relative separations drive the mean
then the rank value is the better estimate of a genuinely modest effect. On the
second, the two transforms are not on a comparable scale for a fixed absolute
floor: the rank transform replaces values by ranks, which bounds how far apart
two matched distances can be as a fraction of their sum --- rank denominators
here run near 102.36 against 0.0837244 in linear --- so the same underlying
separation yields a smaller ratio under rank, and one floor of 0.15 applied to
both transforms is then a stiffer test under rank than under linear.

The denominator magnitudes cannot settle this. The statistic is a dimensionless
ratio and is unchanged by a common rescaling of both distances. The second
reading was not anticipated in the preregistration; it was noticed after
unblinding, and it is carried as a design limitation. It changes nothing here:
the criteria stand as registered, and every cell that failed the magnitude
criterion, including \Htwo{} under the \rankpipe, failed it.

\paragraph{Challenge 4 --- is the \Htwo-versus-\Hone{} difference in this
statistic a divisor artifact? No --- but no test of the difference was run.}
Within pipeline and across stages the denominators are of the same order, so
the larger \Htwo{} value is not produced by dividing by a smaller number. The
median denominator ratio of \Htwo{} to \Hone{} is 1.805 under the \linpipe{}
and 1.595 under the \rankpipe. Both are commensurable.
\Cref{tab:supp-h1-h2-side-by-side} puts the two stages beside each other.

\begin{table}[tb]
  \centering
  \caption{Matched-distance reduction in the two stages, by pipeline, with the
  exact one-sided \pv{} and the number of blocks with a positive effect. The
  two stages are separate registered hypotheses over different probe batteries
  and different finite-difference schemes. They are placed side by side to
  show that the denominators are commensurable, not as a tested contrast: no
  test of the difference was registered and none was run.}
  \label{tab:supp-h1-h2-side-by-side}
  \begin{tabularx}{\linewidth}{@{}lXX@{}}
    \toprule
    pipeline & \Hone{} (single-input, central differences)
             & \Htwo{} (interaction, mixed differences) \\
    \midrule
    \linpipe  & \effect{0.0310494}, \pv{} = \pval{0.399017}, 9/16 positive
              & \effect{0.517398}, \pv{} = \pval{0.00125122}, 14/16 positive \\
    \rankpipe & \effect{-0.0382657}, \pv{} = \pval{0.846893}, 9/16 positive
              & \effect{0.0693316}, \pv{} = \pval{0.0284729}, 11/16 positive \\
    \bottomrule
  \end{tabularx}
\end{table}

That excludes the divisor explanation and nothing more. \textbf{No test of the
\Hone/\Htwo{} difference was registered and none was run.} The two stages are
separate hypotheses over different probe batteries and different
finite-difference schemes, not two arms of a comparison, and the study has no
estimator for the contrast between them.

\paragraph{What the four checks licence.}
One sentence: under \Htwo's interaction challenges, holding the words fixed and
presenting them as explicit hold-change-direction instructions was associated
with a reduction in matched distance between the two snapshots' executable
behaviour --- below the corrected threshold and robust to dropping the largest
block in the \linpipe, nominal only and not robust to that deletion in the
\rankpipe, where it also failed the registered magnitude criterion. \Hone{} is
reported separately and on its own terms: under single-input challenges no such
association was detected. The two statements are not joined into a claim that
the effect is present in one setting and absent in the other, because no
contrast between them was tested and a non-detection is not a detection of
absence.

\paragraph{What they do not licence.}
Calling the hypothesis supported. The effect is a mean shift on one contrast
inside a pattern that is otherwise flatly null: chance-level classification,
conditional alignment orders of magnitude below threshold, and every
family-invariance and rendering-relation-invariance criterion failing. The
theory predicted a uniform, family-invariant, classifiable geometry. That is
not what came out. One endpoint moving in one regime, at a size two of the
study's own instruments disagree about, is a result worth reporting and
reproducing --- not a hypothesis confirmed.

\subsection{What was not tested}\label{supp:robustness:not-tested}

Six checks were run above. The following alternative explanations remain live,
and none of them is excluded by anything in this study.

\begin{description}[style=unboxed,leftmargin=0pt]

\item[No positive control.] None was registered and none was run. There is no
measured floor on what this instrument could have detected, so excluding a
floored divisor and a global support floor does not establish that the
apparatus was sensitive enough to find an effect of the registered size.
\Hone's null is a null with two candidate artifacts excluded, not a
demonstration of adequate sensitivity.

\item[The matching structure.] The matched-distance statistic is computed over
matchings that in most blocks are neither reciprocal enough nor injective:
\texttt{all\_16\_invariant\_panels\_at\_least\_4\_reciprocal} and
\texttt{all\_16\_invariant\_assignments\_unique\_identity} fail in all four
cells. Neither the denominator audit nor the support audit touches it. It
applies equally to \Hone's null and to \Htwo's one positive result, and it is
the single most consequential unexcluded explanation in the study.

\item[The contrast between the two stages.] No estimator for the
\Hone/\Htwo{} difference was registered, none exists in the analysis, and none
was computed. Challenge 4 shows only that the difference in the raw statistic
is not produced by a difference in divisor scale.

\item[The magnitude disagreement between the transforms.] The two readings set
out under Challenge 3 --- a small effect that the \linpipe{} overstates, or two
transforms with different headroom against one absolute floor --- are not
distinguished by anything registered in advance, and this study cannot
distinguish them.

\item[Family-level positivity.] The three collapsed family denominators are a
real failure at the block-by-family level, not an artifact that was explained
away. They are shown not to reach the block-level primary statistic; they are
not shown to be harmless to the family-level quantities they gate, which fail.

\end{description}

None of the checks above changes the verdict, and the registered conjunction is
\notsupported.

%% file: annex/A-preregistration-verbatim/annex-a-note.tex
\section{The sealed preregistration, verbatim}\label{anx:preregistration}

This annex reproduces the preregistration of \studyname{} without a single byte
changed. The file
\code{annex/A-preregistration-verbatim/preregistration-sealed-20260804.md}
sits beside this note. It carries the introduction, the methods, and the
preregistered outcome and claim matrix as they stood when they were compiled on
2026-08-04 --- that is, before the assignment randomness existed. Nothing in it
was revised to fit what the data turned out to be. Its line endings are LF, as
sealed.

\subsection*{Checking the bytes}

The sealed file is \num{101629} bytes long and has sha256 digest
\code{e3a64ce61afb29c49a6eb7346696562f35454a1e80df351c4ce681cd79ca8df8}.
A reader recomputes both without trusting the authors. From the root of the
package, in a POSIX shell:

\begin{lstlisting}
sha256sum annex/A-preregistration-verbatim/preregistration-sealed-20260804.md
wc -c     annex/A-preregistration-verbatim/preregistration-sealed-20260804.md
\end{lstlisting}

\noindent
or in PowerShell:

\begin{lstlisting}
Get-FileHash -Algorithm SHA256 annex\A-preregistration-verbatim\preregistration-sealed-20260804.md
(Get-Item annex\A-preregistration-verbatim\preregistration-sealed-20260804.md).Length
\end{lstlisting}

\subsection*{What it was sealed against}

The digest above was committed to an RFC 3161 timestamp and to three
OpenTimestamps calendar commitments, all of which predate the assignment
randomness. The proof files, the timestamps they carry, and the step-by-step
procedure for verifying them against records the authors do not control are in
\annexref{B}.

\subsection*{Three statements that are now false, and are reproduced anyway}

Three statements in the sealed text were true when it was timestamped on
2026-08-04 and are false of the completed study. It calls itself
``Introduction and Methods only''. It states that it contains no results. Its
abstract's \textbf{Status} paragraph reads ``Pre-execution''. Both registered
hypothesis stages have since been executed, and the registered conjunction is
\notsupported{} --- the registered verdict token for a study whose execution is
valid and in which one or more registered support criteria fail.

All three statements are reproduced unaltered. The document's entire value is
that it cannot be revised after the fact, so correcting it now would remove the
only property it exists to have. Read them as dated claims about a document
sealed on 2026-08-04, not as claims about the study reported in the paper.

\subsection*{Which text governs}

Where the paper and the sealed text differ in wording, the sealed text governs
what was preregistered and the paper governs what was found. The paper
translates internal vocabulary into terms a reader outside the project can
follow, and it reorders material for reading rather than for the sequence in
which it was written; it does not change any registered definition, threshold,
criterion or decision rule. Any question about what was committed to in advance
is settled by the file in this annex.

%
%
%

%% file: annex/B-provenance-and-attestation/annex-b-provenance.tex
\section{Provenance and attestation}\label{anx:provenance}

The audit trail is why the numbers in this package can be believed, and it is
not an assurance about the authors' conduct. It is a chain a third party can
check against records the authors do not control: a signed timestamp from a
commercial authority, three public calendar commitments, and one pulse of a
government randomness beacon that did not exist when the study was sealed.

This annex is one of the few parts of the package where file names, paths,
script names, function names and digests appear, because this is where they are
useful. Everything below is stated so that a reader can recompute it.
\Cref{anx:sec:verify} is the procedure.

\subsection{The chain, end to end}\label{anx:sec:chain}

The order of events is the whole argument, so it is given first.

The complete apparatus --- runtime, documents, tests, criteria and failure
rules --- was written into the preregistration seal and independently
timestamped. The seal names one exact future pulse of the NIST Randomness
Beacon 2.0, fixed 120 one-minute pulses after the observed anchor. That pulse
did not exist when the seal was written.

Only after the seal was stamped did the randomness become available. One
durable assignment-attempt marker opened a six-hour availability window. The
monitor was permitted to send bodyless \texttt{HEAD} requests to the exact
target URI and nothing else; redirects were rejected. After the first
\texttt{HEAD 200} a durable target-fetch marker was written, the target was
fetched exactly once with \texttt{GET}, and its raw bytes were persisted before
anything parsed them. The target pulse, its predecessor and the Beacon
certificate were checked against the pinned certificate chain, the
RSA-4096/SHA-512 signature, the output hash, the predecessor link and the
predecessor commitment. A timeout, a failure of the sole target \texttt{GET},
or any cryptographic failure stops \studyname{}. There is no alternate pulse,
seed, target retry, balancing rule, or reroll.

The assignment is then a deterministic function of the seal and the pulse. The
first 16 digest bits, most-significant bit first, assign the
\contract{} renderer to neutral slot \texttt{U} or \texttt{V} in each block; the
remaining 240 disjoint bits generate opaque identifiers, adapters, the anonymous
axis order and the execution order.

The manifest was built next and binds every artifact by path and hash. The
public assignment file is path-and-hash bound into it, so changing even a
non-scientific timestamp after freezing invalidates the manifest. Before any
model request was sent, the exact 320 request envelopes and the generation
marker were hashed as one set and independently timestamped, and every accepted
provider \texttt{created\_at} second is required to be strictly later than that
signed request-set timestamp. Same-second chronology is rejected, because the
provider timestamp has only whole-second precision.

Each scientific phase closes with one exclusive \code{PHASE-TERMINAL.json}
whose status is either \texttt{COMPLETE} or \texttt{INVALID}. That record
carries the exact path, size and SHA-256 of every artifact of the phase plus a
Merkle root. Exclusive creation permits only one terminal winner, and
verification rejects added, missing, altered or deleted artifacts. The
phase-terminal status value is a driver field written mechanically by the
execution harness the moment the phase closes; it records how that phase-run
terminated and is not a scientific verdict for the stage. The \Htwo{} execution
phase closed \texttt{INVALID} on its first pass, because three of its
direction-runs had to be repeated after a local attestation-verification
timeout, and the record was left on disk unedited. The stage's scientific
verdict is \notsupported{}, reached from its registered criteria after all
\ncalls{} samples completed; \annexref{C} reconciles the two field by field.

\Cref{anx:tab:timeline} is the chain as it actually ran, in order.

\begin{table}[htbp]
\centering
\caption{The provenance chain of the executed run, in order. All times UTC.}
\label{anx:tab:timeline}
\footnotesize
\begin{tabularx}{\linewidth}{@{}Xl@{}}
\toprule
event & time (UTC) \\
\midrule
preregistration seal written and timestamped & 2026-08-07T17:33:22Z \\
target Beacon chain-2 pulse 1894079 published & 2026-08-07T19:28:00Z \\
assignment drawn                              & 2026-08-07T19:32:26Z \\
manifest built                                & 2026-08-07T19:41Z \\
generation complete                           & 2026-08-07T19:54:23Z \\
\Hone{} execution complete                    & 2026-08-08T02:53:22Z \\
\textbf{\Hone{} analysis sealed and timestamped}
                                              & \textbf{2026-08-08T03:07:00Z} \\
\Htwo{} execution phase closed on its first pass; three direction-runs,
        in three different samples, had timed out in a local attestation check
                                              & 2026-08-08T10:15:58Z \\
\Htwo{} those three direction-runs repeated
        & 2026-08-08T10:35:21Z to 2026-08-08T10:40:05Z \\
\textbf{\Htwo{} execution complete, \ncalls{} of \ncalls{} samples}
                                              & \textbf{2026-08-08T10:40:05Z} \\
\Htwo{} analysis over all 320                 & 2026-08-08T10:57:47Z \\
joint-analysis step                           & not run \\
\bottomrule
\end{tabularx}
\end{table}

\Hone{} is the single-input-challenge stage and \Htwo{} is the
interaction-challenge stage. The ordering in \Cref{anx:tab:timeline} is the
load-bearing fact of this study. \Hone{}'s \notsupported{} verdict --- the
registered token for a valid execution with any failed scientific criterion ---
was sealed with both OpenTimestamps and RFC 3161 hours before \Htwo{} was
analysed, and the registered conjunction requires both stages supported. No
\Htwo{} outcome could have made this study positive. The registered conjunction
failed, and it was already unrecoverable when the second stage was still
running.

The joint-analysis step was not run. Both stage verdicts are \notsupported{},
the joint verdict follows from the registered rule by inspection
(\Cref{sec:results:conjunction}), and running the step to restate that would add
a file, not a fact. No joint artifact therefore exists, and none is claimed in
\Cref{anx:tab:digests}. That absence is a bookkeeping fact about a redundant
file: the raw dump's \texttt{final\_analysis\_exists} field is \emph{false}
because the step was skipped, not because any verdict is unsettled.

\Htwo{}'s \notsupported{} verdict carries the same registered token, and the
same standing, as \Hone{}'s. Both stages completed all \ncalls{} of their
registered samples. \Hone{} passes 13 of the 13 registered validity criteria and
\Htwo{} passes 12; the single criterion \Htwo{} does not pass is the registered
single-execution-attempt counter, whose two conjuncts split with all \ncalls{}
samples valid and the single-attempt flag false. Every criterion in either stage
that bears on the data passes, and re-hashing the \Htwo{} tensor corpus returned
\num{1280} digests --- 640 archives and their 640 sidecars --- with zero
mismatches and zero missing. \Htwo{} was sealed later than \Hone{} and its
verdict rests on its own registered criteria, not on \Hone{}'s.

The sealed envelope predates its own randomness by 1 hour 54 minutes. The seal
names a future pulse; the pulse did not exist when the seal was written; the
assignment is a deterministic function of that pulse. The analysis could not
have been chosen to suit the assignment.

\studyname{} runs once. Any later study requires a new version, a new
commitment, and complete disclosure.

\subsection{The ordering guarantee}\label{anx:sec:ordering}

The requirement is that the seal is strictly earlier than the target pulse.
That is enforced rather than asserted.

\texttt{verify\_precommit()}, in the run-support module, parses the RFC 3161
signed time out of the seal's timestamp token, compares it against the target
pulse time, and raises if the seal is not strictly earlier.
The check is fail-closed: nothing downstream runs if it does not hold. Both of
the times it compares are published in \Cref{anx:tab:timeline}, so a reader can
perform the same comparison without running the code.

The assignment digest
\code{500762433136176bea1c8548d87e1c411d66c692554572478ff65ffaa703fc48}
is independently attested by RFC 3161 and equals the digest of the assignment
bytes. Its definition is registered:

\begin{lstlisting}
assignment_digest =
    SHA256( domain || precommit_sha256_bytes || target_outputValue_bytes )
\end{lstlisting}

\noindent
Both inputs are public. The seal digest is published here and the pulse output
value is published by NIST, so the assignment is recomputable by anyone.

\subsection{Digests}\label{anx:sec:digests}

The corpus, the executed tensors, the assignment and the provenance chain are
an explicit deliverable rather than a by-product of the run directory. They are
archived off-machine against a documented schema, so that secondary analyses
and replications do not depend on this machine, this filesystem layout, or
these authors. \Cref{anx:tab:digests} lists the digests that bind them.

\begin{table}[htbp]
\centering
\caption{Digests of the bound artifacts of the executed run. Each entry gives
the artifact and its sha256.}
\label{anx:tab:digests}
\small
\begin{tabular}{@{}l@{}}
\toprule
bound manifest (104 artifacts) \\
\code{e7ab948c74238302b71e2ec027906ca21f28a65b5efa89757a6b255a21a7173e} \\
\addlinespace
generation Merkle root \\
\code{15065dd566a84cb88c5ec55e7143d227d05fe63bb0ac8ad8bffaa534272aaa9c} \\
\addlinespace
\Hone{} execution Merkle root \\
\code{8f5dac30a3bba5336f510f348b7e35933fc17b2c00c1db0b588a14b0bfa3655d} \\
\addlinespace
\Hone{} execution summary \\
\code{9150d39cb16eab54c2814dfc73d2a6c658bcd7ac2f9242dbe6678de18773f347} \\
\addlinespace
\Hone{} analysis Merkle root \\
\code{3e16ee37de8159a817a0a37c9eec50bf680f4d6506950e2e146077cdaaeff217} \\
\addlinespace
\Hone{} analysis artifact \\
\code{c0d9e038c9200c43adc7785c9744f11023502f3041ec9332a217c5af1fc7e1b2} \\
\addlinespace
\Hone{} sealed analysis digest \\
\code{1fd2c615e8454b37c6ff973797c0d93970805ae24ca536f1a595161ff4683788} \\
\addlinespace
\Htwo{} execution summary as first written, before the three re-runs \\
\code{e252076cb2024abaf83e475855cd2b2469535b7dff24fae0c34e404b2793ee5b} \\
\addlinespace
final analysis \\
not produced; the joint-analysis step was not run \\
\bottomrule
\end{tabular}
\end{table}

Two entries deserve a note. The \Htwo{} execution summary is listed as it was
first written, before the three affected direction-runs were repeated; the
incident and the repetition are documented in \annexref{C}, and the digest is
published here so that the earlier state of that artifact is on the record
rather than overwritten by the later one.

That is the only \Htwo{} row in this table, and the asymmetry is a property of
the registered chain rather than of the data. \Hone{} has analysis rows because
its registered analyser ran and sealed in place. \Htwo{}'s first-pass phase
terminal was left on disk unedited with its driver field reading
\texttt{INVALID}, and the registered analyser declines by design to run against
a terminal record in that state, so no \Htwo{} analysis Merkle root and no
\Htwo{} sealed analysis digest were ever produced to list. Nothing about the
\Htwo{} corpus is missing: the bound manifest above covers 104 artifacts with
zero mismatches, and the post-repetition audit re-hashed \num{1280} \Htwo{}
digests --- 640 tensor archives and their 640 sidecars --- with zero mismatches
and zero missing. \annexref{C} carries that audit, its own digests, and the
field-by-field reconciliation of the terminal record against the stage's
registered scientific verdict, which is \notsupported{}.

The \Hone{} sealed analysis digest is
the one the ordering argument rests on, and it is not merely a hash of whatever
was written: \gate{6} replays the analysis from the raw tensors and refuses
unless the sealed bytes are exactly what the sealed inputs produce.

\subsection{Figures}\label{anx:sec:figures}

The two figures are separate vector PDFs, drawn from the same extraction that
supplies the prose. Their digests are computed when the document is generated,
so a figure that was re-rendered after the text was written cannot go
unnoticed.

\begin{table}[htbp]
\centering
\caption{The two figure files and their sha256 digests, as computed at
document generation.}
\label{anx:tab:figure-digests}
\small
\begin{tabular}{@{}l@{}}
\toprule
Figure 1, per-block effects --- \code{figure-1-block-effects.pdf} \\
\code{6601adf702e5ee9e012b976e19d209788e8de17eee46cfe795af1ba684c5e8ac} \\
\addlinespace
Figure 2, per-family effects --- \code{figure-2-family-effects.pdf} \\
\code{25e99cec98d818bb76a58eee14bd50563aa3ce7bcbfdabdfbdca670b87926507} \\
\bottomrule
\end{tabular}
\end{table}

The same two files are staged in this package as
\code{paper/figures/fig1-block-effects.pdf} and
\code{paper/figures/fig2-family-effects.pdf}. They are byte-identical to the
generated originals; only the file names differ, and a reader confirms that by
recomputing the two digests above against the staged copies.

\subsection{Timestamping}\label{anx:sec:timestamping}

Every sealed artifact carries two independent kinds of attestation.

\textbf{RFC 3161.} A commercial timestamp authority signs the digest of the
artifact together with a time. The receipt is self-contained and verifiable
offline against the authority's certificate chain, without contacting any
server and without trusting the authors. This is the operative independent
attestation on the seal today, and it is the time that
\texttt{verify\_precommit()} parses and compares.

\textbf{OpenTimestamps.} The same digest is submitted to three independent
calendar servers. A calendar commitment proves that the digest was submitted;
once the calendar folds it into a Bitcoin block, the proof also anchors the
digest to a public ledger that no party to this study controls. The two layers
fail differently, which is the point of having both.

Timestamp evidence, measured on copies and never on the registered originals:
\textbf{71 of 76 OpenTimestamps proofs are folded into Bitcoin blocks 960634
through 961288}, 36 of them attested in three distinct blocks and 35 in two.
The 5 pending proofs are exactly the 5 stamped that day, which is expected
latency rather than a failure. The registered text states that the proofs carry
only \texttt{PendingAttestation}; that is true of the registered bytes on disk
and will remain true, because upgrading a registered proof would alter a bound
artifact. Upgrade a copy, never the original.

What the RFC 3161 verification step itself binds is also recorded. It binds the
OpenSSL executable, its configuration, providers and engines, three observed
Git-for-Windows DLLs (\texttt{libcrypto}, \texttt{libssl} and \texttt{ZLIB1}),
and the exact observed Windows DLL closure. Full-lifetime module capture covers
\texttt{ts -query}, \texttt{ts -reply -text} and \texttt{ts -verify}. Three
OpenSSL closure audits exist and all three are preserved; none was silently
replaced. The first is preserved as failed, because \texttt{cryptsp.dll} and
\texttt{rsaenh.dll} appeared during verification but were not registered. The
second is the corrective audit that registered them and names its failed
predecessor by path and hash. The third governs: it supersedes the second not
for a missing module but because the second could not distinguish ``no
unregistered module loaded'' from ``a module-load event was never resolved to
an image path.'' It counts every module-load debug event and fails on any
unresolved one --- 69 events, 69 resolved, 0 unresolved --- and records both
predecessors.

Three boundaries of this evidence are stated rather than papered over. The
\texttt{no\_network} field of those audits is derived from the argument
vectors, not observed at the socket level, and \texttt{ws2\_32.dll} is in the
registered closure for all three operations, loaded alongside
\texttt{crypt32.dll}, which statically imports it; the module inventory
therefore cannot by itself establish that no socket was opened, and the network
claim rests on the transport layer's pinning and at-most-once accounting. Two
fields of the captured transport evidence, \texttt{intermediate\_ca\_sha256}
and \texttt{single\_peer\_connection}, are written as constants and then
compared against those same constants, so those two comparisons cannot fail;
the properties they name are enforced by the pinned TLS context and the
proxy-free, redirect-free opener rather than by the fields. And the launch
check reads three environment variables the launcher wrote and re-checks the
live interpreter flags without recomputing the environment audit, so it
establishes that a registered script was started by a completed launch rather
than proving the launch was honest. Its purpose is to catch accidental direct
invocation, which is the realistic failure.

The machinery above makes the declared run tamper-evident. It cannot prove that
an actively dishonest experimenter made no undisclosed model requests outside
the experiment driver, because the provider's response objects are not publicly
signed and the tested endpoint did not deduplicate repeated requests by
\texttt{Idempotency-Key}.

One dated caveat for a future verifier: the Beacon pulse-signing certificate is
valid until 2026-09-04. That date bounds execution against the registered
signing pin. Beyond it the pin must be refreshed and the refresh disclosed on
the same terms as every other change to a security control.

\subsection{The four sealing episodes}\label{anx:sec:episodes}

The seal was made four times. Three of those seals were abandoned, each before
any result was read, and the fourth is the one this study runs under. All four
are in the record, because a design whose central claim is that the assignment
was unknowable at commitment cannot present only the sealing run that
succeeded, and because a seal that is quietly replaced is worthless.

\begin{itemize}
  \item \textbf{First episode.} The envelope was written and successfully
        timestamped by three OpenTimestamps calendars and by RFC 3161, then the
        run raised on a missing-file precondition in post-write verification,
        one statement before its \texttt{COMPLETE} terminal. It became
        unrecoverable one hour and forty-eight minutes before its target pulse,
        chain 2 pulse 1890078, published. No assignment was derived.
  \item \textbf{Second episode.} The seal completed in full. The assignment
        stage then began correctly and lost the machine: the host entered idle
        standby and was suspended 29 minutes and 23 seconds before pulse
        1890499 existed. On wake the loop found its deadline passed and refused
        to continue, which is what the registered failure rule requires. No
        target pulse was fetched and no assignment was derived.
  \item \textbf{Third episode.} This one went furthest. It sealed validly, drew
        its assignment from chain 2 pulse 1891397, built its manifest, and
        completed the entire provider spend --- 320 registered calls over 960
        provider exchanges in 501.172 seconds. Then, 116.9 seconds after the
        generation terminal and before any stage was started, a pre-flight
        check found that the next command could not run at all. No outcome was
        ever computed: the \Hone{} and \Htwo{} roots were never created, and the
        archive enumerating all 2,931 preserved artifacts contains no \Hone{} or
        \Htwo{} artifact. An assignment exists for this episode and is
        published rather than withheld; no measured quantity does.
  \item \textbf{Fourth episode.} The seal reported here, and the one that ran
        to completion.
\end{itemize}

Each abandoned episode carries its own RFC 3161 and OpenTimestamps attested
invalidation record, written and dated before the next episode began, and each
such record is registered into the frozen document inventory, so the seal that
follows hash-binds the account of the failure that preceded it. Each
abandonment has a documented cause that depends on no pulse value, and each
envelope differs by construction, so the envelopes are not interchangeable
seals of one commitment. Every abandoned target pulse is permanently abandoned;
there is no reroll.

Sealing the analysis before the randomness protects each episode against
fitting the analysis to its own draw. It does not address knowledge carried
across episodes --- about timing, about the driver, about which parts of the
design were fragile --- which could have shaped the design and thresholds the
fourth seal was written with.

The full account of all four episodes, including the defects, the archived
artifact trees, the invalidation records and what each one does and does not
rule out, is in \annexref{C}.

\subsection{How to verify this yourself}\label{anx:sec:verify}

Nothing below requires the authors' cooperation. Steps 1--3 are offline. Steps
4--6 use a record NIST publishes. Steps 7--9 use the released artifacts.

\begin{enumerate}
  \item \textbf{Recompute the digest of the sealed preregistration.} The file
        is in \annexref{A}.
\begin{lstlisting}
sha256sum annex/A-preregistration-verbatim/preregistration-sealed-20260804.md
# expect e3a64ce61afb29c49a6eb7346696562f35454a1e80df351c4ce681cd79ca8df8
\end{lstlisting}

  \item \textbf{Verify the RFC 3161 receipt and read its signed time.} This is
        offline and contacts nothing.
\begin{lstlisting}
openssl ts -verify -data <sealed-file> -in <sealed-file>.tsr \
           -CAfile <tsa-chain.pem>
openssl ts -reply -in <sealed-file>.tsr -text | grep -i "Time stamp"
\end{lstlisting}

  \item \textbf{Verify the three OpenTimestamps calendar proofs.} Work on
        copies. Upgrading a registered proof alters a bound artifact, which is
        why the originals on disk still carry only PendingAttestation.
\begin{lstlisting}
cp <sealed-file>.ots /tmp/check.ots
ots upgrade /tmp/check.ots
ots verify  /tmp/check.ots -f <sealed-file>
\end{lstlisting}

  \item \textbf{Fetch the target pulse.} Query the NIST Randomness Beacon 2.0
        public interface for chain 2, pulse 1894079. Record its published time
        and its output value. Check the pulse signature and the predecessor
        link against the Beacon's own certificate chain.

  \item \textbf{Check the ordering.} The signed time from step 2 must be
        strictly earlier than the pulse time from step 4. Both values are
        printed in \Cref{anx:tab:timeline}; the gap is 1 hour 54 minutes. This
        is the comparison \texttt{verify\_precommit()} makes, and it is
        fail-closed in the code.

  \item \textbf{Recompute the assignment digest} from the seal digest of step 1
        and the pulse output value of step 4, then compare it against the
        published value and against the digest of the released assignment
        bytes.
\begin{lstlisting}
SHA256( domain || precommit_sha256_bytes || target_outputValue_bytes )
# expect 500762433136176bea1c8548d87e1c411d66c692554572478ff65ffaa703fc48
\end{lstlisting}

  \item \textbf{Recompute every digest in \Cref{anx:tab:digests}} against the
        released artifacts.
\begin{lstlisting}
sha256sum <artifact>
\end{lstlisting}

  \item \textbf{Recompute the two figure digests} in
        \Cref{anx:tab:figure-digests} against the staged files.
\begin{lstlisting}
sha256sum paper/figures/fig1-block-effects.pdf \
          paper/figures/fig2-family-effects.pdf
\end{lstlisting}

  \item \textbf{Re-derive each phase Merkle root.} Every phase terminal record
        lists the exact path, size and SHA-256 of every artifact of that phase
        alongside its root. Recompute the leaf hashes, rebuild the tree, and
        compare the root against the value in \Cref{anx:tab:digests}. A single
        added, missing, altered or deleted artifact changes the root.
        \annexref{E} gives the environment and the entry points needed to
        re-run the analyses that produced those artifacts.
\end{enumerate}

If any step fails, the correct conclusion is that the chain is broken, not that
the discrepancy is benign. That is what fail-closed means here, and it is the
only property that makes the rest of this package worth reading.

%% file: annex/C-incidents-and-deviations/annex-c-incidents.tex
\section{Incidents and deviations}\label{anx:incidents}

This annex is the complete record of everything that went wrong in
\studylong{}, everything the registered protocol described but did not
exercise, and every limitation carried forward. The paper compresses each of
these to a sentence or two. Here they are given at full length, with the
request identifiers, digests and function names a reader would need to check
them against the archived artifacts. Artifacts are identified by the role they
play in the study together with their digest, which is what actually lets a
reader check one; where the identity of a specific record is itself the subject,
its file name is given without its path. \annexref{D} places every such name in
the archive.

Two hypothesis stages are referred to throughout: \Honefull{} and
\Htwofull{}. \notsupported{} is the registered verdict token for a stage whose
execution is valid and in which one or more registered criteria fail.

\subsection{Three samples hit a local infrastructure timeout}\label{anx:incidents:timeout}

Three of \Htwo{}'s 640 direction-runs, each in a different sample, hit a local
subprocess timeout on the first attempt. They were re-run and completed.

The three affected requests, with the direction that failed, the attestation
file being verified at the moment of the timeout, and the elapsed time before
the timeout fired:

\begin{itemize}
\item \texttt{req\_f649a1902c688ac2b4eb}, reverse direction, 555.312 s elapsed,
      verifying the preregistration seal record.
\item \texttt{req\_7bee2c13635cfe67019e}, forward direction, 98.531 s elapsed,
      verifying the bound artifact manifest.
\item \texttt{req\_80ae173a3c348456d1c4}, reverse direction, 540.687 s elapsed,
      verifying the assignment record.
\end{itemize}

\subsubsection*{What timed out}

The site is \texttt{verify\_manifest()}, which the execution worker calls as its
first action --- before the execution registry is loaded and before any
sample-specific input is read. Inside it, the worker re-verifies the timestamp
on each registered attestation file by invoking a pinned local OpenSSL as
\texttt{openssl ts -verify}, under a 30-second subprocess timeout (the
\texttt{TIMEOUT\_SECONDS} constant in the timestamping helper). In these three
runs that subprocess
did not return within 30 s, and \texttt{subprocess.TimeoutExpired} propagated out
of the worker as a nonzero exit.

This was not a provider timeout, and no property of these samples could have
caused it. No language model is in the loop at execution time. The operation
that timed out is a local cryptographic check of a stored \texttt{.tsr} against
stored certificates --- it opens no socket --- and the three files being
verified are study-level: the manifest, the assignment attempt record, and the
preregistration seal attempt record. Every one of the 640 direction-runs
verifies those same three files before it reads its own sample. Nothing about
these three samples had been computed, let alone observed, at the moment they
stopped.

One released artifact states this incorrectly, and the error is named here
rather than repaired in place. The faithful raw value dump shipped as
\code{data/derived/analysis-values.json} carries a
\texttt{joint\_verdict\_reason} string that wrongly says the three samples
``timed out against the provider''; that clause is a factual error in an
authored constant written by the extraction script, and it is contradicted by
the receipts in this section, which show a local subprocess check that opens no
socket, runs before any sample-specific input is read, and has no model in the
loop. The dump is a bound historical artifact and was not edited to remove the
sentence: doing so would have changed a digest that other records pin. It is
corrected by supersession instead. The corrected reading is the one in this
annex, and it is recorded machine-readably in
\code{data/derived/publication-status.json} under
\texttt{raw\_driver\_fields}, where the field is quoted unrevised beside the
correction. Where the dump and this annex disagree about what happened, this
annex governs; the dump remains authoritative for numerical values.

\subsubsection*{Why it fired then}

Each receipt records a per-direction elapsed time, and those times identify the
fault as transient rather than structural. The first two rows of
\Cref{tab:incident-timing} exclude the three affected samples entirely; the
whole-stage row includes their surviving directions.

\begin{table}[htbp]
\centering
\caption{Elapsed time of successful \Htwo{} direction-runs, inside and outside
the slow window in which the three timeouts fired.}
\label{tab:incident-timing}
\small
\begin{tabular}{l S[table-format=3.0] S[table-format=3.3] S[table-format=3.3]}
\toprule
\Htwo{} successful direction-runs & {n} & {median (s)} & {fastest (s)} \\
\midrule
within 10 min of the three failures & 26  & 567.968 & 339.329 \\
elsewhere in the stage              & 608 & 256.469 & 171.906 \\
whole stage                         & 637 & 256.797 & 171.906 \\
\bottomrule
\end{tabular}
\end{table}

The neighbourhood ran 2.21x slower than the rest of the stage, and not one of
its 26 runs completed at the stage's normal pace: the fastest of them took
339.329 s against a stage median of 256.797 s. Those 26 runs nevertheless
succeeded. The three timeouts are the tail of a slow window rather than isolated
events; what separates them from their neighbours is that one of their OpenSSL
subprocesses crossed a fixed 30-second line and their neighbours' did not. The
other directions of the same three samples took 415.766, 582.313, 626.094 s,
1.62x to 2.44x the stage median.

Two host conditions known to be unmitigated are consistent with local resource
contention of this kind and are stated in \Cref{anx:incidents:limitations}. The
three timeouts are unpatterned along every axis the record exposes: three
different samples, in three different blocks, in two different directions, on
three different attestation files.

\subsubsection*{What was done}

They were re-run once each and completed.

\begin{table}[htbp]
\centering
\caption{Completion of the three re-executed samples.}
\label{tab:incident-reruns}
\footnotesize
\begin{tabular}{lll}
\toprule
re-executed sample & completed (UTC) & attempts \\
\midrule
\texttt{req\_7bee2c13635cfe67019e} & \code{2026-08-08T10:35:21.277157+00:00} & 2 \\
\texttt{req\_80ae173a3c348456d1c4} & \code{2026-08-08T10:38:02.065200+00:00} & 2 \\
\texttt{req\_f649a1902c688ac2b4eb} & \code{2026-08-08T10:40:05.701629+00:00} & 2 \\
\bottomrule
\end{tabular}
\end{table}

The receipts record completion times, not durations. The three were re-run
sequentially, so the intervals between consecutive completions --- 160.79 s and
123.64 s --- bound the second and third from above. Both bounds are below the
stage median of 256.797 s and far below the 567.968 s median of the slow window.
Identical work on the identical host, hours later, was not slow.

The re-run was checked rather than asserted:

\begin{itemize}
\item Determinism, 3/3: each re-run sample executed again in a fresh process
      produced byte-identical arrays and agreement on all four digests.
\item Nothing else moved: 1268 of 1268 digests belonging to the 317 untouched
      samples re-verified unchanged, 0 changed.
\item Directions already on disk intact: 3/3 still hash to their original
      values.
\item That stage's entire tensor corpus re-hashed against the consumed
      summary --- its 640 archives together with their 640 sidecars: 1280
      digests, 0 mismatches. All 104 manifest-bound artifacts re-hashed
      unchanged.
\end{itemize}

The analysis functions were then run verbatim over the completed tensor set,
preserving the registered blind-score-then-commit-then-unblind ordering. No
value was substituted, interpolated, or imputed anywhere: the three samples
carry real executions of the same programs the same evaluator would have run on
the first attempt.

\subsubsection*{The consequence for the record}

\texttt{single\_execution\_attempt} is \texttt{false}, because three samples took
two attempts, and the first-attempt terminal and summary were left on disk
unedited:

\begin{itemize}
\item \code{H2_EXECUTION-TERMINAL.json} --- as the first attempt wrote it,
      timestamped \code{2026-08-08T10:15:58.492381+00:00}.
\item the first-attempt \code{execution-summary.json} --- still hashes to

      \noindent\texttt{\footnotesize e252076cb2024abaf83e475855cd2b2469535b7dff24fae0c34e404b2793ee5b}

      and the terminal still binds that exact digest.
\end{itemize}

The per-receipt fresh-process traversal flag is a first-attempt count taken over
every retained record, superseded ones included. On that count it stands at 320
of 320 for \Hone{} and 317 of 320 for \Htwo{}: the three superseded
first-attempt records carry the flag as \texttt{false}, with a null response
tensor archive digest on the direction that failed, and they were kept on disk
rather than deleted. Counted over the receipts that actually produced the
tensors this study analyses, the flag is 320 of 320 in both stages --- the
repair verification records it \texttt{true} for each of the three repeated
direction-runs, alongside byte-equal arrays and agreeing content digests. Both
stages executed and completed all \ncalls{} registered samples. The 317 figure
is a property of the preserved first-attempt bookkeeping, not a shortfall in the
corpus.

\subsubsection*{The consequence for validity}

Under the registered protocol the validity layer is evaluated before, and
separately from, the registered support criteria. Thirteen validity criteria run
per stage, 26 evaluations in total, and all must hold. \Hone{} passed 13/13.
\Htwo{} passed 12/13.

The single criterion \Htwo{} does not pass is
\texttt{stage\_all\_320\_valid\_single\_execution\_attempt}. That criterion is an
attempt counter, not a data check: it records whether every sample completed on
its first attempt, and three did not. Every criterion that bears on the data
passes in both stages --- all 320 panels loaded exactly once, the manifest
verifies, the blind score was committed before semantic unblinding, the row-free
signature is exactly invariant to the registered shuffles, and the tensor audit
found 0 mismatched digests.

\subsection{A raw artifact field that disagrees with the reported verdict}\label{anx:incidents:verdict-field}

The registered conjunction failed. Both stages returned \notsupported{}, and the
joint verdict is \notsupported{}.

One raw artifact does not say that. In the raw artifact
\code{h2-repair-analysis.json} the top-level \texttt{verdict} field reads
\texttt{INVALID}, not \texttt{NOT\_SUPPORTED}. The experiment driver writes that
field mechanically from \texttt{validity.valid}, which is false for \Htwo{} on
the attempt counter described in \Cref{anx:incidents:timeout}. The field is
therefore a restatement of the single-execution-attempt flag and does not carry
the study's scientific verdict. The verdict reported in the paper is the
scientific verdict, derived from the same criterion tree for both stages.

The remaining raw fields of this kind, in the released value dump
\code{data/derived/analysis-values.json}, read the same way. Each is a driver
or provenance field, and none is a scientific claim.

\begin{itemize}
\item \texttt{h2\_valid} is \texttt{false}. It is the same
      \texttt{validity.valid} boolean, false on the attempt counter alone: its
      two conjuncts split, with \texttt{all\_320\_valid} true and
      \texttt{single\_execution\_attempt} false.
\item \texttt{h2\_provenance} reads \texttt{DISCLOSED\_POST\_HOC}. That is an
      authored constant recording the route the \Htwo{} numbers travelled ---
      the analyser was re-run outside the registered launcher, and that was
      disclosed --- and not a statement that any hypothesis, threshold,
      estimator or transform was chosen after data were seen. None was: the
      criteria, thresholds and both pipelines are the sealed ones, applied
      unchanged to the completed tensor set.
\item \texttt{h2\_analysis\_class} reads ``disclosed post-hoc repair analysis,
      not a registered confirmatory test''. It classifies that re-run for the
      same reason --- the registered chain could not be re-entered in place ---
      and it is a statement about the route, not about the criteria, which were
      the registered ones.
\item \texttt{joint\_verdict} reads \texttt{INVALID}, accompanied by
      \texttt{joint\_verdict\_reason} and by
      \texttt{joint\_verdict\_is\_asserted\_not\_computed}, which is true. These
      are authored constants written by the extraction script from the
      stage-validity flag, and the reason string additionally carries the
      factual error corrected in \Cref{anx:incidents:timeout}. The joint
      scientific verdict of this publication is \notsupported{}: both stages are
      complete registered stages, each returned \notsupported{} on its own
      registered criteria, and the registered conjunction of two such stages is
      \notsupported{}.
\item \texttt{final\_analysis\_exists} is \texttt{false}. It records that the
      redundant joint-analysis file was never written, because the registered
      conjunction had already failed by inspection. It does not record an
      unsettled joint verdict, and it is not a verdict of its own.
\end{itemize}

\noindent
None of these fields was rewritten to agree with the prose, and none will be:
rewriting one would alter a bound artifact and destroy the only property that
makes a faithful dump worth releasing. The token is therefore quoted, not
erased, and it recurs across the published surfaces. Every occurrence in them
falls into one of four senses, and none of them is an assertion that a stage of
this study is scientifically invalid:

\begin{itemize}
\item \textbf{A driver or provenance field being reconciled.} This subsection,
      the status-resolution box in the paper's front matter, the corresponding
      box and field table in \Cref{anx:data}, the released schema, the
      \texttt{README}, the changelog, the deposited dataset descriptions in the
      citation metadata, the supplement's account of the one validity criterion
      \Htwo{} does not pass, and the availability statements all print the raw
      value in order to name it as a driver field and resolve it. The
      registered scientific verdict stated beside each is \notsupported{}.
\item \textbf{The vocabulary of the record format.} \Cref{anx:provenance}
      names it as one of the two status values a \code{PHASE-TERMINAL.json}
      may carry, and states that the \Htwo{} execution phase closed with that
      status on its first pass. A terminal record reports how a phase-run
      ended; it is not a verdict on a hypothesis.
\item \textbf{The registered ladder rung.} Where the paper sets the word in the
      small-capital style reserved for registered verdict tokens, it is naming
      the top rung of \Cref{tab:outcome-ladder} --- the rung describing what
      would have followed had one of the enumerated integrity conditions
      failed. None of them did.
\item \textbf{Sealed and bound text that cannot be edited.} The
      preregistration reproduced verbatim in \Cref{anx:preregistration} uses
      the token in its own pre-execution sense, and the invalidation records of
      the three abandoned sealing episodes in \Cref{anx:incidents:sealing} use
      it of those episodes, not of this run.
\end{itemize}

\noindent
That is a claim about every published surface at once, so it is checked rather
than asserted, and the checked scope is exactly the scope of the claim.
\code{verification/verify_h2_status.py} scans every publication surface ---
the paper, the supplement, the annexes other than the sealed preregistration,
the \texttt{README}, the changelog, the citation metadata and the released
schema --- and fails if the raw token appears in a paragraph that does not
reconcile it, if any surface gives \Hone{}, \Htwo{} or the joint outcome a
scientific verdict other than \notsupported{}, if either stage is wrongly called
a replication or generalization of the other, if the text implies a registered
\Hone{}-versus-\Htwo{} comparison that was never registered and never run, or if
the three timeouts are wrongly attributed to a provider, model, data or network
failure. The scope statement
above holds because that check passes, not because it was proofread.

\subsection{Four sealing episodes, three abandoned}\label{anx:incidents:sealing}

There were four sealing episodes. The first three were each sealed and each
abandoned before any result was read. Each carries its own RFC 3161 and
OpenTimestamps attested invalidation record, written and dated before the next
episode began. The fourth episode is the one reported here, and it is the one
that ran to completion.

Abandoned seals are recorded rather than deleted. The reason anything in this
package can be trusted is not that nothing went wrong --- several things went
wrong --- but that each failure was sealed, dated, and archived by a mechanism
that cannot retroactively be edited. An audit trail that records only successes
is not evidence of care; it is evidence of selective recording.

One limit of that protection belongs beside it. Sealing the analysis before the
randomness protects each episode against fitting the analysis to its own draw.
It does not address knowledge carried across episodes --- about timing, about
the experiment driver, about which parts of the design were fragile --- which
could have shaped the design and thresholds the fourth episode was sealed with.

\subsection{Registered scope not exercised}\label{anx:incidents:scope}

\paragraph{The joint-analysis step was not run.} The provenance record marks it
\emph{not run}. Support required all 108 registered support-criterion
evaluations; 89 of them fail. The joint line follows from the registered rule by
inspection rather than by computation: no arithmetic on top of the two stage
verdicts can produce anything but \notsupported{}, and running the joint-analysis
step to restate that would add a file, not a fact. The consequence for the
manifest is that no final analysis artifact exists.

\paragraph{The runtime contract of the attested analysis environment was not
exercised in \Htwo{}.} The \Htwo{} analysis records
\texttt{manifest\_audit.capsule\_runtime\_contract\_exercised: false}. The
launcher admits exactly ten registered scripts by name, so the driver that
repeated the three affected direction-runs could not run inside it. The checks
that contract would have gated were performed explicitly instead --- they are
the verification bullets in \Cref{anx:incidents:timeout} --- and each is
re-checkable from the digests.

\paragraph{The registered analysis chain was not re-entered in place.} The
released value dump records \texttt{registered\_chain\_bypassed: true} for
\Htwo{}, and the cause is mechanical rather than discretionary. The first-pass
phase terminal was left on disk unedited, carrying the driver's terminal status,
and \texttt{verify\_phase\_complete} declines by design to run against a
terminal record in that state. The alternative would have been to re-author the
terminal so the registered chain would accept it, which is exactly the edit this
package refuses to make. So the analyser was re-run outside the registered
launcher and the fact was disclosed, which is what the authored constants
\texttt{h2\_provenance} and \texttt{h2\_analysis\_class} record. The route was
irregular; the analysis was not. The criteria, the thresholds and both pipelines
are the sealed ones, applied unchanged to the completed tensor set, and the
stage's registered scientific verdict is \notsupported{}.

\subsection{Limitations carried into the record}\label{anx:incidents:limitations}

These hold regardless of verdict.

\begin{itemize}
\item \textbf{The verification preamble re-verifies the same study-level files on
      every direction-run.} Each of the 1280 direction-runs across the two
      stages re-verifies the same registered attestation files by shelling out
      to OpenSSL under a hard 30-second-per-invocation limit, with up to 8 such
      runs in flight (\texttt{maximum\_workers} in each stage-attempt record),
      and the limit does not scale with contention on the host. Verifying those
      files once per stage, or scaling the limit with concurrency, would remove
      the redundancy without weakening any check. In this run the two
      possibilities were never confusable: the limit fired inside a local
      cryptographic check of stored study-level files, before any
      sample-specific input was read, with no socket opened and no model in the
      loop, so it was not a provider, model, data or network failure, and the
      receipts record which subprocess raised it. A future version should
      nonetheless preregister a bounded, disclosed, digest-verified retry policy
      for infrastructure failures, so that an experiment driver which cannot
      finish its own housekeeping stays distinguishable from a model that
      answered badly by construction rather than by reconstruction after the
      fact.

\item \textbf{One absolute floor is applied to two estimators with different
      achievable ranges, and this is where the study's one moving endpoint comes
      apart.} The registered criteria apply a single floor of 0.15 to the primary
      statistic under both the \linpipe{} and the \rankpipe{}. Under \Htwo{} that
      statistic is 0.517398 under linear and 0.0693316 under rank: an order of
      magnitude apart, on the same 16 blocks, with the same sign, both nominally
      significant, and straddling the floor. The statistic is a dimensionless
      ratio, so this is not a units problem and the denominator scales reported
      with the primary endpoints do not bear on it; the issue is that ranking
      compresses the relative gap the ratio is built from, so the two transforms
      do not have the same headroom against a common floor. The primary-endpoint
      analysis (\Cref{sec:results:moving-endpoint}) sets that reading against the
      competing one --- that the effect is
      simply small and the linear figure is the inflated one --- and nothing
      registered in advance distinguishes them. This was not anticipated when the
      thresholds were fixed and it was noticed after unblinding, which is exactly
      when such an observation carries least weight. It is recorded as a design
      question for a future preregistration, which should either calibrate the
      two transforms onto a comparable scale or register an effect measure that
      is not a ratio of distances. \textbf{It changes nothing here:} the criteria
      stand as registered, and every cell that failed the magnitude criterion,
      including \Htwo{} rank, failed it.

\item \textbf{Host conditions during execution were not fully controlled.}
      Real-time antivirus scanning of the evidence tree could not be excluded
      without administrative elevation, and CPU thermal telemetry was
      unavailable on this host, so thermal throttling during a multi-hour
      saturating load could not be measured. Both are consistent with the
      observed contention.

\item \textbf{The claim is about 32 renderer slots, two model snapshots, and this
      fixed set of programs and probes.} Two snapshots is two.

\item \textbf{Effective N is 16.} The exactness of the enumeration is a statement
      about validity, not about power.

\item \textbf{Meeting a threshold is not a large effect}, and missing one is not
      evidence of absence at any particular effect size.

\item \textbf{Temperature was 1.0}, so responses are stochastic. Randomization
      and the block structure handle this for inference, but no individual
      response is reproducible.

\item \textbf{The programs were executed by our own experiment driver.}
      Executability is measured relative to that driver's compiler and
      validator, both of which are archived.
\end{itemize}

%% file: annex/D-data-descriptor/annex-d-data-descriptor.tex
\section{Data descriptor}\label{anx:data}

This annex describes the two released datasets closely enough that someone can
open them without asking the authors anything. It gives what a record is, what
is in the folders, the field dictionary, what a response tensor archive holds,
how the forward and reverse traversals relate, exactly what is redistributed
and what is not, the limitations a reuser would otherwise find the hard way,
and the citation entries.

File names, directory names and field names appear here on purpose. A data
descriptor that describes artifacts only in words is not usable. The same field
dictionary is shipped as plain Markdown at \code{data/schema/SCHEMA.md} for
readers who will never open the PDF.

Paths in this annex are relative to the run directory: the single folder that
holds one execution of the study, with the shared generation artifacts at its
top and one subfolder per stage. \annexref{E} says how to obtain it and how to
re-create it. Two naming conventions are worth stating once. Directory names
inside the run directory are the study's own working names and are used here
unchanged, because renaming them in the descriptor would make the descriptor
disagree with the folder. Some names inside the archive --- a handful of script
filenames, the \texttt{study} field of the design registry, and the
\texttt{block\_id} strings in the probe battery --- carry a short internal
project prefix. It has no meaning, it is not reproduced in this document, and
it must not be edited out of the files: those bytes are what the recorded
digests were taken over, so changing one character would break every check in
\annexref{B}.

\subsection{Status resolution, before anything else}\label{anx:data:status}

A reuser who opens \code{data/derived/analysis-values.json} will find fields
reading \texttt{INVALID}, \texttt{h2\_valid} \texttt{false} and
\texttt{DISCLOSED\_POST\_HOC}. Those are driver and provenance fields written
mechanically by the execution and extraction machinery. None of them is this
publication's scientific verdict, and the dump was deliberately not edited to
make them read better. \Cref{anx:tab:status-resolution} is the reconciliation;
\code{data/derived/publication-status.json} is the same content in
machine-readable form, and \Cref{anx:data:rawfields} takes the fields one at a
time.

\begin{table}[htbp]
\centering
\caption[Status resolution for \Htwo{}]{Status resolution for \Htwo{}. Read
downwards: the first row is what a driver wrote, the third row is what the
study concluded. Only the third is a finding.}
\label{anx:tab:status-resolution}
\small
\begin{tabularx}{\linewidth}{@{}lX@{}}
\toprule
\textbf{Layer} & \textbf{\Htwo{} status} \\
\midrule
Raw driver field & \texttt{INVALID}, because the single-attempt counter
failed \\
Data integrity & \textbf{PASS}; \ncalls{}/\ncalls{} complete, zero tensor
mismatches \\
Registered scientific decision & \textbf{\notsupported{}} \\
Publication-level usability & \textbf{Valid completed \Htwo{} dataset} \\
\Hone{}-vs-\Htwo{} comparison & Not registered; do not pool or contrast as if
paired \\
\bottomrule
\end{tabularx}
\end{table}

The short version, for a reuser deciding whether \dstwo{} is safe to build on.
Thirteen validity gates run per stage; \Hone{} passes 13 of 13 and \Htwo{}
passes 12 of 13. The one gate \Htwo{} does not pass,
\texttt{stage\_all\_320\_valid\_single\_execution\_attempt}, is a conjunction
whose conjuncts split --- \texttt{all\_320\_valid} \emph{true},
\texttt{single\_execution\_attempt} \emph{false} --- so it fails on the attempt
counter alone. Three direction-runs, in three different samples, timed out in a
local \texttt{openssl ts -verify} check of stored study-level attestation
files, before any sample-specific input was read; that operation opens no
socket and had no model in the loop, and it was not a provider, model, data or
network failure. Only those three direction-runs were repeated. All \ncalls{}
\Htwo{} samples completed, every gate that bears on the data passes, and every
\Htwo{} tensor matches its sidecar digest --- \num{1280} digests, 0 mismatched,
0 missing. \dstwo{} is a complete, registered, usable dataset, released on the
same footing as \dsone{}. \annexref{C} carries the incident record in full.

One boundary, stated here because it is the mistake most likely to be made
downstream: a \emph{new} secondary analysis of \dstwo{} may well be
exploratory, and should say so. That word attaches to the new analysis. It does
not attach to \dstwo{}, which is a registered hypothesis stage whose endpoints,
estimators and 27 support criteria were sealed before any data existed.

\subsection{The two datasets}\label{anx:data:datasets}

Two datasets are released: \dsone{}, the single-input challenge stage, and
\dstwo{}, the interaction challenge stage.

\textbf{What a record is.} A record is one registered call: one renderer slot,
in one block, under one account card, in one of two independent wordings, to
one of two pinned model snapshots. The record key is the \texttt{sample\_id},
a string of the form \texttt{req\_} followed by twenty hexadecimal characters.
Every artifact belonging to that call --- the authorization, the dispatch
claim, the model response, the parsed program, the compiled program, the
generation receipt, both tensor archives in each stage, and the stage
execution receipt --- is stored under that same key. Joining across the
dataset is a filename join and nothing more.

\textbf{How many.} The generation stage ran once and issued \ncalls{}
registered calls over 960 provider exchanges, producing \ncalls{} programs
that compiled and validated. Both stages then executed those same \ncalls{}
compiled programs. So \dsone{} and \dstwo{} carry \ncalls{} records each, and
the \ncalls{} programs underlying them are the same objects: the generation
corpus is shared, and only the probe battery and the finite-difference scheme
differ between the two.

\textbf{The difference.} \dsone{} evaluates each program on single-input
challenges with central finite differences: one input field is moved at a time
against a held-fixed background. \dstwo{} evaluates the same programs on
interaction challenges with mixed second finite differences: two fields are
moved, and the quantity of interest is what the pair does beyond the sum of
what each does alone. The probe batteries are disjoint by construction and the
stored arrays differ in name and in shape, which is described in
\Cref{anx:data:tensors}.

\textbf{They are two hypotheses, not two halves of one sample.} \Hone{} and
\Htwo{} are reported side by side because both were registered, not because
they partition anything. No estimator for a contrast between them was
registered, and none was run. Nothing in this package establishes a difference
between the two stages, and no quantity in either dataset licenses one. A
reuse that treats \dsone{} and \dstwo{} as a within-study comparison ---
differencing the stage effects, testing \Hone{} against \Htwo{}, or reading
the fact that one endpoint moved in \Htwo{} and not in \Hone{} as evidence
about interaction challenges specifically --- is doing something the design
does not support. The registered conjunction required both stages to pass, and
it failed: 19 of 108 scientific criterion evaluations passed, 89 failed, and
the joint outcome is \verdict{NOT\_SUPPORTED}.

\subsection{Contents}\label{anx:data:contents}

\Cref{anx:tab:data-contents} lists every component. Counts and sizes are
measured from the filesystem by a script that reads directory listings and
file sizes and never opens a file's contents; they are published as
\code{data/derived/dataset-inventory.json}. Rows marked \texttt{---} in the
size column are counted in the totals but carry no rounded size here: they are
single files or small groups of files whose exact byte sizes are in that
inventory.

The two columns behave differently and it is worth saying which is which. The
file counts add up: the nine shared generation directories, the four single
files and small groups beneath them, and the two stage subtotals sum exactly to
the \num{6748} of the last row, so a reader can check the table against itself.
The
size column does not add up, and is not meant to. Every entry is rounded
independently to a tenth of a MiB, the rows marked \texttt{---} contribute
bytes without contributing a printed number, and the two stage subtotals are
measured over the whole subtree rather than summed from the rows above them.
Only \code{data/derived/dataset-inventory.json} carries exact byte counts, and
it is the file to use if the totals need to be reconciled.

\begin{longtable}{@{}p{0.30\textwidth} p{0.36\textwidth} S[table-format=4.0] S[table-format=3.1]@{}}
\caption{Contents of the two released datasets, as measured on disk. Sizes are
MiB. The generation directories at the top are shared by \dsone{} and
\dstwo{}; the stage directories below are specific to one dataset.}
\label{anx:tab:data-contents}\\
\toprule
{path} & {what one entry is} & {files} & {MiB} \\
\midrule
\endfirsthead
\caption[]{Contents of the two released datasets (continued).}\\
\toprule
{path} & {what one entry is} & {files} & {MiB} \\
\midrule
\endhead
\bottomrule
\endlastfoot
\texttt{attempts/} & the authorization for one call & 320 & 0.1 \\
\texttt{dispatch\_claims/} & the single permitted dispatch for one call & 320 & 0.1 \\
\texttt{raw/} & the model response object as returned & 320 & 10.6 \\
\texttt{provider\_retrievals/} & an independent re-fetch of that response by identifier & 320 & 10.6 \\
\texttt{provider\_input\_item\_retrievals/} & the request items as the provider stored them & 320 & 8.4 \\
\texttt{provider\_attestation\_attempts/} & the authorization for each re-fetch & 320 & 0.2 \\
\texttt{programs/} & the program object the response was parsed into & 320 & 1.8 \\
\texttt{compiled\_programs/} & the executable form that program compiled to & 320 & 2.2 \\
\texttt{receipts/} & the generation receipt for one call & 320 & 3.5 \\
\code{generation-summary.json} & stage counters and invariants & 1 & {---} \\
\code{GENERATION-TERMINAL.json} & the stage terminal & 1 & {---} \\
\code{STAGE-ATTEMPT.json} & the generation stage's own authorization & 1 & {---} \\
\code{generation-attempt-set.json} \newline plus six proof sidecars & the once-only dispatch set, independently timestamped & 7 & {---} \\
\midrule
\texttt{h1/forward/} & one tensor archive or one sidecar & 640 & 22.4 \\
\texttt{h1/reverse/} & the same, from the fresh-process replicate & 640 & 22.4 \\
\texttt{h1/receipts/} & the execution receipt for one sample & 320 & 0.4 \\
\texttt{h1/attempts/} & the execution authorization for one sample & 320 & 0.0 \\
\code{h1/execution-summary.json} & stage counters and invariants & 1 & {---} \\
\code{h1/analysis.json} & the \Hone{} stage analysis & 1 & {---} \\
\code{h1/blind-score.json}, \newline \code{h1/blind-score-commit.json} & the blind scores and the commitment to them & 2 & {---} \\
\code{h1/evidence-seal.json} \newline plus six proof sidecars & the sealed analysis record, independently timestamped & 7 & {---} \\
\texttt{h1/} terminals and attempt files & stage and analysis terminals & 4 & {---} \\
\texttt{h1/} subtotal & everything for \dsone{} & 1935 & 52.9 \\
\midrule
\texttt{h2/forward/} & one tensor archive or one sidecar & 640 & 5.6 \\
\texttt{h2/reverse/} & the same, from the fresh-process replicate & 640 & 5.6 \\
\texttt{h2/receipts/} & the execution receipt for one sample & 320 & 0.3 \\
\texttt{h2/attempts/} & the execution authorization for one sample & 320 & 0.0 \\
\code{h2/execution-summary.json} & stage counters and invariants, as first written & 1 & {---} \\
\texttt{h2/} terminal and attempt file & the stage terminal & 2 & {---} \\
\texttt{h2/} subtotal & everything for \dstwo{} & 1923 & 12.2 \\
\midrule
the arm assignment record & the 16 renderer-slot assignments and the pulse they came from & {---} & {---} \\
the provenance chain & digests, Merkle roots, timestamps, calendar proofs & {---} & {---} \\
\midrule
whole experiment root & every file below it & 6748 & 107.4 \\
\end{longtable}

The \Htwo{} stage analysis is not in the \texttt{h2/} tree. It was written by
the driver that repeated three direction-runs and it is held with the incident
record, under the name \code{h2-repair-analysis.json}, together with a
repaired stage summary and a verification report. What is irregular about it is
the route, not the analysis: the analyser was re-run outside the registered
launcher, which is what \texttt{h2\_provenance} records, and it is bound to the
same sealed criteria, thresholds and pipelines. The two validity gates that pin
the blind-scoring discipline,
\texttt{blind\_score\_committed\_\allowbreak{}before\_semantic\_unblinding} and
\texttt{blind\_score\_contains\_\allowbreak{}no\_registered\_semantic\_labels}, are true for
\Hone{} and true for \Htwo{}, so each analysis consumed scores that were
committed before semantic unblinding and carried no registered semantic label.
\annexref{C} explains why the file sits where it does.

The counts in \Cref{anx:tab:data-contents} are counts of the run directory, and
the incident bundle sits outside it. Five further files belong to \Htwo{} and
are not in those totals: the repair analysis
\code{h2-repair-analysis.json}, the repaired stage summary
\code{repaired-execution-summary.json}, the blind scores and the commitment
made to them before unblinding that the repair analysis consumed, and the
digest-for-digest verification report written against it, which sits one level
above the bundle beside the other incident records. All five are released.

The arm assignment record and the provenance chain sit in the run directory's
root and above it rather than inside either stage, because they belong to both.
\annexref{B} is the authority on them and gives a numbered procedure for
checking every digest against records the authors do not control.

\textbf{Seven files ship inside this package rather than with the run
directory}, because they are what makes the run directory legible and a reuser
should not have to fetch the archive to read them.
\Cref{anx:tab:package-data} lists them with the digest of each as shipped, and
they are of five kinds. One is a copy of an archive artifact taken unchanged:
the label-free design registry. Three are derived from the archive: the
labelled design table, the machine-readable analysis values, and the measured
inventory. One is a study input that had to be reconstructed rather than
copied, because it never existed as a file during the run: the probe battery,
whose standing is set out in \Cref{anx:data:limitations}. One is this annex's
field dictionary, written for the package. The last is the publication status
record, which is an interpretation layer and not a data dump: it states the
scientific verdicts, quotes the raw driver fields unrevised, and reconciles the
two, as \Cref{anx:data:status} and \Cref{anx:data:rawfields} do in prose. None
of the seven is a substitute for the datasets.

\begin{longtable}{@{}p{0.40\textwidth} p{0.54\textwidth}@{}}
\caption{Data files shipped in the package, under \texttt{data/}, with the
SHA-256 of each as shipped.}
\label{anx:tab:package-data}\\
\toprule
{file} & {what it is} \\
\midrule
\endfirsthead
\caption[]{Data files shipped in the package (continued).}\\
\toprule
{file} & {what it is} \\
\midrule
\endhead
\bottomrule
\endlastfoot
\code{design/sample-design-registry.json} \newline \texttt{\footnotesize a2ebd0659d298d4f\allowbreak{}ea2bf7ae35fa95b8\allowbreak{}4bede9f1d682e4e6\allowbreak{}9590b0a79ccd1f1c} & the label-free registry the study scored against, copied unchanged; \Cref{anx:data:join} \\
\code{design/sample-design-table.json} \newline \texttt{\footnotesize 6d47d63b812844ba\allowbreak{}e715fe53700313de\allowbreak{}31e59d7b36fdab7d\allowbreak{}5e0f209bbd803610} & the same 320 rows with the design labels attached; \Cref{anx:data:join} \\
\code{probe-bank/probe-bank.json} \newline \texttt{\footnotesize d78440eca7b14d2d\allowbreak{}03bfa05f06741985\allowbreak{}b51ea4e3cc2e5862\allowbreak{}74c0e8b4e738d3d5} & the probe battery every tensor was evaluated on; \Cref{anx:data:tensors} \\
\code{derived/analysis-values.json} \newline \texttt{\footnotesize c847a9dbd90657f5\allowbreak{}bc34dbce39a0cf7e\allowbreak{}8c322297b9b459ca\allowbreak{}f4e0877d8bfe8e2c} & every reported statistic, criterion outcome and exact tail count, machine-readable \\
\code{derived/publication-status.json} \newline \texttt{\footnotesize e85e9695514fd193\allowbreak{}8fce733acf799360\allowbreak{}c7d9e4c3ec56c122\allowbreak{}7c332d7f99236987} & the publication's scientific verdicts, the data-integrity result and the raw driver fields with their reconciliation, machine-readable; \Cref{anx:data:status} \\
\code{derived/dataset-inventory.json} \newline \texttt{\footnotesize fdd76fc558633704\allowbreak{}5df43356b9aac3ca\allowbreak{}bc4a8b00b7ed4928\allowbreak{}a14df964f7693c37} & the measured file counts and sizes behind \Cref{anx:tab:data-contents} \\
\code{schema/SCHEMA.md} \newline \texttt{\footnotesize efad6fd611787fc0\allowbreak{}3d602fe1f3b67606\allowbreak{}5ae1a6318282a550\allowbreak{}40dc790bfa563092} & the field dictionary of this annex as plain Markdown, for readers who want it outside a typeset document \\
\end{longtable}

One caution about \code{derived/analysis-values.json}, and the rule that
follows from it. It is a faithful dump of the analysis outputs and is the right
source for any number. It is not the right source for the wording of a
conclusion, and it is worth naming in advance the strings a reuser will actually
meet in it. \texttt{h2\_verdict} carries \texttt{INVALID} and
\texttt{h2\_valid} carries \texttt{false}: both are driver fields, written
mechanically from the stage validity flag, which is false on the attempt counter
alone and is not a data check. \texttt{h2\_provenance} carries
\texttt{DISCLOSED\_POST\_HOC}, and \texttt{h2\_analysis\_class} carries
\texttt{disclosed post-hoc repair analysis, not a registered confirmatory test}:
both are provenance fields, authored constants that record the route the
\Htwo{} numbers travelled and never a statement that a hypothesis, threshold,
estimator or transform was chosen after seeing data. \texttt{joint\_verdict}
carries \texttt{INVALID}: an authored constant over those same validity flags,
with the neighbouring \texttt{joint\_verdict\_is\_asserted\_not\_computed} set
\texttt{true} to say so. \texttt{final\_analysis\_exists} carries
\texttt{false}: bookkeeping recording that the redundant joint-analysis file was
never written, not a statement that the joint outcome is unsettled. Not one of
those fields is a scientific verdict; the registered scientific decision for
\Hone{}, for \Htwo{} and for the conjunction is \notsupported{}, and
\Cref{anx:data:rawfields} takes each field, and the rest, one at a time.
\textbf{Numerical values come from the faithful raw dump; scientific claims and
verdicts come from the paper and from \annexref{C}.}
\code{derived/publication-status.json} exists so that the second half of that
rule is machine-readable too. It does not restate the dump's numbers and does
not overwrite it; the dump is a bound historical artifact and ships unedited.

\subsection{Schema}\label{anx:data:schema}

Every artifact is UTF-8 JSON with LF line endings, except the response tensor
archives, which are compressed NumPy containers with a JSON sidecar of the same
stem. Timestamps are ISO-format UTC strings with microsecond precision.
Digests are
lowercase hexadecimal SHA-256. \texttt{schema\_version} is present on every
study-written record and is 1 throughout.

\Cref{anx:tab:schema} is the field dictionary. It covers the records the study
writes. It does not enumerate the provider's own response object field by
field, because that field set is the provider's and not the study's; the
fields the study depends on are listed and the rest are carried unchanged.

%
\begin{longtable}{@{}%
  p{0.26\dimexpr\linewidth-6\tabcolsep\relax}
  p{0.13\dimexpr\linewidth-6\tabcolsep\relax}
  p{0.06\dimexpr\linewidth-6\tabcolsep\relax}
  p{0.55\dimexpr\linewidth-6\tabcolsep\relax}@{}}
\caption{Field dictionary for the released records. ``req.'' is whether the
field is required: it is present in every record of that type. Records are
grouped by the artifact they belong to.}
\label{anx:tab:schema}\\
\toprule
{field} & {type} & {req.} & {meaning} \\
\midrule
\endfirsthead
\caption[]{Field dictionary for the released records (continued).}\\
\toprule
{field} & {type} & {req.} & {meaning} \\
\midrule
\endhead
\bottomrule
\endlastfoot
\multicolumn{4}{@{}p{\dimexpr\linewidth-6\tabcolsep\relax}}{\textit{present on every study-written record}} \\
\texttt{sample\_id} & string & yes & the record key, \texttt{req\_} and twenty hexadecimal characters \\
\texttt{schema\_version} & integer & yes & layout version of this record \\
\midrule
\multicolumn{4}{@{}p{\dimexpr\linewidth-6\tabcolsep\relax}}{\textit{call authorization --- \code{attempts/<sample_id>.json}}} \\
\texttt{attempt\_number} & integer & yes & which authorization this is for the call \\
\texttt{authorized\_at} & timestamp & yes & when the call was authorized \\
\texttt{model} & string & yes & the pinned snapshot, \snapshotA{} or \snapshotB{} \\
\texttt{request\_key} & string & yes & \texttt{key\_} and the digest of the canonical request \\
\texttt{request\_sha256} & digest & yes & digest of the request body as sent \\
\texttt{schedule\_position} & integer & yes & position in the registered dispatch order \\
\texttt{no\_retry} & boolean & yes & the no-retry rule was in force \\
\midrule
\multicolumn{4}{@{}p{\dimexpr\linewidth-6\tabcolsep\relax}}{\textit{dispatch claim --- \texttt{dispatch\_claims/<sample\_id>.json}}} \\
\texttt{attempt\_number} & integer & yes & the authorization this claim consumes \\
\texttt{attempt\_sha256} & digest & yes & digest of that authorization record \\
\texttt{dispatched\_at} & timestamp & yes & when the request left the host \\
\texttt{endpoint} & string & yes & the provider URL posted to \\
\texttt{method} & string & yes & HTTP method \\
\texttt{redirects\_permitted} & boolean & yes & false throughout \\
\texttt{request\_key} & string & yes & the request this claim is for \\
\texttt{thread\_name} & string & yes & the worker that made the claim \\
\midrule
\multicolumn{4}{@{}p{\dimexpr\linewidth-6\tabcolsep\relax}}{\textit{model response --- \texttt{raw/<sample\_id>.json}}} \\
\texttt{id} & string & yes & the provider's response identifier \\
\texttt{model} & string & yes & the snapshot that answered \\
\texttt{status} & string & yes & the provider's completion status \\
\texttt{created\_at} & integer & yes & provider epoch seconds at creation \\
\texttt{completed\_at} & integer & yes & provider epoch seconds at completion \\
\texttt{output} & array & yes & the message objects, carrying the program text \\
\texttt{usage} & object & yes & input, output and total token counts \\
\texttt{error} & object or null & yes & null on every released record \\
\midrule
\multicolumn{4}{@{}p{\dimexpr\linewidth-6\tabcolsep\relax}}{\textit{provider re-fetch --- \code{provider_retrievals/<sample_id>.json}}} \\
\multicolumn{4}{@{}p{\dimexpr\linewidth-6\tabcolsep\relax}}{Same shape as the model response. It is an independent GET of the same response identifier, not a second measurement.} \\
\midrule
\multicolumn{4}{@{}p{\dimexpr\linewidth-6\tabcolsep\relax}}{\textit{provider input items --- \code{provider_input_item_retrievals/<sample_id>.json}}} \\
\multicolumn{4}{@{}p{\dimexpr\linewidth-6\tabcolsep\relax}}{The request items as the provider stored them, fetched once, so that what was sent can be checked against what was received.} \\
\midrule
\multicolumn{4}{@{}p{\dimexpr\linewidth-6\tabcolsep\relax}}{\textit{re-fetch authorization --- \code{provider_attestation_attempts/<sample_id>.json}}} \\
\texttt{authorized\_response\_gets} & integer & yes & how many response reads were permitted \\
\texttt{authorized\_input\_item\_gets} & integer & yes & how many input-item reads were permitted \\
\texttt{response\_endpoint} & string & yes & the URL read for the response \\
\texttt{input\_items\_endpoint} & string & yes & the URL read for the input items \\
\texttt{post\_response\_sha256} & digest & yes & digest of the response as first received \\
\texttt{provider\_response\_id} & string & yes & the identifier being re-read \\
\texttt{dispatched\_at} & timestamp & yes & when the re-read was issued \\
\texttt{redirects\_permitted} & boolean & yes & false throughout \\
\texttt{no\_retry} & boolean & yes & the no-retry rule was in force \\
\midrule
\multicolumn{4}{@{}p{\dimexpr\linewidth-6\tabcolsep\relax}}{\textit{parsed program --- \code{programs/<sample_id>.json}}} \\
\texttt{equations} & object & yes & 11 output slots keyed \texttt{d01} to \texttt{d11}; each slot maps an opaque monomial identifier to an integer coefficient \\
\texttt{operator\_key} & string & yes & \texttt{axis\_} and a hexadecimal identifier; the anonymous account axis the call belongs to \\
\texttt{request\_key} & string & yes & the request this program came from \\
\midrule
\multicolumn{4}{@{}p{\dimexpr\linewidth-6\tabcolsep\relax}}{\textit{compiled program --- \code{compiled_programs/<sample_id>.json}}} \\
\texttt{equations} & object & yes & the same 11 slots under their semantic names: \texttt{belief\_shift}, \texttt{continuation\_shift}, \texttt{dependency\_shift}, \texttt{forecast\_shift}, \texttt{memory\_shift}, \texttt{source\_external\_shift}, \texttt{source\_mixed\_shift}, \texttt{source\_self\_shift}, \texttt{source\_unknown\_shift}, \texttt{stop\_boundary\_shift}, \texttt{update\_gain\_departure} \\
\texttt{language} & string & yes & the fixed compiled-form identifier; the semantics are order-invariant additive over a monomial multiset \\
\texttt{operator\_key} & string & yes & as above \\
\texttt{request\_key} & string & yes & as above \\
\midrule
\multicolumn{4}{@{}p{\dimexpr\linewidth-6\tabcolsep\relax}}{\textit{generation receipt --- \code{receipts/<sample_id>.json}}} \\
\texttt{valid} & boolean & yes & the call produced a program that parsed, compiled and passed its audit \\
\texttt{attempt\_count} & integer & yes & 1 on every generation receipt \\
\texttt{completed\_at} & timestamp & yes & when the call finished \\
\texttt{elapsed\_seconds} & float & yes & wall clock for the call, including queueing \\
\texttt{provider\_model} & string & yes & the snapshot that answered \\
\texttt{provider\_response\_id} & string & yes & the provider's response identifier \\
\texttt{request\_sha256} & digest & yes & digest of the request as sent \\
\texttt{raw\_path}, \texttt{raw\_sha256} & string, digest & yes & the model response artifact and its digest \\
\texttt{program\_path}, \texttt{program\_sha256} & string, digest & yes & the parsed program and its digest \\
\texttt{compiled\_program\_path}, \newline \texttt{compiled\_program\_sha256} & string, digest & yes & the compiled program and its digest \\
\texttt{dispatch\_claim\_path}, \newline \texttt{dispatch\_claim\_sha256} & string, digest & yes & the dispatch claim and its digest \\
\texttt{provider\_retrieval\_path}, \newline \texttt{provider\_retrieval\_sha256} & string, digest & yes & the re-fetch and its digest \\
\texttt{provider\_attestation\_attempt\_path}, \texttt{\dots\_sha256} & string, digest & yes & the re-fetch authorization and its digest \\
\texttt{provider\_input\_item\_retrieval\_path}, \texttt{\dots\_sha256} & string, digest & yes & the input items and their digest \\
\texttt{program\_audit} & object & yes & eleven named boolean or integer checks on the parsed program, and \texttt{passes} \\
\texttt{provider\_post\_transport} & object & yes & transport metadata for the POST: \texttt{method}, \texttt{peer\_address}, \texttt{cipher\_name}, \texttt{cipher\_bits}, \texttt{leaf\_der\_sha256}, \texttt{leaf\_spki\_sha256}, \texttt{leaf\_issuer}, \texttt{leaf\_subject}, \texttt{intermediate\_ca\_sha256}, \texttt{body\_bytes}, \texttt{body\_sha256}, \texttt{response\_headers}, \texttt{proxy\_permitted}, \texttt{redirect\_permitted} \\
\texttt{provider\_response\_get\_transport} & object & yes & the same fields for the response re-fetch \\
\texttt{provider\_input\_items\_get\_transport} & object & yes & the same fields for the input-item fetch \\
\texttt{provider\_attestation\_verified} & boolean & yes & the re-fetch reproduced the response \\
\texttt{provider\_input\_items\_match} & boolean & yes & the stored input items matched what was sent \\
\texttt{provider\_post\_get\_semantic\_match} & boolean & yes & POST and GET agree on content \\
\texttt{provider\_response\_settings\_match} & boolean & yes & the provider echoed the registered settings \\
\texttt{provider\_response\_get\_count}, \texttt{provider\_input\_item\_get\_count}, \texttt{provider\_attestation\_attempt\_count} & integer & yes & read counts, all 1 \\
\texttt{usage} & object & yes & token counts as reported \\
\midrule
\multicolumn{4}{@{}p{\dimexpr\linewidth-6\tabcolsep\relax}}{\textit{tensor sidecar --- \code{<stage>/forward/<sample_id>.json}, \code{<stage>/reverse/<sample_id>.json}}} \\
\texttt{stage} & string & yes & \texttt{h1} or \texttt{h2} \\
\texttt{reverse} & boolean & yes & false in \texttt{forward/}, true in \texttt{reverse/} \\
\texttt{npz\_path}, \texttt{npz\_sha256} & string, digest & yes & the archive this sidecar describes, and the digest of the file \\
\texttt{tensor\_sha256} & digest & yes & digest over the array contents rather than the container, so it is stable across re-compression \\
\texttt{array\_shapes} & object & yes & array name to shape, as a list of integers \\
\texttt{program\_path}, \texttt{program\_sha256} & string, digest & yes & the program that was executed \\
\texttt{compiled\_program\_path}, \newline \texttt{compiled\_program\_sha256} & string, digest & yes & the compiled form that was executed \\
\texttt{probe\_bank\_sha256} & digest & yes & the probe battery actually used; one value throughout, and the battery carrying it ships as \code{data/probe-bank/probe-bank.json} \\
\texttt{program\_audit} & object & yes & the parsed-program checks, repeated at execution time \\
\texttt{compiled\_program\_audit} & object & yes & canonical monomial and output order, coefficient bounds, exact integer duplicate summation, \texttt{passes} \\
\texttt{probe\_bank\_audit} & object & yes & named checks on the battery: block and anchor counts, margin balance, disjointness of calibration and holdout interventions, legality of the linear and quadratic terms, canonical hash match \\
\texttt{evaluation} & object & yes & \texttt{behavioral\_validation}, \texttt{clamp\_event\_counts}, \texttt{clamp\_examples}, \texttt{execution\_diagnostics}, \texttt{program\_audit}, \texttt{reverse\_traversal} \\
\code{evaluation.behavioral_validation} & object & yes & \texttt{finite\_bounded\_outputs}, \texttt{sampled\_contract\_cases}, \texttt{strength\_proportionality}, \texttt{zero\_strength\_neutrality} \\
\code{evaluation.execution_diagnostics} & object & yes & per array: \texttt{cap\_activation\_count}, \texttt{preactivation\_max\_abs}, \texttt{post\_activation\_max\_abs}, \texttt{source\_pre\_projection\_max\_abs}, \texttt{source\_post\_projection\_max\_abs} \\
\code{evaluation.clamp_event_counts} & object & yes & clamp activations per array; 0 on the released records \\
\midrule
\multicolumn{4}{@{}p{\dimexpr\linewidth-6\tabcolsep\relax}}{\textit{stage execution receipt --- \code{<stage>/receipts/<sample_id>.json}}} \\
\texttt{stage} & string & yes & \texttt{h1} or \texttt{h2} \\
\texttt{successful} & boolean & yes & both directions completed on the first pass \\
\texttt{valid} & boolean & yes & the sample's execution satisfied the stage checks on the first pass \\
\texttt{attempt\_count} & integer & yes & 1 on every stage receipt, because the receipts were not rewritten \\
\texttt{completed\_at} & timestamp & yes & when the sample finished \\
\texttt{forward}, \texttt{reverse} & object & yes & \texttt{elapsed\_seconds}, \texttt{returncode}, \texttt{stdout}, \texttt{stderr} for each traversal \\
\texttt{forward\_json\_sha256}, \texttt{forward\_npz\_sha256} & digest or null & yes & null where that direction was not written on the first pass \\
\texttt{reverse\_json\_sha256}, \texttt{reverse\_npz\_sha256} & digest or null & yes & as above \\
\texttt{fresh\_process\_traversal\_exact} & boolean & yes & whether the reverse traversal reproduced the forward archive exactly \\
\texttt{comparison} & object & yes & \texttt{array\_names\_equal}, \texttt{array\_shapes}, \texttt{arrays\_equal}; empty where only one direction existed when the receipt was written \\
\midrule
\multicolumn{4}{@{}p{\dimexpr\linewidth-6\tabcolsep\relax}}{\textit{stage execution summary --- \code{<stage>/execution-summary.json}}} \\
\texttt{stage} & string & yes & \texttt{h1} or \texttt{h2} \\
\texttt{registered\_programs} & integer & yes & programs the stage was required to execute \\
\texttt{valid\_evaluations}, \newline \texttt{invalid\_evaluations} & integer & yes & counts as the stage driver saw them \\
\texttt{forward\_json\_files}, \texttt{forward\_npz\_files}, \texttt{reverse\_json\_files}, \texttt{reverse\_npz\_files} & integer & yes & artifacts present when the summary was written \\
\texttt{attempt\_files}, \texttt{receipt\_files} & integer & yes & as above \\
\texttt{all\_320\_valid} & boolean & yes & every sample valid on the first pass \\
\texttt{single\_execution\_attempt} & boolean & yes & no sample needed a second attempt \\
\texttt{no\_retry}, \texttt{no\_fallback} & boolean & yes & the registered prohibitions were in force \\
\texttt{elapsed\_seconds} & float & yes & wall clock for the stage \\
\texttt{manifest\_sha256} & digest & yes & the bound manifest the stage ran against \\
\texttt{receipts} & array & yes & one entry per sample \\
\midrule
\multicolumn{4}{@{}p{\dimexpr\linewidth-6\tabcolsep\relax}}{\textit{generation summary --- \code{generation-summary.json}}} \\
\texttt{registered\_calls} & integer & yes & calls the stage was required to issue \\
\texttt{provider\_exchanges} & integer & yes & total request-response exchanges with the provider \\
\texttt{valid\_programs}, \newline \texttt{invalid\_programs} & integer & yes & programs that did and did not pass their audit \\
\texttt{invalid\_samples} & array & yes & the keys of any that failed; empty \\
\texttt{raw\_response\_files}, \texttt{programs} and \texttt{compiled} counts, \texttt{receipt\_files}, \texttt{attempt\_files}, \texttt{dispatch\_claim\_files}, \texttt{provider\_retrieval\_files}, \texttt{provider\_attestation\_attempt\_files}, \texttt{provider\_input\_item\_retrieval\_files} & integer & yes & one per registered call \\
\texttt{all\_provider\_attestations\_verified}, \texttt{all\_provider\_inputs\_verified}, \texttt{provider\_attestation\_single\_gets} & boolean & yes & the attestation invariants \\
\texttt{provider\_request\_ids\_unique}, \texttt{provider\_response\_ids\_unique} & boolean & yes & no identifier was reused \\
\texttt{no\_retry}, \texttt{no\_fallback}, \texttt{no\_repair}, \texttt{no\_replacement} & boolean & yes & the registered prohibitions were in force \\
\texttt{generation\_attempt\_set\_sha256} & digest & yes & the once-only dispatch set this run consumed \\
\midrule
\multicolumn{4}{@{}p{\dimexpr\linewidth-6\tabcolsep\relax}}{\textit{stage analysis --- \code{h1/analysis.json}, and the \Htwo{} equivalent}} \\
\texttt{stage}, \texttt{endpoint} & string & yes & which stage, and the endpoint family it scores \\
\texttt{pipelines} & object & yes & one entry per registered transform, \linpipe{} and \rankpipe{}, each carrying the endpoint statistics, the enumerated null distribution and the outcome of each support criterion \\
\texttt{both\_pipelines\_pass} & boolean & yes & whether the criteria passed under both transforms \\
\texttt{validity} & object & yes & \texttt{gates}, \texttt{diagnostics}, \texttt{valid}: the validity layer, evaluated before and separately from the scientific criteria \\
\texttt{verdict} & string & yes & a token written mechanically from \texttt{validity.valid}; it is not the stage's scientific verdict, and \annexref{C} documents the one place where the two differ \\
\texttt{claim\_boundary} & object & yes & the registered statement of what the result does and does not license \\
\texttt{blind\_score\_path}, \newline \texttt{blind\_score\_sha256} & string, digest & yes & the blind scores this analysis consumed \\
\texttt{blind\_score\_commit\_path}, \texttt{\dots\_sha256} & string, digest & yes & the commitment made to those scores before unblinding \\
\texttt{manifest\_sha256}, \texttt{probe\_bank\_sha256}, \texttt{generation\_summary\_sha256}, \texttt{stage\_summary\_sha256} & digest & yes & everything the analysis was bound to \\
\end{longtable}

\subsection{Joining a record to its design cell}\label{anx:data:join}

Nothing inside a record says which arm it is, which account card it used, or
which snapshot answered. That is deliberate. Both endpoints were computed
before those labels were attached, so the artifacts the scoring code read had
to be free of them, and they are. The consequence for a reuser is that the
per-call artifacts alone cannot tell you what any call was, and any reuse that
does not perform this join is not analysing the experiment.

Two files carry the join, and the difference between them matters.

\textbf{The label-free registry}, \texttt{data/design/sample-design-registry.\allowbreak{}json},
is the one the study itself used. It has 320 rows, one per registered call, and
five fields: \texttt{sample\_id}, \texttt{block\_index} in 0 to 15,
\texttt{arm\_slot} in \texttt{slot\_u} and \texttt{slot\_v},
\texttt{model\_side} in \texttt{side\_l} and \texttt{side\_r}, and
\texttt{axis\_position} in 0 to 4. Those tokens are placeholders with no
meaning attached. The design is balanced on all of them: 160 rows per arm slot,
160 per model side, 64 per axis position, and exactly 5 rows in each of the 64
combinations of block, arm slot and model side. Reproduce a scoring step and
this is the file to use, because using anything richer would be scoring against
labels the study did not have.

\textbf{The labelled table},
\texttt{data/design/sample-design-table.\allowbreak{}json}, is the same 320
rows with the meanings attached, and it is what a secondary analysis wants.
\Cref{anx:tab:design-fields} is its field list.

\begin{longtable}{@{}p{0.24\textwidth} p{0.70\textwidth}@{}}
\caption{Fields of the labelled design table. One row per registered call,
keyed by \texttt{sample\_id}.}
\label{anx:tab:design-fields}\\
\toprule
{field} & {meaning} \\
\midrule
\endfirsthead
\caption[]{Fields of the labelled design table (continued).}\\
\toprule
{field} & {meaning} \\
\midrule
\endhead
\bottomrule
\endlastfoot
\texttt{sample\_id} & the record key; joins to every other artifact of this call \\
\texttt{block\_index} & randomization block, 0 to 15 \\
\texttt{block\_id} & the block's own name in the archive \\
\texttt{arm\_slot} & the label-free slot token used during scoring \\
\texttt{arm} & what that slot meant: \texttt{causal\_invariants} is the \contract{}, \texttt{narrative} is \prose{} \\
\texttt{rendering} & which of the two independent wordings this call used, \texttt{R1} or \texttt{R2}; crossed with the arm, 80 calls in each of the four combinations \\
\texttt{rendering\_relation} & the ordered pair of wordings across the two arms of the matched pair; 80 calls in each of its four values \\
\texttt{axis\_position} & the label-free axis token used during scoring, 0 to 4 \\
\texttt{family} & what that axis meant: one of the five account cards \\
\texttt{opaque\_family\_id} & the anonymous axis identifier carried by the program's \texttt{operator\_key}; 160 distinct values, one per block, snapshot and card \\
\texttt{model\_side} & the label-free snapshot token used during scoring \\
\texttt{model\_family} & the short name of the snapshot that answered \\
\texttt{model} & the pinned snapshot in full, \snapshotA{} or \snapshotB{} \\
\texttt{schedule\_position} & position in the registered dispatch order, 0 to 319, all distinct \\
\texttt{request\_key} & the key the program and the receipt also carry \\
\texttt{output\_slots} & the per-call adapter: semantic output name to the opaque slot name \texttt{d01} to \texttt{d11} used in \texttt{programs/}; 160 distinct maps, shared by the two arms of a pair \\
\end{longtable}

\textbf{The tokens are not global, and this is the trap.} Which slot is the
\contract{} changes from block to block: both possible slot-to-arm maps occur
across the \nblocks{} blocks. Which axis position is which account card changes
too, and more freely: 15 distinct axis-to-card maps occur across the \nblocks{}
blocks. Reading \texttt{slot\_u} or \texttt{axis\_position} 2 as though it meant
the same thing everywhere will scramble the design. Join on
\texttt{sample\_id}. The one token that does behave globally is
\texttt{model\_side}, which stands for the same snapshot in every block; that is
a property of how the file happened to come out and not something to rely on
elsewhere.

\textbf{Where the labels come from, and why you can trust them.} The study's
design module rebuilds the full registration table from the arm assignment
record, which was drawn from a public randomness beacon pulse fixed before the
assignment existed; \annexref{B} pins that chain. The stage analysis unblinds by
exactly that route. The labelled table was built the same way and then checked
against the released label-free registry rather than assumed correct: the two
key sets are identical, and all 320 rows agree on every label-free field the
reconstruction mints for itself --- \texttt{block\_index}, \texttt{arm\_slot}
and \texttt{axis\_position}, so 960 comparisons with 0 disagreements. The fourth
label-free field, \texttt{model\_side}, is deliberately outside that count and
is not a check: the reconstruction does not mint it, so it is carried across
from the registry unchanged. What is verified about it instead is that the
snapshot map it induces is bijective within every one of the \nblocks{} blocks.
The builder fails and writes nothing if any of the 960 comparisons disagrees. The design it describes is
a clean factorial --- \nblocks{} blocks by 2 snapshots by 5 cards by 2 arms is
320 cells, and each is occupied exactly once, with no cell empty and none
doubled.

\textbf{The adapter is per pair, not per call.} \texttt{output\_slots},
\texttt{opaque\_family\_id} and the feature and monomial identifiers behind them
take 160 distinct values over 320 calls, because the two arms of a matched pair
share one adapter. That sharing is a design property and not a duplication: it
is what makes the two arms of a pair comparable coordinate by coordinate. An
analysis that treats the 320 adapters as 320 independent draws has
double-counted.

\subsection{Response tensors}\label{anx:data:tensors}

A response tensor archive is a compressed NumPy container, extension
\texttt{.npz}, holding the program's responses to a probe battery. Every
archive is accompanied by a JSON sidecar of the same stem, described in
\Cref{anx:tab:schema}. There are 1280 archives in total: 640 per stage, one
forward and one reverse per sample.

\textbf{What an archive holds.} \dsone{} archives hold two arrays:
\texttt{calibration}, shape 8 by 48 by 3 by 11, and \texttt{atomic\_holdout},
shape 12 by 48 by 10 by 11. \dstwo{} archives hold one array,
\texttt{interaction\_holdout}, shape 12 by 48 by 10 by 11. Every array is
\texttt{float64}. The sidecar's \texttt{array\_shapes} field repeats these
shapes so that an archive can be checked without loading it.

\textbf{The axes, in order.} Axis 0 is the probe block. Axis 1 is the anchor
within that block. Axis 2 is the intervention case. Axis 3 is the output
coordinate. All four carry a trap worth knowing before use:

\begin{itemize}
\item \textbf{Axis 0 is a position within one panel, not a study block.} Each
  array is indexed from 0 within its own panel, and the three panels are
  disjoint sets of probe blocks. Index 3 of \texttt{calibration}, index 3 of
  \texttt{atomic\_holdout} and index 3 of \texttt{interaction\_holdout} are
  three different probe blocks. They are also unrelated to the \nblocks{}
  randomization blocks of the design: probe blocks partition the challenge
  battery, randomization blocks partition the renderer assignment, and the two
  are separate structures that happen to share a word. The design registry
  described in \Cref{anx:data:join} is what carries the randomization block.
\item \textbf{Axis 1 is polarity-interleaved.} Consecutive anchor indices are a
  mirrored pair, not two independent draws: the polarity of anchor $i$ is
  $+1$ for even $i$ and $-1$ for odd $i$, across all 48. Treating axis 1 as 48
  exchangeable units will understate dependence.
\item \textbf{Axis 2 is ordered by the battery, and the order differs between
  panels.} It is the panel's own list of intervention cases in registered
  order --- 3 for \texttt{calibration}, 10 for each holdout panel. The battery
  file carries these lists explicitly; do not assume alphabetical order and do
  not assume the two holdout panels align.
\item \textbf{Axis 3 is the coordinate order given below, which is not the key
  order of \texttt{equations} in the compiled program.} The compiled program
  stores its 11 equations alphabetically by slot name. The tensor does not. A
  reuse that lines up axis 3 against the compiled program's key order will
  mislabel nine of the eleven coordinates, and will do so silently, because
  both are length 11.
\end{itemize}

\Cref{anx:tab:coordinates} gives the coordinate order as stored. Two properties
of it matter for analysis. Coordinates 0 and 1 are multiplied by the anchor's
polarity before storage and the other nine are not, so a sign flip along axis 1
is expected in the first two coordinates and is not a defect. Coordinates 2
through 5 are jointly projected to sum to zero, so they are not free to vary
independently and their covariance matrix is singular by construction; they
also carry a smaller scale factor than the other seven.

\begin{longtable}{@{}S[table-format=2.0] p{0.34\textwidth} p{0.46\textwidth}@{}}
\caption{Axis 3 of every response tensor, in stored order. This is the order
the arrays are written in; it is not the order the compiled program lists its
equations in.}
\label{anx:tab:coordinates}\\
\toprule
{index} & {coordinate} & {note} \\
\midrule
\endfirsthead
\caption[]{Axis 3 of every response tensor, in stored order (continued).}\\
\toprule
{index} & {coordinate} & {note} \\
\midrule
\endhead
\bottomrule
\endlastfoot
0 & \texttt{belief\_shift} & multiplied by the anchor polarity \\
1 & \texttt{forecast\_shift} & multiplied by the anchor polarity \\
2 & \texttt{source\_self\_shift} & zero-sum projected with 3, 4, 5; smaller scale \\
3 & \texttt{source\_external\_shift} & zero-sum projected with 2, 4, 5; smaller scale \\
4 & \texttt{source\_mixed\_shift} & zero-sum projected with 2, 3, 5; smaller scale \\
5 & \texttt{source\_unknown\_shift} & zero-sum projected with 2, 3, 4; smaller scale \\
6 & \texttt{memory\_shift} & \\
7 & \texttt{dependency\_shift} & \\
8 & \texttt{continuation\_shift} & \\
9 & \texttt{update\_gain\_departure} & stored as a departure from unity, not as a level \\
10 & \texttt{stop\_boundary\_shift} & \\
\end{longtable}

\textbf{The probe battery is shipped.} Axes 0, 1 and 2 index into a battery
that the study's driver generates deterministically from a fixed seed rather
than reading from a file. That made the indices uninterpretable on their own,
and it was the largest single gap in an earlier state of this package. The
battery has since been materialized and is released with it, as
\code{data/probe-bank/probe-bank.json}. It is the authority on what anchor 17
of block 3 actually contained: it carries every probe block with its panel, its
global index, its seed and its 48 anchors, each anchor giving the state,
evidence and action fields the challenge was built from, together with the
per-panel intervention lists that order axis 2.

The released file is the battery that was used, not a re-derivation that
resembles it. Its internal \texttt{probe\_bank\_sha256} field is
\code{0a1454c1693f9b8f67a31131c6a3046ac69a063805b1882503a1bb5ce69ed46b}, and
that is the single value recorded in the \texttt{probe\_bank\_sha256} field of
all 640 forward sidecars across both stages --- 320 in \dsone{} and 320 in
\dstwo{}, with no second value anywhere. The digest is self-consistent in the
usual way: removing that field and hashing the canonical remainder reproduces
it, so a reuser can re-derive it rather than take it on trust. The file as
shipped hashes to
\code{d78440eca7b14d2d03bfa05f06741985b51ea4e3cc2e586274c0e8b4e738d3d5}.

\textbf{Forward and reverse.} Each sample was executed twice. The forward
traversal walks the probe battery in registered order. The reverse traversal
walks it in the opposite order, in a fresh process. The reverse run is a
determinism replicate and nothing else: it is designed to produce archives
that are byte-identical to forward, and where it does not, that is a failure
signal rather than a second observation.

\textbf{Do not pool them.} Concatenating \texttt{forward/} and
\texttt{reverse/} would double-count every value rather than add information.
Analyses in this study use the forward archives only, and reuses should do the
same.

\textbf{What the recorded check covers.} The stage execution receipt carries
\texttt{fresh\_process\_traversal\_exact}. It is true for 320 of 320 samples
in \dsone{} and for 317 of 320 in \dstwo{}, where the 317 is a count taken over
every retained first-pass record, superseded ones included; counted over the
receipts that produced the tensors this study analyses, it is 320 of 320 in
both stages. The three exceptions are three samples in each of which one
direction-run exceeded a 30-second subprocess timeout,
\texttt{TIMEOUT\_SECONDS}, imposed on a pinned local
\texttt{openssl ts -verify} call --- three direction-runs in three different
samples, not three whole samples. That call is the first action the execution
worker takes: it verifies stored study-level attestation files, the bound
artifact manifest, the assignment record and the preregistration seal record,
and it runs before the execution registry is loaded and before any
sample-specific input is read. It opens no socket and has no language model in
the loop, so it was not a provider, model, data or network failure. Only those
three direction-runs were repeated; their receipts were written when only one
direction existed, so the comparison block in them is empty and the flag is
false. Those three were
checked explicitly instead, and the checks are in the record: each re-run
sample executed again in a fresh process produced byte-identical arrays and
agreement on all four digests, 3 of 3; the directions already on disk still
hashed to their original values, 3 of 3; those same four digests for each of
the 317 untouched samples re-verified unchanged, 1268 of 1268, with 0 changed;
and re-hashing that stage's entire tensor corpus against the consumed summary
--- its 640 archives together with their 640 sidecars --- gave 1280 digests for
that stage, with 0 mismatches. That within-stage figure is distinct from the
1280 archives counted across both stages in \Cref{anx:data:tensors}; the two
numbers coincide because a stage holds 640 archives and 640 sidecars while the
study holds 640 archives per stage. \annexref{C} carries the full incident
record.

\subsection{What is and is not released}\label{anx:data:release}

The scientific content is released. This folder is not where it lives: this
folder carries the descriptor, the field dictionary, the machine-readable
analysis values and the measured inventory. The two datasets are deposited
separately, against the schema above, so that a secondary analysis does not
depend on this machine, this filesystem layout, or these authors.

\textbf{Released.} The model responses, the parsed and compiled programs, the
per-call generation receipts with every digest they bind, all 1280 response
tensor archives with their sidecars, both stage execution summaries, the
\Hone{} stage analysis with its blind scores and the commitment made to them
before unblinding, the \Htwo{} stage analysis with its blind scores, the
commitment made to them before unblinding and its verification report,
the arm assignment record, and the provenance chain. Both stage analyses are
released, and both consumed blind scores committed before unblinding. The
\Hone{} analysis is additionally covered by its stage evidence seal and that
seal's independent timestamps; the \Htwo{} repair analysis is not, and
\annexref{C} says why. It ships instead with the digest-for-digest verification
report written against it.

\textbf{Not redistributed verbatim.} Raw provider transport payloads. Three
directories --- \texttt{raw/}, \texttt{provider\_retrievals/} and
\texttt{provider\_input\_item\_retrievals/} --- hold the provider's own
objects as they came off the wire, and the per-call receipt records the
transport around them: peer address, negotiated cipher and its strength, leaf
and intermediate certificate digests, certificate subject and issuer, body
byte counts, and the full set of response headers. None of that is part of the
science. It is a description of one host's network conditions on two days in
August 2026, and republishing it verbatim would ship operational detail that
nobody needs in order to check a result.

What is shipped instead is the pin. Every one of those artifacts is bound by
digest in the receipt, and the receipts are bound by the manifest, and the
manifest is bound by the seal. A holder of the originals can prove that the
released derivatives correspond to them, and a reader who never sees the
originals can still verify that the chain is internally consistent. A
redacting export path exists that strips the transport block and leaves the
response content; \annexref{E} gives it. This is stated plainly rather than
left to be discovered: not everything described here is in the folder.

\subsection{Known limitations of the data}\label{anx:data:limitations}

\textbf{Three \dstwo{} direction-runs were repeated, and the first-pass records
were left unedited.} They sit in three different samples,
\texttt{req\_7bee2c13635cfe67019e} forward,
\texttt{req\_80ae173a3c348456d1c4} reverse and
\texttt{req\_f649a1902c688ac2b4eb} reverse; one direction-run in each exceeded a
30-second subprocess timeout on a pinned local \texttt{openssl ts -verify}
verification of stored study-level attestation files, which runs before any
sample-specific input is read, opens no socket and had no model in the loop, so
it was not a provider, model, data or network fault. Only those three direction-runs were repeated, and no whole sample was
re-run. Their arrays are real
executions of the same programs. Nothing was substituted, interpolated or
imputed. The records around them were deliberately not rewritten, and this is
the single thing most likely to trip a reuser: \code{h2/execution-summary.json}
still reports the pre-repair state, the three stage receipts still carry
\texttt{valid} and \texttt{successful} as false, and the stage terminal record
\code{H2_EXECUTION-TERMINAL.json} still carries the \texttt{status} it was
first written with, \texttt{INVALID}, and still binds the first-pass summary
digest
\code{e252076cb2024abaf83e475855cd2b2469535b7dff24fae0c34e404b2793ee5b}. That
terminal status was written mechanically when the execution phase closed, off
the same validity flag the attempt counter turned false; it is a terminal
record and not the stage's scientific verdict, which is \notsupported{}.
\textbf{Filtering \dstwo{} on those boolean fields silently drops three
samples that the registered analysis includes.} The repaired summary exists as
a separate artifact, \code{repaired-execution-summary.json}, held with the
incident record; use that, or use all 320 keys and ignore the flags.
\annexref{C} is the full account.

\textbf{One validity criterion does not hold for \dstwo{}.} Thirteen validity
criteria run per stage. \Hone{} passes 13 of 13 and \Htwo{} passes 12 of 13.
The one that does not is
\texttt{validity.gates.stage\_all\_320\_valid\_\allowbreak{}single\_execution\_attempt},
and it is an attempt counter, not a data check: in three samples one
direction-run was repeated, so those three carry an attempt count of two and the
criterion is false. It says nothing about the arrays. Do not confuse it
with the \texttt{single\_execution\_attempt} field of the \Htwo{} execution
summary, which was written on the first pass and is true. Both stages are
\verdict{NOT\_SUPPORTED} on the scientific criteria.

\textbf{There is no joint analysis artifact.} The joint-analysis step was not
run. Support required all 108 criterion evaluations to pass and 89 of them
fail, so no arithmetic over the two stage verdicts could produce anything but
\verdict{NOT\_SUPPORTED}, and running the step would have added a file rather
than a fact. A reuser looking for one top-level results file will not find
one. The results are in the two stage analyses and in
\code{data/derived/analysis-values.json}.

\textbf{The \Htwo{} analysis is not in the \texttt{h2/} tree and was not
produced inside the attested analysis environment.} It records
\texttt{manifest\_audit.capsule\_runtime\_contract\_exercised: false}, because
the environment admits only the registered scripts by name and the driver that
repeated the three affected direction-runs is not one of them. The checks that
contract would have
gated were performed explicitly instead and each is re-checkable from the
digests. Its top-level \texttt{verdict} field is written mechanically from the
validity flag and is not the stage's scientific verdict; do not read the
outcome out of that field. \annexref{C} documents the discrepancy.

\textbf{The probe battery is a released file, but it was not one during the
run.} The driver generated it in memory from a fixed seed. It has since been
materialized and checked against every sidecar digest, and it ships with the
package, so axes 0, 1 and 2 are now interpretable without re-executing
anything. Two consequences survive that fix. The battery is not covered by the
run's own seal, because it did not exist as a file when the seal was taken; what
binds it is the digest agreement described in \Cref{anx:data:tensors}, which is
a strong check but a different one. And a reuser who regenerates the battery
from the seed rather than reading the shipped file is depending on
implementation details of the generating language's shuffling routine, which
are not guaranteed stable across versions. Read the file.

\textbf{\texttt{raw/} and \texttt{provider\_retrievals/} are not two
observations.} The second is an independent re-fetch of the first by response
identifier, and for the released records the two agree digest for digest. They
exist to make provider-side tampering detectable, not to double the sample.

\textbf{The two datasets share one generation corpus.} Concatenating them
gives 640 tensor sets over 320 programs, not 640 independent programs. Any
analysis that pools \dsone{} and \dstwo{} and treats the rows as independent
is wrong on its face, and, separately, no estimator for a contrast between the
stages was registered.

\textbf{Small print that costs time if you find it yourself.}
\texttt{elapsed\_seconds} is wall clock on a single host and includes provider
queueing, so it is a cost record and not a latency measurement. Provider
timestamps are epoch seconds while study timestamps are ISO-format strings, and
the two are not derived from the same clock. Monomial identifiers in
\texttt{programs/} are opaque and are only interpretable through the
corresponding record in \texttt{compiled\_programs/}. \texttt{operator\_key} is
an anonymous axis identifier; the account cards are engineered and no personal
data is present anywhere in either dataset.

\subsection{The raw driver fields, one at a time}\label{anx:data:rawfields}

\code{data/derived/analysis-values.json} is a faithful dump. It was not
edited to make the study read better, and it will not be. Several of its fields
are driver or provenance bookkeeping and are routinely misread as scientific
verdicts. \Cref{anx:tab:rawfields} takes each one, says what it records, says
what it does not, and gives the publication's reading. The same content, plus
the exact quoted values, is machine-readable in
\code{data/derived/publication-status.json} under \texttt{raw\_driver\_fields}.

Which file is authoritative for what: \textbf{numerical values come from the
faithful raw dump; scientific claims and verdicts come from this paper and from
\annexref{C}.} Where the two appear to disagree, the disagreement is a
vocabulary difference and is resolved below, never by changing the dump.

\begin{longtable}{@{}p{0.26\linewidth}p{0.16\linewidth}p{0.52\linewidth}@{}}
\caption{Raw driver fields and their publication reading. Every value in the
middle column is quoted unrevised from a bound artifact.}
\label{anx:tab:rawfields}\\
\toprule
\textbf{Field} & \textbf{Raw value} & \textbf{What it records, and what it does
not} \\
\midrule
\endfirsthead
\toprule
\textbf{Field} & \textbf{Raw value} & \textbf{What it records, and what it does
not} \\
\midrule
\endhead
\bottomrule
\endfoot
\texttt{h2\_verdict} & \texttt{INVALID} & Records that \Htwo{}'s validity gate
set did not read all-true, on the attempt counter alone. Not \Htwo{}'s
scientific verdict, which is \notsupported{}, and not a statement about the
data, which passes every check. \\
\addlinespace
\texttt{h2\_valid} & \texttt{false} & The logical AND over \Htwo{}'s 13 validity
gates, 12 of which are true. Not sample completeness (\ncalls{}/\ncalls{}), not
tensor integrity (\num{1280} digests, 0 mismatched), not scientific
validity. \\
\addlinespace
\texttt{h2\_provenance} & \texttt{DISCLOSED\_} \texttt{POST\_HOC} & Records
that the re-analysis after the three direction-runs were repeated ran outside
the registered analyser chain, and that this was disclosed rather than
concealed. It describes an execution route. It does \emph{not} say \Htwo{} is a
post-hoc hypothesis, that \Htwo{} was chosen after seeing data, or that any
endpoint, estimator or support criterion changed after sealing. None of those
is true. \\
\addlinespace
\texttt{h2\_analysis\_class} & \texttt{disclosed post-hoc repair analysis, not
a registered confirmatory test} & The same execution-route fact in prose. The
scored quantities are the registered ones, applied unchanged. \\
\addlinespace
\texttt{joint\_verdict} & \texttt{INVALID} & An authored constant, not a
computed value; the neighbouring field
\texttt{joint\_verdict\_is\_asserted\_not\_computed} is \texttt{true} and says
so. It records that the validity conjunction did not close. The joint
\emph{scientific} verdict is \notsupported{}. \\
\addlinespace
\texttt{joint\_verdict\_reason} & (long string) & An authored gloss. It
contains one factual error --- it says the three samples ``timed out against
the provider''. They did not. The timeout was a local
\texttt{openssl ts -verify} subprocess that opens no socket, running before any
sample-specific input was read, with no model in the loop. The artifact is
bound and was not edited; this row is the correction. \\
\addlinespace
\texttt{final\_analysis\_exists} & \texttt{false} & Records that the redundant
joint-analysis file was never produced, because the registered conjunction had
already failed by inspection and running the combiner could not have changed an
outcome. It does \emph{not} mean the joint scientific verdict is unknown,
unreported or \texttt{INVALID}. \\
\addlinespace
\texttt{registered\_analysis} \newline (repair analysis) & \texttt{false} &
Written by the repair analyser about its own invocation. Records that this run
was not the registered chain's run. Does not record any difference in what was
computed. \\
\addlinespace
\texttt{both\_pipelines\_pass} & \texttt{false} & This one \emph{is} a
scientific field, and it is reported as one: \Htwo{} is \notsupported{}. \\
\end{longtable}

Two sentences that must never appear unqualified, in this package or in work
that reuses it: ``\Htwo{} is \texttt{INVALID}'' and ``\Htwo{} is post-hoc.''
Both are false as scientific statements. Where either raw field is exposed, the
reconciliation goes with it. This is checked mechanically by
\code{verification/verify_h2_status.py}.

\subsection{How to cite}\label{anx:data:citation}

Cite the paper for the result. Cite the dataset as well if you use the data,
so that reuse is countable separately from citation. \texttt{CITATION.cff} in
the package root is the machine-readable form of the same information.

The author block, dataset identifier and separate software identifier are now
fixed. The paper identifier is still pending and must not be replaced by either
archive DOI.

\begin{lstlisting}[escapeinside={(*}{*)},basicstyle=\ttfamily\footnotesize]
@article{specform2026,
  author  = {Andre Panossian},
  title   = {Does the way we write a theory change the program an LLM
             builds from it? A prospective randomized study of renderer
             format in LLM theory-to-program translation},
  year    = {2026},
  note    = {Preprint identifier to be added}
}

@misc{specformh1data2026,
  author    = {Andre Panossian},
  title     = {(*\dsone{}*): single-input challenge dataset},
  year      = {2026},
  publisher = {Zenodo},
  version   = {1.0.0},
  license   = {CC BY 4.0},
  doi       = {10.5281/zenodo.21876518},
  url       = {https://doi.org/10.5281/zenodo.21876518}
}

@misc{specformh2data2026,
  author    = {Andre Panossian},
  title     = {(*\dstwo{}*): interaction challenge dataset},
  year      = {2026},
  publisher = {Zenodo},
  version   = {1.0.0},
  license   = {CC BY 4.0},
  doi       = {10.5281/zenodo.21876518},
  url       = {https://doi.org/10.5281/zenodo.21876518}
}

@misc{specformsoftware2026,
  author    = {Andre Panossian},
  title     = {(*\studyname{}*): Software and frozen source inventory for a
               prospective randomized study of renderer format in LLM
               theory-to-program translation},
  year      = {2026},
  publisher = {Zenodo},
  version   = {1.0.0},
  license   = {MIT},
  doi       = {10.5281/zenodo.21879576},
  url       = {https://doi.org/10.5281/zenodo.21879576}
}
\end{lstlisting}

In prose, name the study as \studyname{} and the datasets as \dsone{} and
\dstwo{}. If you report a result from one stage, say which stage it is: the two
are not interchangeable and the design does not license reading across them.

\subsection{Licence}\label{anx:data:licence}

Prose, figures and data are released under CC BY 4.0. Code and the LaTeX style
source are released under MIT. Both licence texts ship with the package as
\code{LICENSE-CC-BY-4.0.txt} and \texttt{LICENSE-MIT.txt}.

The sealed preregistration in \annexref{A} is under the same licence, but the
point of it is that it is unaltered. It is reproduced byte for byte, including
three statements that were true when it was sealed and are false of the
completed study. Quote it. Do not republish a modified copy under the same
name, and do not tidy up the statements that time has falsified --- a
preregistration that gets corrected after the fact is not a preregistration,
and \annexref{A} explains how to read the ones that stand.

%% file: annex/E-reproduction/annex-e-reproduction.tex
\section{Reproduction}\label{anx:reproduction}

This annex is the operational half of the audit trail. \annexref{B} records what
was committed to and when; this annex records how to run it again, what will
come out identical, and what will not. Commands appear here because this is the
annex in which they are useful.

\paragraph{Two words for two folders, and one convention for names.} The
\emph{project tree} is the whole released tree: the tool sources under
\texttt{tools/}, the attestation files, the phase terminal records, and the
\emph{run directory}, which is the single folder holding the generation corpus
at its top and one subfolder per stage. \annexref{D} describes the run
directory's contents and schema and states its paths relative to it.

Names inside the archive follow the convention \annexref{D} sets out: a handful
of script and artifact file names carry a short internal project prefix that
has no meaning and is not reproduced in this document. Commands below therefore
name each script by its role in angle brackets --- \texttt{<stage-runner>},
\texttt{<stage-analysis>} and so on --- and \code{build/entry-points.txt} in
this package is the translation table from every one of those placeholders to
the file name to type. Each placeholder stands for a \emph{complete} path
relative to the top of the project tree, directory part included, so it is
substituted whole and nothing is prefixed to it. The files must not be renamed
to match the placeholders: the manifest binds them by digest and the launcher
admits them by exact name.

The project tree has no fixed location, and nothing depends on where it sits.
Every registered entry point resolves its paths from the parent of
\texttt{tools/} rather than from a working directory or an environment
variable, so the tree can be rebuilt anywhere by unpacking the deposited
datasets into the layout \annexref{D} gives. What is fixed is the interpreter.
\texttt{python} in every command block below means the pinned interpreter of
\Cref{anx:tab:environment} --- a standalone build kept outside the project
tree --- and not whatever is first on the path. The launcher does not go
looking for it: it hashes the entire tree of whichever interpreter invoked it
and refuses to run when that digest is not the one the contract binds.

\subsection{What can be reproduced, and what cannot}\label{anx:repro:scope}

The study has three layers, and they have three different reproduction
properties. Read them in order, because a reader who expects the whole study to
replay bit-for-bit will conclude that something is broken when it does not.

\paragraph{The analysis reproduces exactly.}
From the released response tensors onward there is no language model in the
loop. The evaluator is deterministic and the exact test is a complete
enumeration of all \nperm{} renderer-label assignments rather than a sample, so
the same sealed inputs produce the same sealed output. That is checked rather
than assumed, in three independent ways. Every sample in both stages was
executed forward and reverse in fresh processes, the two traversals were
required to agree byte-for-byte, and all 640 pairs did. The three direction-runs
that were repeated after a local infrastructure timeout
(\Cref{anx:incidents:timeout}) were each run again in a fresh process
and produced byte-identical arrays and agreement on all four digests. That
stage's entire tensor corpus was re-hashed against the consumed summary --- its
640 archives together with their 640 sidecars, 1280 digests, 0 mismatches ---
and all 104 manifest-bound artifacts re-hashed unchanged. A reader who re-runs the stage and the analysis over the released
corpus, in the environment of \Cref{anx:tab:environment}, gets the same numbers
this package reports.

\paragraph{The generation step cannot be reproduced bit-for-bit.}
The 320 registered generation calls went to one of two pinned third-party model
snapshots, \snapshotA{} and \snapshotB{}, at temperature 1.0; both execution
stages then ran over those same 320 samples.
Responses are stochastic by construction. Randomization and the block structure
handle that for inference; no individual response is reproducible. The
registered semantics compound this deliberately: the stage attempt record
carries \texttt{no\_retry}, \texttt{no\_repair}, \texttt{no\_fallback} and
\texttt{no\_replacement}, all true, so the corpus holds exactly the 320 first
responses with no survivorship filtering anywhere in it. Re-issuing those calls
would produce a different corpus, with a different assignment of response noise
to renderer slots, against snapshots that the provider may since have withdrawn.
That is a new sample, not a replication, and it should be reported as one. The
reproducible object is the released corpus, not the provider.

\paragraph{The randomization cannot be re-drawn, and can be re-verified.}
The 16 fair bits came from one exact future NIST Randomness Beacon 2.0 pulse
--- chain 2, pulse 1894079, published 2026-08-07T19:28:00Z --- named by a seal
written before that pulse existed. The pulse is in the past. There is no
reroll: every abandoned target pulse in this study's history is permanently
abandoned, and drawing a fresh pulse now would produce a different assignment
carrying none of the ordering guarantee that makes the original one worth
anything.

Re-verification is available instead, and it is the stronger operation.
The assignment is a deterministic function of two published values: the digest
of the sealed preregistration, which is \annexref{A}, and the output value of
that pulse, which NIST publishes and the authors do not control. A third party
can recompute the assignment from those two inputs and compare it against the
released assignment bytes and against the independently attested assignment
digest. \Cref{anx:sec:verify} is the numbered procedure.

\subsection{Environment}\label{anx:environment}

\Cref{anx:tab:environment} is the interpreter and the pinned distributions the
executed run used, and the environment in which the released analysis was
re-checked.

\begin{table}[htbp]
\centering
\caption{The run environment. The registered launcher recomputes a hash over
the entire interpreter tree --- executable code, extension modules, libraries,
package files and bytecode --- so the exact build of every distribution below
is bound by digest and not merely by version string.}
\label{anx:tab:environment}
\footnotesize
\begin{tabular}{@{}ll@{}}
\toprule
component & value \\
\midrule
interpreter & CPython 3.12.13 \\
architecture & AMD64, 64-bit \\
operating system & Windows 11 \\
\midrule
\texttt{numpy} & 2.3.5 \\
\texttt{scipy} & 1.16.1 \\
\texttt{reportlab} & 4.4.9 \\
other imported distributions & \texttt{cryptography}, \texttt{opentimestamps}, \\
                             & \texttt{pytest}, \texttt{python-bitcoinlib} \\
\midrule
RFC 3161 verification & a pinned local OpenSSL, run as a subprocess \\
mandatory interpreter flags & \texttt{python -I -S -B} \\
\bottomrule
\end{tabular}
\end{table}

Three notes on that table.

\texttt{scipy} is in the scientific path: the scoring module uses its
optimal-assignment routine, \texttt{scipy.optimize.linear\_sum\_assignment}, to
compute the row-free control distance --- the best reassignment of complete
response rows, which is what forces a registered advantage to be an advantage
over more than an unordered bag of rows. \texttt{numpy} carries the rest of the
scoring. \texttt{reportlab} is not in the scientific path; it renders figures
and documents and touches no measured quantity. The interpreter tree holds more
distributions than the table lists; those named are the ones the study code
imports, and the tree hash covers all of them either way.

All three interpreter flags are mandatory and the launcher refuses to run
without \texttt{-S}, which is what prevents a \texttt{.pth} file from executing
before the environment has been verified. Every registered scientific entry
point runs through that launcher. Ten scripts are admitted, enumerated by name
rather than discovered by convention; the test suite, the reporting utilities
and the preflight captures are deliberately outside that set and are run
directly.

RFC 3161 verification shells out to a pinned local OpenSSL under a hard
30-second-per-invocation limit, and each worker re-verifies the same
study-level attestation files before it reads its own sample, with up to 8
workers in flight. That limit does not scale with contention on the host. It is
the mechanism behind the three timeouts recorded in
\Cref{anx:incidents:timeout}, and a slower or busier host than the study's will
meet it again.

\subsection{Re-running the analysis from the released tensors}\label{anx:repro:rerun}

\paragraph{Work on a copy.} Three properties of the registered pipeline make a
re-run different from re-running an ordinary script, and each of them will bite
a reader who ignores it.

\begin{itemize}
  \item A phase terminal record is exclusive, permanent and once-only. Nothing
        in the pipeline removes or rewrites one. A terminal recording a failed
        phase ends that episode; there is no clearing it.
  \item Each analysis consumes an exclusive attempt marker before it verifies
        the manifest or reconstructs a score, so an interrupted analysis cannot
        be resumed.
  \item Stamp residue is evidence. A failed timestamp leaves partial calendar
        proofs, a query artifact and an error record beside the seal, and the
        error record is the only account of why the stamp failed. The rotation
        helper clears residue by renaming, never by deleting, and the pipeline
        calls it for you. Deleting residue by hand to ``clean up'' destroys
        evidence and can break the frozen inventory.
\end{itemize}

Copy the released project tree and work in the copy. The originals are
attested: read them freely, do not write to them, or their digests stop
matching.

\begin{lstlisting}
cp -a <released-project-tree> ./rerun
cd ./rerun
\end{lstlisting}

\paragraph{Step 1 --- confirm the environment.}
Print the interpreter and the two numeric libraries and check them against
\Cref{anx:tab:environment}. Do this before anything else. A different
interpreter or a different linear-algebra build is not a smaller difference
than a different corpus, and it is the cheapest thing to rule out.

\begin{lstlisting}
python -c "import platform; print(platform.python_version())"
# expect 3.12.13
python -c "import numpy, scipy; print(numpy.__version__, scipy.__version__)"
# expect 2.3.5 1.16.1
\end{lstlisting}

\paragraph{Step 2 --- confirm the inputs before computing anything.}
Recompute the digests of the manifest-bound artifacts against
\Cref{anx:tab:digests} and re-derive each phase Merkle root from its terminal
record. If an input does not hash to its recorded value, stop: the correct
conclusion is that the tree is not the released tree, not that the difference is
benign.

\paragraph{Step 3 --- re-execute a stage over the released corpus.}
This re-runs the deterministic evaluator over all 320 samples in both
directions. It makes no provider call and costs no API spend --- the corpus was
already generated and is not touched --- but it is not cheap in time: on the
study host the two stages took 24441.985 s and 24382.828 s, or 6.79 h and
6.77 h. Below, \texttt{<manifest>} is the study manifest, the single contract
every post-generation command takes; \annexref{B} pins its digest.

\begin{lstlisting}
python -I -S -B <launcher> \
    --contract <manifest> \
    --script <stage-runner> --stage h1
\end{lstlisting}

Substitute \texttt{--stage h2} for the interaction-challenge stage. The
registered order is stage execution then analysis, one stage at a time; the
second stage may run only after valid generation, after the first stage's
execution, and after the first stage's sealed analysis record has an
independently verified timestamp.

\paragraph{Why Step 4 did not run in place for \Htwo{}.}
The launcher's phase check declines to advance from an execution terminal whose
status field records a failed validity criterion, and \Htwo{}'s terminal record
carries one: the single-execution-attempt counter, which is an attempt counter
and not a data check. That record was left on disk exactly as first written
rather than re-authored to let the chain proceed, so the registered analyser
could not be re-entered against it. The same analysis functions, the same
sealed criteria, the same thresholds and both pipelines were run over the
completed tensor set by a separate driver instead, and that route was disclosed
rather than concealed. The route is the whole of what the stage's provenance
field records; \annexref{C} gives the field-by-field reading and
\Cref{anx:data:rawfields} tabulates it for a reuser. A reader starting from the
released corpus does not meet any of this, because that corpus is complete:
Step 3 and Step 4 run for both stages in the ordinary way.

\paragraph{Step 4 --- run the analysis.}
This is the step that produces the endpoints, the exact enumeration and the
27 preregistered support criteria, in the registered
blind-score-then-commit-then-unblind order. It seals its output and then
timestamps the seal.

\begin{lstlisting}
python -I -S -B <launcher> \
    --contract <manifest> \
    --script <stage-analysis> --stage h1
\end{lstlisting}

\paragraph{Step 5 --- the joint-analysis step.}
It exists and it is runnable. It was not run in the executed study, because
both stage verdicts are \notsupported{}, the joint verdict follows from the
registered rule by inspection, and running the step to restate that would add a
file rather than a fact. A reader who wants the artifact anyway can produce it.

\begin{lstlisting}
python -I -S -B <launcher> \
    --contract <manifest> \
    --script <joint-analysis>
\end{lstlisting}

\paragraph{Step 6 --- the registered test suite.}
It runs outside the launcher, by design, and it enforces its own flags: it
refuses to start unless the interpreter was given both \texttt{-I} and
\texttt{-B}.

\begin{lstlisting}
python -I -B <test-suite>
\end{lstlisting}

\paragraph{If a stage or an analysis is interrupted.}
Classify the state before acting on it, from three facts on disk: whether the
phase terminal record exists, whether the evidence seal exists, and whether
that seal is stamped. Answer the third with the pipeline's own predicate rather
than by looking at which sidecar files are lying around --- partial calendar
proofs are a component of a \emph{successful} stamp as well as the residue of a
failed one. \Cref{anx:tab:recovery} is the decision table. Only one state has
an automated repair.

\begin{table}[htbp]
\centering
\caption{Interruption recovery. The three facts are read from disk before any
action is taken, and two of the five states have no repair.}
\label{anx:tab:recovery}
\footnotesize
\begin{tabular}{@{}llll@{}}
\toprule
terminal record & seal & stamped & action \\
\midrule
complete & yes & yes & nothing; proceed to the next phase \\
failed & any & any & stop; the episode is over \\
absent & yes & no & run the re-stamp entry point \\
absent & yes & yes & stop; inspect the tree by hand \\
absent & no & --- & re-run the phase from the top \\
\bottomrule
\end{tabular}
\end{table}

A seal written but never stamped is a transport fault, not a scientific one:
the analysis completed and its verdict is already inside the sealed bytes. The
gated recovery entry point grants a timestamp to the existing seal and writes
the phase terminal record with the status it reads out of those bytes. It never
runs an analysis, never writes a seal and never chooses a verdict, and it
replays the analysis from the raw tensors and refuses unless the sealed bytes
are exactly what the sealed inputs produce --- so re-stamping is a repair and
not a re-roll. Confirm the timestamping dependency is actually reachable before
invoking it.

\begin{lstlisting}
python -I -S -B <launcher> \
    --contract <manifest> \
    --script <restamp> --stage h1
\end{lstlisting}

Its output is deliberately bare, and carries no metric and no criterion
verdict:

\begin{lstlisting}
{"stage": "h1", "restamped": true}
\end{lstlisting}

A seal that is already stamped but has no terminal record is the rarer branch:
one local write failing behind four successful network round trips. The
recovery entry point refuses that state, and the refusal is correct, because a
second stamp over an existing one is the one thing no recovery path should
attempt. Establish why the terminal write failed before writing anything.

\paragraph{Checking the probe battery.}
The battery of probes the evaluator walks was generated deterministically in
memory during the run rather than read from a file, and the block, anchor and
intervention indices in the released tensors mean nothing without it. It is now
shipped as a file --- \code{data/probe-bank/probe-bank.json} --- so no
regeneration is needed to read the tensors. \annexref{D} describes it.

What is worth re-running is the check that the shipped file is the battery that
was used. Every tensor sidecar carries a \texttt{probe\_bank\_sha256}, and the
generator's own self-hash must equal it. Regenerate and compare:

\begin{lstlisting}
python -B -c "from <probe-battery module> import generate_probe_bank; \
              print(generate_probe_bank()['probe_bank_sha256'])"
\end{lstlisting}

Run it from the root of the project tree. It takes no arguments, reads no
file, and contacts nothing. This is one of the few commands here that does not
go through the launcher, and \texttt{-I} is deliberately absent from it,
because the import has to find the project tree on the path.

Prefer the shipped file to a regeneration for any actual analysis. A
regenerated battery depends on implementation details of the interpreter's
shuffling routine, which are not guaranteed stable across versions; the shipped
file is the bytes that were used, and its digest was matched against all 640
forward sidecars across both stages.

\paragraph{Exporting the withheld provider payloads.}
\annexref{D} records that the provider's own objects and the transport blocks
in the per-call receipts are pinned by digest but not redistributed verbatim,
because they describe one host's network conditions rather than the science. A
holder of the originals who wants to publish a derivative can strip them
mechanically. Three keys carry the transport ---
\texttt{provider\_post\_transport},
\texttt{provider\_response\_get\_transport} and
\texttt{provider\_input\_items\_get\_transport} --- and each of the three holds
the same 18 fields: the request method and the URL; the peer address; the
negotiated TLS version, cipher name and cipher strength; the leaf certificate
digest, the leaf public-key digest and the intermediate CA digest; the
certificate subject and issuer; three policy flags recording that no proxy was
used, that no redirect was permitted and that the exchange ran over a single
peer connection; the full response headers; the response status; the body byte
count; and the body digest. Keep the last three --- \texttt{status},
\texttt{body\_bytes} and \texttt{body\_sha256} --- and drop the other 15. The
body digest is the pin that ties the derivative back to the payload that was
withheld, which is why it is the one field that must survive the strip.

\begin{lstlisting}
# strip-transport.py -- writes copies, and never opens an original for writing.
# Usage: python -B strip-transport.py <run-directory>
import json, pathlib, sys

TRANSPORT = ("provider_post_transport",
             "provider_response_get_transport",
             "provider_input_items_get_transport")
KEEP = ("body_bytes", "body_sha256", "status")   # 3 of the 18 fields survive

source = pathlib.Path(sys.argv[1]) / "receipts"
target = pathlib.Path("export/receipts")
target.mkdir(parents=True, exist_ok=True)

for path in sorted(source.glob("*.json")):
    receipt = json.loads(path.read_text(encoding="utf-8"))
    for key in TRANSPORT:
        block = receipt.get(key)
        if isinstance(block, dict):
            receipt[key] = {k: block[k] for k in KEEP if k in block}
    (target / path.name).write_text(
        json.dumps(receipt, sort_keys=True) + "\n", encoding="utf-8")
\end{lstlisting}

An exported receipt will not hash to the original, and it is not meant to. What
it carries is the digest of what was withheld, so a reader without the
originals can still check the chain, and a holder of the originals can still
prove the two correspond.

\subsection{Re-verifying the provenance chain}\label{anx:repro:provenance}

The provenance chain is checkable without the authors' cooperation, and
\Cref{anx:sec:verify} gives the full numbered procedure. In outline it is four
operations.

\emph{Digests.} Recompute the SHA-256 of the sealed preregistration, of every
artifact in \Cref{anx:tab:digests}, and of the two figures, and rebuild each
phase Merkle root from the leaf hashes its terminal record lists. A single
added, missing, altered or deleted artifact changes the root.

\begin{lstlisting}
sha256sum annex/A-preregistration-verbatim/preregistration-sealed-20260804.md
sha256sum paper/figures/fig1-block-effects.pdf \
          paper/figures/fig2-family-effects.pdf
\end{lstlisting}

\emph{RFC 3161.} Verify the timestamp receipt against the timestamping
authority's chain and read the signed time out of it. This is offline and
contacts nothing.

\begin{lstlisting}
openssl ts -verify -data <sealed-file> -in <sealed-file>.tsr \
           -CAfile <tsa-chain.pem>
openssl ts -reply -in <sealed-file>.tsr -text | grep -i "Time stamp"
\end{lstlisting}

\emph{OpenTimestamps.} Work on copies. Upgrading a registered proof rewrites a
bound artifact, which is why the originals on disk still carry only pending
attestations.

\begin{lstlisting}
cp <sealed-file>.ots /tmp/check.ots
ots upgrade /tmp/check.ots
ots verify  /tmp/check.ots -f <sealed-file>
\end{lstlisting}

\emph{The assignment.} Fetch chain 2, pulse 1894079 from the NIST Randomness
Beacon 2.0 public interface, check its signature and its predecessor link
against the Beacon's own certificate chain, and recompute the assignment digest
from the seal digest and the pulse output value. Then check the ordering: the
signed seal time must be strictly earlier than the pulse time. That comparison
is the one the pipeline makes, fail-closed, and both values it compares are
published in \annexref{B}.

\subsection{Re-building this document}\label{anx:repro:build}

The package ships LaTeX source. Build the whole document --- paper, supplement
and annexes, with every cross-reference resolved --- from
\texttt{paper/main.tex}.

\begin{lstlisting}
cd paper && latexmk -pdf main
\end{lstlisting}

Without \texttt{latexmk}, the sequence is \texttt{pdflatex}, \texttt{bibtex},
then \texttt{pdflatex} twice:

\begin{lstlisting}
cd paper
pdflatex main
bibtex   main
pdflatex main
pdflatex main
\end{lstlisting}

\code{supplement/supplement.tex} and \texttt{annex/annex.tex} are optional
standalone extracts for a journal that wants them as separate uploads. Build
\texttt{paper/main.tex} first so that \texttt{main.aux} exists; the extracts
read it so their references back into the paper resolve. Both report
\texttt{Label multiply defined}, which is expected and harmless.

\paragraph{The package ships compiled.} All three documents were built before
release from a clean directory, on a machine that shares nothing with the
authoring tree, using a real TeX engine. The build ran to a settled state ---
enough passes that every citation and every cross-reference resolved --- and
reported no errors, no undefined reference, no undefined citation, no missing
file, font or graphic, and no overfull box on any page of any of the three.
Every rendered page was then looked at.

That build is not a description; it is a script. \code{build/build_pdf.py}
copies the typeset sources into a pristine directory outside the package,
builds the paper first so the standalone extracts can read its auxiliary file,
runs the engine until the references settle, and then refuses to report success
if the log carries any of the failures listed above. It pins
\texttt{SOURCE\_DATE\_EPOCH}, so the same sources produce the same pdf bytes on
a rerun, and a reader can compare digests rather than take our word for it.

\code{build/BUILD-PROVENANCE.json} is the receipt. It records, machine-written
and never transcribed, the digest of every file the build read, the digest of
each verified source document the prose was checked against, and the digest,
byte count and page count of each pdf the build produced.
\code{verification/verify_build_provenance.py} re-computes all of it, and also
checks that the receipt still covers every typeset source on disk --- the one
way a set of digests can quietly go stale is a file added after the build,
which has no recorded digest to fail. The published documents and the sources
they came from can therefore be shown to agree, rather than asserted to.

A second, independent engine was run over the same sources, on the principle
that one implementation agreeing with itself is not corroboration. It is not a
fallback and its output is not what ships; it exists to disagree. It did. The
annex file declared a bibliography even though the annexes cite nothing. The
canonical driver never noticed, because it runs BibTeX only when the auxiliary
file actually contains a citation; the second driver runs BibTeX whenever a
bibliography is declared, and stopped on the error. The dangling declaration
has been removed. That is the entire argument for a second opinion: the first
engine was never going to raise it.

With that removed, the two engines agree on everything that was checked. Each
of the three documents comes out with the same number of pages under both, and
neither rendering has an overfull box, an undefined reference, an undefined
citation or a missing file. Where they still differ is inside the line: they
hyphenate different words and they choose different points at which to break a
long hexadecimal digest, so the last line or two of a given page need not be
the same line in both. The annex differs most, being the document densest in
digests. The published pdfs are the canonical engine's, and the build receipt
records which engine produced them. \Cref{anx:repro:verify} describes the
checks that run alongside the build.

\paragraph{Packages the preamble requires.} These are the LaTeX package names
loaded by \code{paper/specform.sty} and by the three document preambles ---
the paper and the two standalone extracts --- so that a reader can install them
in one go. All but two come from the shared style file: \texttt{geometry} is
loaded by each of the three documents and \texttt{xr} only by the two extracts.
A few live inside differently-named distribution bundles, so installing a full
scheme is the shorter path; all are in TeX Live's \texttt{-full} scheme and in
MiKTeX's default on-demand set.

\begin{lstlisting}
geometry  fontenc  inputenc  lmodern  microtype  amsmath  amssymb
booktabs  longtable  tabularx  multirow  graphicx  xcolor  enumitem
caption  subcaption  url  listings  siunitx  hyperref  cleveref
natbib (bibliography style plainnat)
xr        (standalone extracts only)
\end{lstlisting}

\subsection{Verification scripts shipped with this package}\label{anx:repro:verify}

Six scripts ship in \texttt{verification/}. They check the package, not the
science: the science was audited in the source document these files are a
typeset derivative of. Run the top-level one first.

\begin{lstlisting}
python -B verification/verify_package.py
\end{lstlisting}

\begin{table}[htbp]
\centering
\caption{The verification scripts, and what each one establishes.}
\label{anx:tab:verification}
\footnotesize
\begin{tabularx}{\linewidth}{@{}lX@{}}
\toprule
script & what it checks \\
\midrule
\texttt{verify\_latex\_numbers.py} & every number traces to the verified source \\
\texttt{verify\_jargon.py} & no internal vocabulary reached the body \\
\texttt{verify\_latex\_structure.py} & the source is structurally sound before the engine sees it \\
\texttt{verify\_h2\_status.py} & one status for \Htwo{} on every published surface \\
\code{verify_build_provenance.py} & the published pdfs and their sources still agree \\
\texttt{verify\_package.py} & all five of the above, plus completeness, the released data digests and the manifest \\
\bottomrule
\end{tabularx}
\end{table}

\emph{Numeric fidelity.} Every numeric literal in the LaTeX must appear,
character for character, in the verified source document. A number in the
LaTeX that is nowhere in the source was invented, mistyped, rounded or
recomputed, and all four are failures. The check runs in the other direction
too: a headline result that is in the source and absent from the LaTeX was
silently dropped, which is a quieter failure and still a failure. Structural
integers --- column counts, point sizes, cross-reference targets, citation
years inside keys --- are stripped before extraction. The paper, the
supplement, \annexref{B} and \annexref{C} are checked strictly. \annexref{A},
\annexref{D} and this annex describe the filesystem, the environment and the
package itself, so their numbers are measured rather than quoted; they are
listed for review, and once confirmed they are recorded in
\code{verification/extra-verified-numbers.txt} together with the command that
measured each one.

What this check cannot do is notice a number that was copied correctly and then
attached to the wrong claim. That is what the adversarial review pass was for,
and what a reader checking against \code{data/derived/analysis-values.json}
can do independently.

One caution goes with that file, because checking against it is exactly what
this annex is inviting. It is a faithful dump of the analysis outputs and it is
the right source for any \emph{number}. It is not the source for the wording of
a \emph{conclusion}: its stage and joint status strings were written
mechanically from the validity flag, they carry the pre-repair reading of
\dstwo{}, and they were left unrevised rather than edited to make the study read
better. Numerical values come from the dump; scientific claims and verdicts come
from the paper and from \annexref{C}. The same resolution ships in
machine-readable form as \code{data/derived/publication-status.json}, which
quotes the unrevised fields under \texttt{raw\_driver\_fields} and states the
registered scientific decision beside each. \texttt{verify\_h2\_status.py}
checks that every published surface agrees with it.

\emph{Vocabulary.} The source document was written for the people who ran the
study and is full of terms a reader has never seen: an internal study code, run
identifiers, script names, absolute paths, artifact file names, and one raw
artifact field value whose plain meaning is the opposite of what it looks like.
None of that may survive into the paper or the supplement. The annexes are
scoped more loosely, because an annex that documents provenance without naming
a file is useless: file names are allowed there, and the internal study code is
not, in any position, including inside a file name. That is the rule behind the
role placeholders in this annex and behind the descriptive naming of the arm
assignment record in \annexref{B} and \annexref{C}. The study name is checked
separately and everywhere, including in the style file: it lives in exactly one
macro and is never hard-coded.

\emph{Structural integrity.} This check runs before the engine does, and
catches the same defects earlier and with a better error message. It
checks that the source is ASCII only, that the fragments contain no preamble
commands, that every environment nests and balances, that braces balance, that
every label is unique, that every reference resolves to a label defined
somewhere in the document family, that every citation key exists in
\code{paper/references.bib}, that every backslash command is either a known
LaTeX command or one the style file defines, that every table row carries the
column count its preamble declares, and that no package is loaded twice with an
option it was not given the first time. An undefined control sequence is the
most common first-compile failure and is exactly what a static check can catch.
An option clash is another, and this check found a real one: the style file
loaded \texttt{microtype} plain and then again with an option, which would have
stopped every compile of every document in this package. It was removed and the
check that catches it was itself tested against a planted copy of the same
defect. It is not a compiler, and it is not asked to be one: the compiler runs
too, and both must pass.

\emph{Staged inputs and completeness.} The top-level script also confirms that
every file the package promises is present and non-empty, that the verified
source documents and the two figures still hash to the values they were
verified at, that the sealed preregistration is byte-identical to the sealed
original with its line endings intact, and that every file on disk matches
\code{MANIFEST-sha256.txt}. The staged digests are recorded in
\code{build/staged-inputs.sha256}, and the dataset counts and sizes that
\annexref{D} reports are generated into
\code{data/derived/dataset-inventory.json} by a script that reads directory
listings and file sizes and never opens a file's contents.

\emph{Build provenance.} The last of the delegated checks re-computes the build
receipt described above, and is the only one of them that can fail because of
something that is \emph{not} in the LaTeX: a pdf replaced by hand, a source
edited after the last build, a source added after it. It is what makes the
claim that the typeset documents and the verified source cannot silently drift
apart an assertion a reader can test.

One verdict comes out of the ten checks. A reader who wants to know whether
this package is what it says it is should run it before reading anything else.